\documentclass[longauth]{aa}

\usepackage{graphicx}
\usepackage{txfonts}
\usepackage{hyperref}
\usepackage{natbib,twoopt}
\usepackage{amsmath}
\usepackage{amssymb}
\usepackage{longtable}
\usepackage{booktabs}
\usepackage{multirow}

\usepackage{color}
\usepackage{tabularx}
\usepackage{mathtools}
\usepackage{nicefrac}
\usepackage{physics}
\usepackage{url}
\usepackage{epstopdf}

\usepackage{textcomp}
\usepackage{gensymb}
\usepackage{xspace}
\usepackage{relsize}
\usepackage{latexsym,amsfonts}
\usepackage[T1]{fontenc}
\usepackage{placeins}
\usepackage{afterpage}

\usepackage{orcidlink}
\let\orcid\orcidlink
\makeatletter
\renewcommand*\maketitle{
  \thispagestyle{firstpage}
\begingroup
    \if@wideboxfn
    \setlength\bibindent{1.4\parindent}
    \else
    \setlength\bibindent{\parindent}
    \fi
    \renewcommand*\thefootnote{\@fnsymbol\c@footnote}
    \renewcommand\@makefntext[1]{
    \ifaa@longfn\hsize\textwidth\fi
    \noindent
    \hb@xt@\bibindent{\hss\@makefnmark\enspace}##1}
  \ifaa@twocolumn
  \begin{aa@strip}
    \aa@maketitle
    \@thanks
  \end{aa@strip}
  \else
    \begingroup
      \let\thanks\footnote
      \aa@maketitle
    \endgroup
  \fi
\endgroup
  \setcounter{footnote}{0}
}

\makeatother

\usepackage[encapsulated]{CJK}
\usepackage{ucs}

\usepackage[utf8]{inputenc}

\newcommand{\cntext}[1]{}

\bibpunct{(}{)}{;}{a}{}{,}

\def\sgra{Sgr~A$^*$\xspace}
\def\m87{M87$^*$\xspace}

\DeclareGraphicsExtensions{.pdf,.png,.jpeg}
\begin{document}

\title{Full-polarization millimeter wavelength variability of Sagittarius~A$^*$ during the 2018 EHT campaign}

\author{\small
Ezequiel Albentosa-Ruiz \orcid{0000-0002-7816-6401} \inst{\ref{4}} \and
Jasmin E. Washington \orcid{0000-0002-7046-0470} \inst{\ref{14}} \and
Nicola Marchili \orcid{0000-0002-5523-7588} \inst{\ref{110}, \ref{6}} \and
Iván Martí-Vidal \orcid{0000-0003-3708-9611} \inst{\ref{4}, \ref{12}} \and
Ciriaco Goddi \orcid{0000-0002-2542-7743} \inst{\ref{31}, \ref{71}, \ref{72}, \ref{73}} \and
Maciek Wielgus \orcid{0000-0002-8635-4242} \inst{\ref{5}} \and
Alejandro Mus \orcid{0000-0003-0329-6874} \inst{\ref{71}, \ref{110}} \and
Angelo Ricarte \orcid{0000-0001-5287-0452} \inst{\ref{3}, \ref{10}} \and
Daniel P. Marrone \orcid{0000-0002-2367-1080} \inst{\ref{14}} \and
León D. S. Salas \orcid{0000-0003-1979-6363} \inst{\ref{117}} \and
Yuhei Iwata \orcid{0000-0002-9255-4742}\inst{\ref{81}}\and
Douglas F. Carlos \orcid{0000-0002-1340-7702} \inst{\ref{31}} \and
Alexandra J. Tetarenko \orcid{0000-0003-3906-4354} \inst{\ref{150}} \and
Kotaro Moriyama \orcid{0000-0003-1364-3761} \inst{\ref{47}, \ref{81}} \and
Vedant Dhruv \orcid{0000-0001-6765-877X} \inst{\ref{17}} \and
Kazunori Akiyama \orcid{0000-0002-9475-4254} \inst{\ref{1}, \ref{2}, \ref{3}} \and
Antxon Alberdi \orcid{0000-0002-9371-1033} \inst{\ref{5}} \and
Walter Alef \inst{\ref{6}} \and
Juan Carlos Algaba \orcid{0000-0001-6993-1696} \inst{\ref{7}} \and
Richard Anantua \orcid{0000-0003-3457-7660} \inst{\ref{8}, \ref{9}, \ref{3}, \ref{10}} \and
Keiichi Asada \orcid{0000-0001-6988-8763} \inst{\ref{11}} \and
Rebecca Azulay \orcid{0000-0002-2200-5393} \inst{\ref{4}, \ref{12}, \ref{6}} \and
Uwe Bach \orcid{0000-0002-7722-8412} \inst{\ref{6}} \and
Anne-Kathrin Baczko \orcid{0000-0003-3090-3975} \inst{\ref{13}, \ref{6}} \and
David Ball \inst{\ref{14}} \and
Mislav Baloković \orcid{0000-0003-0476-6647} \inst{\ref{15}} \and
Bidisha Bandyopadhyay \orcid{0000-0002-2138-8564} \inst{\ref{16}} \and
John Barrett \orcid{0000-0002-9290-0764} \inst{\ref{1}} \and
Michi Bauböck \orcid{0000-0002-5518-2812} \inst{\ref{17}} \and
Bradford A. Benson \orcid{0000-0002-5108-6823} \inst{\ref{18}, \ref{19}} \and
Dan Bintley \inst{\ref{20}, \ref{21}} \and
Lindy Blackburn \orcid{0000-0002-9030-642X} \inst{\ref{3}, \ref{10}} \and
Raymond Blundell \orcid{0000-0002-5929-5857} \inst{\ref{10}} \and
Katherine L. Bouman \orcid{0000-0003-0077-4367} \inst{\ref{22}} \and
Geoffrey C. Bower \orcid{0000-0003-4056-9982} \inst{\ref{20}, \ref{21}, \ref{39}, \ref{24}} \and
Michael Bremer \inst{\ref{25}} \and
Roger Brissenden \orcid{0000-0002-2556-0894} \inst{\ref{3}, \ref{10}} \and
Silke Britzen \orcid{0000-0001-9240-6734} \inst{\ref{6}} \and
Avery E. Broderick \orcid{0000-0002-3351-760X} \inst{\ref{26}, \ref{27}, \ref{28}} \and
Dominique Broguiere \orcid{0000-0001-9151-6683} \inst{\ref{25}} \and
Thomas Bronzwaer \orcid{0000-0003-1151-3971} \inst{\ref{29}} \and
Sandra Bustamante \orcid{0000-0001-6169-1894} \inst{\ref{30}} \and
John E. Carlstrom \orcid{0000-0002-2044-7665} \inst{\ref{32}, \ref{19}, \ref{33}, \ref{34}} \and
Andrew Chael \orcid{0000-0003-2966-6220} \inst{\ref{35}} \and
Chi-kwan Chan \orcid{0000-0001-6337-6126} \inst{\ref{14}, \ref{36}, \ref{37}} \and
Dominic O. Chang \orcid{0000-0001-9939-5257} \inst{\ref{3}, \ref{10}} \and
Koushik Chatterjee \orcid{0000-0002-2825-3590} \inst{\ref{3}, \ref{10}} \and
Shami Chatterjee \orcid{0000-0002-2878-1502} \inst{\ref{38}} \and
Ming-Tang Chen \orcid{0000-0001-6573-3318} \inst{\ref{39}} \and
Yongjun Chen (\cntext{陈永军}) \orcid{0000-0001-5650-6770} \inst{\ref{40}, \ref{41}} \and
Xiaopeng Cheng \orcid{0000-0003-4407-9868} \inst{\ref{42}} \and
Pierre Christian \orcid{0000-0001-6820-9941} \inst{\ref{44}} \and
Ilje Cho \orcid{0000-0001-6083-7521} \inst{\ref{42}, \ref{43}, \ref{5}} \and
Nicholas S. Conroy \orcid{0000-0003-2886-2377} \inst{\ref{45}, \ref{10}} \and
John E. Conway \orcid{0000-0003-2448-9181} \inst{\ref{13}} \and
Thomas M. Crawford \orcid{0000-0001-9000-5013} \inst{\ref{19}, \ref{32}} \and
Geoffrey B. Crew \orcid{0000-0002-2079-3189} \inst{\ref{1}} \and
Alejandro Cruz-Osorio \orcid{0000-0002-3945-6342} \inst{\ref{46}, \ref{47}} \and
Yuzhu Cui (\cntext{崔玉竹}) \orcid{0000-0001-6311-4345} \inst{\ref{48}} \and
Brandon Curd \orcid{0000-0002-8650-0879} \inst{\ref{8}, \ref{3}, \ref{10}} \and
Rohan Dahale \orcid{0000-0001-6982-9034} \inst{\ref{5}} \and
Jordy Davelaar \orcid{0000-0002-2685-2434} \inst{\ref{49}, \ref{50}} \and
Mariafelicia De Laurentis \orcid{0000-0002-9945-682X} \inst{\ref{51}, \ref{52}} \and
Roger Deane \orcid{0000-0003-1027-5043} \inst{\ref{53}, \ref{54}, \ref{55}} \and
Jessica Dempsey \orcid{0000-0003-1269-9667} \inst{\ref{20}, \ref{21}, \ref{56}} \and
Gregory Desvignes \orcid{0000-0003-3922-4055} \inst{\ref{6}, \ref{57}} \and
Jason Dexter \orcid{0000-0003-3903-0373} \inst{\ref{58}} \and
Indu K. Dihingia \orcid{0000-0002-4064-0446} \inst{\ref{59}} \and
Sheperd S. Doeleman \orcid{0000-0002-9031-0904} \inst{\ref{3}, \ref{10}} \and
Sergio A. Dzib \orcid{0000-0001-6010-6200} \inst{\ref{6}} \and
Ralph P. Eatough \orcid{0000-0001-6196-4135} \inst{\ref{60}, \ref{6}} \and
Razieh Emami \orcid{0000-0002-2791-5011} \inst{\ref{10}} \and
Heino Falcke \orcid{0000-0002-2526-6724} \inst{\ref{29}} \and
Joseph Farah \orcid{0000-0003-4914-5625} \inst{\ref{61}, \ref{62}} \and
Vincent L. Fish \orcid{0000-0002-7128-9345} \inst{\ref{1}} \and
Edward Fomalont \orcid{0000-0002-9036-2747} \inst{\ref{63}} \and
H. Alyson Ford \orcid{0000-0002-9797-0972} \inst{\ref{14}} \and
Marianna Foschi \orcid{0000-0001-8147-4993} \inst{\ref{5}} \and
Antonio Fuentes \orcid{0000-0002-8773-4933} \inst{\ref{5}} \and
Raquel Fraga-Encinas \orcid{0000-0002-5222-1361} \inst{\ref{29}} \and
William T. Freeman \inst{\ref{64}, \ref{65}} \and
Per Friberg \orcid{0000-0002-8010-8454} \inst{\ref{20}, \ref{21}} \and
Christian M. Fromm \orcid{0000-0002-1827-1656} \inst{\ref{66}, \ref{47}, \ref{6}} \and
Peter Galison \orcid{0000-0002-6429-3872} \inst{\ref{3}, \ref{67}, \ref{68}} \and
Charles F. Gammie \orcid{0000-0001-7451-8935} \inst{\ref{17}, \ref{45}, \ref{69}} \and
Roberto García \orcid{0000-0002-6584-7443} \inst{\ref{25}} \and
Olivier Gentaz \orcid{0000-0002-0115-4605} \inst{\ref{25}} \and
Gertie Geertsema \orcid{0000-0003-3933-0069} \inst{\ref{70}} \and
Boris Georgiev \orcid{0000-0002-3586-6424} \inst{\ref{14}} \and
Roman Gold \orcid{0000-0003-2492-1966} \inst{\ref{74}, \ref{75}, \ref{76}} \and
José L. Gómez \orcid{0000-0003-4190-7613} \inst{\ref{5}} \and
Arturo I. Gómez-Ruiz \orcid{0000-0001-9395-1670} \inst{\ref{77}, \ref{78}} \and
Minfeng Gu (\cntext{顾敏峰}) \orcid{0000-0002-4455-6946} \inst{\ref{40}, \ref{79}} \and
Mark Gurwell \orcid{0000-0003-0685-3621} \inst{\ref{10}} \and
Kazuhiro Hada \orcid{0000-0001-6906-772X} \inst{\ref{80}, \ref{81}} \and
Daryl Haggard \orcid{0000-0001-6803-2138} \inst{\ref{82}, \ref{83}} \and
Ronald Hesper \orcid{0000-0003-1918-6098} \inst{\ref{84}} \and
Dirk Heumann \orcid{0000-0002-7671-0047} \inst{\ref{14}} \and
Luis C. Ho (\cntext{何子山}) \orcid{0000-0001-6947-5846} \inst{\ref{85}, \ref{86}} \and
Paul Ho \orcid{0000-0002-3412-4306} \inst{\ref{11}, \ref{21}, \ref{20}} \and
Mareki Honma \orcid{0000-0003-4058-9000} \inst{\ref{81}, \ref{87}, \ref{88}} \and
Chih-Wei L. Huang \orcid{0000-0001-5641-3953} \inst{\ref{11}} \and
Lei Huang (\cntext{黄磊}) \orcid{0000-0002-1923-227X} \inst{\ref{40}, \ref{79}} \and
David H. Hughes \inst{\ref{77}} \and
Shiro Ikeda \orcid{0000-0002-2462-1448} \inst{\ref{2}, \ref{89}, \ref{90}, \ref{91}} \and
C. M. Violette Impellizzeri \orcid{0000-0002-3443-2472} \inst{\ref{92}, \ref{63}} \and
Makoto Inoue \orcid{0000-0001-5037-3989} \inst{\ref{11}} \and
Sara Issaoun \orcid{0000-0002-5297-921X} \inst{\ref{10}, \ref{50}} \and
David J. James \orcid{0000-0001-5160-4486} \inst{\ref{93}, \ref{94}} \and
Buell T. Jannuzi \orcid{0000-0002-1578-6582} \inst{\ref{14}} \and
Michael Janssen \orcid{0000-0001-8685-6544} \inst{\ref{29}, \ref{6}} \and
Britton Jeter \orcid{0000-0003-2847-1712} \inst{\ref{11}} \and
Wu Jiang (\cntext{江悟}) \orcid{0000-0001-7369-3539} \inst{\ref{40}} \and
Alejandra Jiménez-Rosales \orcid{0000-0002-2662-3754} \inst{\ref{29}} \and
Michael D. Johnson \orcid{0000-0002-4120-3029} \inst{\ref{3}, \ref{10}} \and
Adam C. Jones \inst{\ref{19}} \and
Svetlana Jorstad \orcid{0000-0001-6158-1708} \inst{\ref{95}} \and
Abhishek V. Joshi \orcid{0000-0002-2514-5965} \inst{\ref{17}} \and
Taehyun Jung \orcid{0000-0001-7003-8643} \inst{\ref{42}, \ref{96}} \and
Ramesh Karuppusamy \orcid{0000-0002-5307-2919} \inst{\ref{6}} \and
Tomohisa Kawashima \orcid{0000-0001-8527-0496} \inst{\ref{97}} \and
Garrett K. Keating \orcid{0000-0002-3490-146X} \inst{\ref{10}} \and
Mark Kettenis \orcid{0000-0002-6156-5617} \inst{\ref{98}} \and
Dong-Jin Kim \orcid{0000-0002-7038-2118} \inst{\ref{99}} \and
Jae-Young Kim \orcid{0000-0001-8229-7183} \inst{\ref{100}, \ref{6}} \and
Jongsoo Kim \orcid{0000-0002-1229-0426} \inst{\ref{42}} \and
Junhan Kim \orcid{0000-0002-4274-9373} \inst{\ref{101}} \and
Motoki Kino \orcid{0000-0002-2709-7338} \inst{\ref{2}, \ref{102}} \and
Jun Yi Koay \orcid{0000-0002-7029-6658} \inst{\ref{11}} \and
Prashant Kocherlakota \orcid{0000-0001-7386-7439} \inst{\ref{47}} \and
Yutaro Kofuji \inst{\ref{81}, \ref{88}} \and
Patrick M. Koch \orcid{0000-0003-2777-5861} \inst{\ref{11}} \and
Shoko Koyama \orcid{0000-0002-3723-3372} \inst{\ref{103}, \ref{11}} \and
Carsten Kramer \orcid{0000-0002-4908-4925} \inst{\ref{25}} \and
Joana A. Kramer \orcid{0009-0003-3011-0454} \inst{\ref{6}} \and
Michael Kramer \orcid{0000-0002-4175-2271} \inst{\ref{6}} \and
Thomas P. Krichbaum \orcid{0000-0002-4892-9586} \inst{\ref{6}} \and
Cheng-Yu Kuo \orcid{0000-0001-6211-5581} \inst{\ref{104}, \ref{11}} \and
Noemi La Bella \orcid{0000-0002-8116-9427} \inst{\ref{29}} \and
Sang-Sung Lee \orcid{0000-0002-6269-594X} \inst{\ref{42}} \and
Aviad Levis \orcid{0000-0001-7307-632X} \inst{\ref{22}} \and
Zhiyuan Li (\cntext{李志远}) \orcid{0000-0003-0355-6437} \inst{\ref{105}, \ref{106}} \and
Rocco Lico \orcid{0000-0001-7361-2460} \inst{\ref{110}, \ref{5}} \and
Greg Lindahl \orcid{0000-0002-6100-4772} \inst{\ref{108}} \and
Michael Lindqvist \orcid{0000-0002-3669-0715} \inst{\ref{13}} \and
Mikhail Lisakov \orcid{0000-0001-6088-3819} \inst{\ref{109}} \and
Jun Liu (\cntext{刘俊}) \orcid{0000-0002-7615-7499} \inst{\ref{6}} \and
Kuo Liu \orcid{0000-0002-2953-7376} \inst{\ref{40}, \ref{41}} \and
Elisabetta Liuzzo \orcid{0000-0003-0995-5201} \inst{\ref{110}} \and
Wen-Ping Lo \orcid{0000-0003-1869-2503} \inst{\ref{11}, \ref{111}} \and
Andrei P. Lobanov \orcid{0000-0003-1622-1484} \inst{\ref{6}} \and
Laurent Loinard \orcid{0000-0002-5635-3345} \inst{\ref{112}, \ref{3}, \ref{113}} \and
Colin J. Lonsdale \orcid{0000-0003-4062-4654} \inst{\ref{1}} \and
Amy E. Lowitz \orcid{0000-0002-4747-4276} \inst{\ref{14}} \and
Ru-Sen Lu (\cntext{路如森}) \orcid{0000-0002-7692-7967} \inst{\ref{40}, \ref{41}, \ref{6}} \and
Nicholas R. MacDonald \orcid{0000-0002-6684-8691} \inst{\ref{6}} \and
Jirong Mao (\cntext{毛基荣}) \orcid{0000-0002-7077-7195} \inst{\ref{114}, \ref{115}, \ref{116}} \and
Sera Markoff \orcid{0000-0001-9564-0876} \inst{\ref{117}, \ref{118}} \and
Alan P. Marscher \orcid{0000-0001-7396-3332} \inst{\ref{95}} \and
Satoki Matsushita \orcid{0000-0002-2127-7880} \inst{\ref{11}} \and
Lynn D. Matthews \orcid{0000-0002-3728-8082} \inst{\ref{1}} \and
Lia Medeiros \orcid{0000-0003-2342-6728} \inst{\ref{49}, \ref{50}} \and
Karl M. Menten \orcid{0000-0001-6459-0669} \inst{\ref{6}} \dag \and
Izumi Mizuno \orcid{0000-0002-7210-6264} \inst{\ref{20}, \ref{21}} \and
Yosuke Mizuno \orcid{0000-0002-8131-6730} \inst{\ref{59}, \ref{120}, \ref{47}} \and
Joshua Montgomery \orcid{0000-0003-0345-8386} \inst{\ref{83}, \ref{19}} \and
James M. Moran \orcid{0000-0002-3882-4414} \inst{\ref{3}, \ref{10}} \and
Monika Moscibrodzka \orcid{0000-0002-4661-6332} \inst{\ref{29}} \and
Wanga Mulaudzi \orcid{0000-0003-4514-625X} \inst{\ref{117}} \and
Hendrik Müller \orcid{0000-0002-9250-0197} \inst{\ref{6}} \and
Cornelia Müller \orcid{0000-0002-2739-2994} \inst{\ref{6}, \ref{29}} \and
Gibwa Musoke \orcid{0000-0003-1984-189X} \inst{\ref{117}, \ref{29}} \and
Ioannis Myserlis \orcid{0000-0003-3025-9497} \inst{\ref{121}} \and
Hiroshi Nagai \orcid{0000-0003-0292-3645} \inst{\ref{2}, \ref{87}} \and
Neil M. Nagar \orcid{0000-0001-6920-662X} \inst{\ref{16}} \and
Dhanya G. Nair \orcid{0000-0001-5357-7805} \inst{\ref{16}, \ref{6}} \and
Masanori Nakamura \orcid{0000-0001-6081-2420} \inst{\ref{122}, \ref{11}} \and
Gopal Narayanan \orcid{0000-0002-4723-6569} \inst{\ref{30}} \and
Iniyan Natarajan \orcid{0000-0001-8242-4373} \inst{\ref{10}, \ref{3}} \and
Antonios Nathanail \orcid{0000-0002-1655-9912} \inst{\ref{123}, \ref{47}} \and
Santiago Navarro Fuentes \inst{\ref{121}} \and
Joey Neilsen \orcid{0000-0002-8247-786X} \inst{\ref{124}} \and
Chunchong Ni \orcid{0000-0003-1361-5699} \inst{\ref{27}, \ref{28}, \ref{26}} \and
Michael A. Nowak \orcid{0000-0001-6923-1315} \inst{\ref{125}} \and
Junghwan Oh \orcid{0000-0002-4991-9638} \inst{\ref{98}} \and
Hiroki Okino \orcid{0000-0003-3779-2016} \inst{\ref{81}, \ref{88}} \and
Héctor Raúl Olivares Sánchez \orcid{0000-0001-6833-7580} \inst{\ref{126}} \and
Tomoaki Oyama \orcid{0000-0003-4046-2923} \inst{\ref{81}} \and
Feryal Özel \orcid{0000-0003-4413-1523} \inst{\ref{127}} \and
Daniel C. M. Palumbo \orcid{0000-0002-7179-3816} \inst{\ref{3}, \ref{10}} \and
Georgios Filippos Paraschos \orcid{0000-0001-6757-3098} \inst{\ref{6}} \and
Jongho Park \orcid{0000-0001-6558-9053} \inst{\ref{128}, \ref{11}} \and
Harriet Parsons \orcid{0000-0002-6327-3423} \inst{\ref{20}, \ref{21}} \and
Nimesh Patel \orcid{0000-0002-6021-9421} \inst{\ref{10}} \and
Ue-Li Pen \orcid{0000-0003-2155-9578} \inst{\ref{11}, \ref{26}, \ref{129}, \ref{130}, \ref{131}} \and
Dominic W. Pesce \orcid{0000-0002-5278-9221} \inst{\ref{10}, \ref{3}} \and
Vincent Piétu \inst{\ref{25}} \and
Aleksandar PopStefanija \inst{\ref{30}} \and
Oliver Porth \orcid{0000-0002-4584-2557} \inst{\ref{117}, \ref{47}} \and
Ben Prather \orcid{0000-0002-0393-7734} \inst{\ref{17}} \and
Giacomo Principe \orcid{0000-0003-0406-7387} \inst{\ref{132}, \ref{133}, \ref{110}} \and
Dimitrios Psaltis \orcid{0000-0003-1035-3240} \inst{\ref{127}} \and
Hung-Yi Pu \orcid{0000-0001-9270-8812} \inst{\ref{134}, \ref{135}, \ref{11}} \and
Venkatessh Ramakrishnan \orcid{0000-0002-9248-086X} \inst{\ref{16}, \ref{136}, \ref{137}} \and
Ramprasad Rao \orcid{0000-0002-1407-7944} \inst{\ref{10}} \and
Mark G. Rawlings \orcid{0000-0002-6529-202X} \inst{\ref{138}, \ref{20}, \ref{21}} \and
Luciano Rezzolla \orcid{0000-0002-1330-7103} \inst{\ref{47}, \ref{139}, \ref{140}} \and
Bart Ripperda \orcid{0000-0002-7301-3908} \inst{\ref{129}, \ref{141}, \ref{130}, \ref{26}} \and
Jan Röder \orcid{0000-0002-2426-927X} \inst{\ref{5}} \and
Freek Roelofs \orcid{0000-0001-5461-3687} \inst{\ref{29}} \and
Cristina Romero-Cañizales \orcid{0000-0001-6301-9073} \inst{\ref{11}} \and
Eduardo Ros \orcid{0000-0001-9503-4892} \inst{\ref{6}} \and
Arash Roshanineshat \orcid{0000-0002-8280-9238} \inst{\ref{14}} \and
Helge Rottmann \inst{\ref{6}} \and
Alan L. Roy \orcid{0000-0002-1931-0135} \inst{\ref{6}} \and
Ignacio Ruiz \orcid{0000-0002-0965-5463} \inst{\ref{121}} \and
Chet Ruszczyk \orcid{0000-0001-7278-9707} \inst{\ref{1}} \and
Kazi L. J. Rygl \orcid{0000-0003-4146-9043} \inst{\ref{110}} \and
Salvador Sánchez \orcid{0000-0002-8042-5951} \inst{\ref{121}} \and
David Sánchez-Argüelles \orcid{0000-0002-7344-9920} \inst{\ref{77}, \ref{78}} \and
Miguel Sánchez-Portal \orcid{0000-0003-0981-9664} \inst{\ref{121}} \and
Mahito Sasada \orcid{0000-0001-5946-9960} \inst{\ref{142}, \ref{81}, \ref{143}} \and
Kaushik Satapathy \orcid{0000-0003-0433-3585} \inst{\ref{14}} \and
Saurabh \orcid{0000-0001-7156-4848} \inst{\ref{6}} \and
Tuomas Savolainen \orcid{0000-0001-6214-1085} \inst{\ref{144}, \ref{137}, \ref{6}} \and
F. Peter Schloerb \inst{\ref{30}} \and
Jonathan Schonfeld \orcid{0000-0002-8909-2401} \inst{\ref{10}} \and
Karl-Friedrich Schuster \orcid{0000-0003-2890-9454} \inst{\ref{25}} \and
Lijing Shao \orcid{0000-0002-1334-8853} \inst{\ref{86}, \ref{6}} \and
Zhiqiang Shen (\cntext{沈志强}) \orcid{0000-0003-3540-8746} \inst{\ref{40}, \ref{41}} \and
Sasikumar Silpa \orcid{0000-0003-0667-7074} \inst{\ref{16}} \and
Des Small \orcid{0000-0003-3723-5404} \inst{\ref{98}} \and
Bong Won Sohn \orcid{0000-0002-4148-8378} \inst{\ref{42}, \ref{96}, \ref{43}} \and
Jason SooHoo \orcid{0000-0003-1938-0720} \inst{\ref{1}} \and
Kamal Souccar \orcid{0000-0001-7915-5272} \inst{\ref{30}} \and
Joshua S. Stanway \orcid{0009-0003-7659-4642} \inst{\ref{146}} \and
He Sun (\cntext{孙赫}) \orcid{0000-0003-1526-6787} \inst{\ref{147}, \ref{148}} \and
Fumie Tazaki \orcid{0000-0003-0236-0600} \inst{\ref{149}} \and
Paul Tiede \orcid{0000-0003-3826-5648} \inst{\ref{10}, \ref{3}} \and
Remo P. J. Tilanus \orcid{0000-0002-6514-553X} \inst{\ref{14}, \ref{29}, \ref{92}, \ref{151}} \and
Michael Titus \orcid{0000-0001-9001-3275} \inst{\ref{1}} \and
Kenji Toma \orcid{0000-0002-7114-6010} \inst{\ref{152}, \ref{153}} \and
Pablo Torne \orcid{0000-0001-8700-6058} \inst{\ref{121}, \ref{6}} \and
Teresa Toscano \orcid{0000-0003-3658-7862} \inst{\ref{5}} \and
Efthalia Traianou \orcid{0000-0002-1209-6500} \inst{\ref{5}, \ref{6}} \and
Tyler Trent \inst{\ref{14}} \and
Sascha Trippe \orcid{0000-0003-0465-1559} \inst{\ref{154}, \ref{155}} \and
Matthew Turk \orcid{0000-0002-5294-0198} \inst{\ref{45}} \and
Ilse van Bemmel \orcid{0000-0001-5473-2950} \inst{\ref{56}} \and
Huib Jan van Langevelde \orcid{0000-0002-0230-5946} \inst{\ref{98}, \ref{92}, \ref{156}} \and
Daniel R. van Rossum \orcid{0000-0001-7772-6131} \inst{\ref{29}} \and
Jesse Vos \orcid{0000-0003-3349-7394} \inst{\ref{29}} \and
Jan Wagner \orcid{0000-0003-1105-6109} \inst{\ref{6}} \and
Derek Ward-Thompson \orcid{0000-0003-1140-2761} \inst{\ref{146}} \and
John Wardle \orcid{0000-0002-8960-2942} \inst{\ref{157}} \and
Jonathan Weintroub \orcid{0000-0002-4603-5204} \inst{\ref{3}, \ref{10}} \and
Robert Wharton \orcid{0000-0002-7416-5209} \inst{\ref{6}} \and
Kaj Wiik \orcid{0000-0002-0862-3398} \inst{\ref{158}, \ref{136}, \ref{137}} \and
Gunther Witzel \orcid{0000-0003-2618-797X} \inst{\ref{6}} \and
Michael F. Wondrak \orcid{0000-0002-6894-1072} \inst{\ref{29}, \ref{159}} \and
George N. Wong \orcid{0000-0001-6952-2147} \inst{\ref{160}, \ref{35}} \and
Qingwen Wu (\cntext{吴庆文}) \orcid{0000-0003-4773-4987} \inst{\ref{161}} \and
Nitika Yadlapalli \orcid{0000-0003-3255-4617} \inst{\ref{22}} \and
Paul Yamaguchi \orcid{0000-0002-6017-8199} \inst{\ref{10}} \and
Aristomenis Yfantis \orcid{0000-0002-3244-7072} \inst{\ref{29}} \and
Doosoo Yoon \orcid{0000-0001-8694-8166} \inst{\ref{117}} \and
André Young \orcid{0000-0003-0000-2682} \inst{\ref{29}} \and
Ziri Younsi \orcid{0000-0001-9283-1191} \inst{\ref{162}, \ref{47}} \and
Wei Yu (\cntext{于威}) \orcid{0000-0002-5168-6052} \inst{\ref{10}} \and
Feng Yuan (\cntext{袁峰}) \orcid{0000-0003-3564-6437} \inst{\ref{163}} \and
Ye-Fei Yuan (\cntext{袁业飞}) \orcid{0000-0002-7330-4756} \inst{\ref{164}} \and
Ai-Ling Zeng (\cntext{曾艾玲}) \orcid{0009-0000-9427-4608} \inst{\ref{5}} \and
J. Anton Zensus \orcid{0000-0001-7470-3321} \inst{\ref{6}} \and
Shuo Zhang \orcid{0000-0002-2967-790X} \inst{\ref{165}} \and
Guang-Yao Zhao \orcid{0000-0002-4417-1659} \inst{\ref{5}, \ref{6}} \and
Shan-Shan Zhao (\cntext{赵杉杉}) \orcid{0000-0002-9774-3606} \inst{\ref{40}}
}
\institute{\small
Departament d'Astronomia i Astrofísica, Universitat de València, C. Dr. Moliner 50, E-46100 Burjassot, València, Spain \label{4} \and
Steward Observatory and Department of Astronomy, University of Arizona, 933 N. Cherry Ave., Tucson, AZ 85721, USA \label{14} \and
INAF-Istituto di Radioastronomia \& Italian ALMA Regional Centre, Via P. Gobetti 101, I-40129 Bologna, Italy \label{110} \and
Max-Planck-Institut für Radioastronomie, Auf dem Hügel 69, D-53121 Bonn, Germany \label{6} \and
Observatori Astronòmic, Universitat de València, C. Catedrático José Beltrán 2, E-46980 Paterna, València, Spain \label{12} \and
Instituto de Astronomia, Geofísica e Ciências Atmosféricas, Universidade de São Paulo, R. do Matão, 1226, São Paulo, SP 05508-090, Brazil \label{31} \and
Dipartimento di Fisica, Università degli Studi di Cagliari, SP Monserrato-Sestu km 0.7, I-09042 Monserrato (CA), Italy \label{71} \and
INAF - Osservatorio Astronomico di Cagliari, via della Scienza 5, I-09047 Selargius (CA), Italy \label{72} \and
INFN, sezione di Cagliari, I-09042 Monserrato (CA), Italy \label{73} \and
Instituto de Astrofísica de Andalucía-CSIC, Glorieta de la Astronomía s/n, E-18008 Granada, Spain \label{5} \and
Black Hole Initiative at Harvard University, 20 Garden Street, Cambridge, MA 02138, USA \label{3} \and
Center for Astrophysics $|$ Harvard \& Smithsonian, 60 Garden Street, Cambridge, MA 02138, USA \label{10} \and
Anton Pannekoek Institute for Astronomy, University of Amsterdam, Science Park 904, 1098 XH, Amsterdam, The Netherlands \label{117} \and
Mizusawa VLBI Observatory, National Astronomical Observatory of Japan, 2-12 Hoshigaoka, Mizusawa, Oshu, Iwate 023-0861, Japan \label{81} \and
Department of Physics and Astronomy, University of Lethbridge, Lethbridge, Alberta T1K 3M4, Canada \label{150} \and
Institut für Theoretische Physik, Goethe-Universität Frankfurt, Max-von-Laue-Straße 1, D-60438 Frankfurt am Main, Germany \label{47} \and
Department of Physics, University of Illinois, 1110 West Green Street, Urbana, IL 61801, USA \label{17} \and
Massachusetts Institute of Technology Haystack Observatory, 99 Millstone Road, Westford, MA 01886, USA \label{1} \and
National Astronomical Observatory of Japan, 2-21-1 Osawa, Mitaka, Tokyo 181-8588, Japan \label{2} \and
Department of Physics, Faculty of Science, Universiti Malaya, 50603 Kuala Lumpur, Malaysia \label{7} \and
Department of Physics \& Astronomy, The University of Texas at San Antonio, One UTSA Circle, San Antonio, TX 78249, USA \label{8} \and
Physics \& Astronomy Department, Rice University, Houston, TX 77005-1827, USA \label{9} \and
Institute of Astronomy and Astrophysics, Academia Sinica, 11F of Astronomy-Mathematics Building, AS/NTU No. 1, Sec. 4, Roosevelt Rd., Taipei 106216, Taiwan, R.O.C. \label{11} \and
Department of Space, Earth and Environment, Chalmers University of Technology, Onsala Space Observatory, SE-43992 Onsala, Sweden \label{13} \and
Yale Center for Astronomy \& Astrophysics, Yale University, 52 Hillhouse Avenue, New Haven, CT 06511, USA \label{15} \and
Astronomy Department, Universidad de Concepción, Casilla 160-C, Concepción, Chile \label{16} \and
Fermi National Accelerator Laboratory, MS209, P.O. Box 500, Batavia, IL 60510, USA \label{18} \and
Department of Astronomy and Astrophysics, University of Chicago, 5640 South Ellis Avenue, Chicago, IL 60637, USA \label{19} \and
East Asian Observatory, 660 N. A'ohoku Place, Hilo, HI 96720, USA \label{20} \and
James Clerk Maxwell Telescope (JCMT), 660 N. A'ohoku Place, Hilo, HI 96720, USA \label{21} \and
California Institute of Technology, 1200 East California Boulevard, Pasadena, CA 91125, USA \label{22} \and
Department of Physics and Astronomy, University of Hawaii at Manoa, 2505 Correa Road, Honolulu, HI 96822, USA \label{24} \and
Institut de Radioastronomie Millimétrique (IRAM), 300 rue de la Piscine, F-38406 Saint Martin d'Hères, France \label{25} \and
Perimeter Institute for Theoretical Physics, 31 Caroline Street North, Waterloo, ON N2L 2Y5, Canada \label{26} \and
Department of Physics and Astronomy, University of Waterloo, 200 University Avenue West, Waterloo, ON N2L 3G1, Canada \label{27} \and
Waterloo Centre for Astrophysics, University of Waterloo, Waterloo, ON N2L 3G1, Canada \label{28} \and
Department of Astrophysics, Institute for Mathematics, Astrophysics and Particle Physics (IMAPP), Radboud University, P.O. Box 9010, 6500 GL Nijmegen, The Netherlands \label{29} \and
Department of Astronomy, University of Massachusetts, Amherst, MA 01003, USA \label{30} \and
Kavli Institute for Cosmological Physics, University of Chicago, 5640 South Ellis Avenue, Chicago, IL 60637, USA \label{32} \and
Department of Physics, University of Chicago, 5720 South Ellis Avenue, Chicago, IL 60637, USA \label{33} \and
Enrico Fermi Institute, University of Chicago, 5640 South Ellis Avenue, Chicago, IL 60637, USA \label{34} \and
Princeton Gravity Initiative, Jadwin Hall, Princeton University, Princeton, NJ 08544, USA \label{35} \and
Data Science Institute, University of Arizona, 1230 N. Cherry Ave., Tucson, AZ 85721, USA \label{36} \and
Program in Applied Mathematics, University of Arizona, 617 N. Santa Rita, Tucson, AZ 85721, USA \label{37} \and
Cornell Center for Astrophysics and Planetary Science, Cornell University, Ithaca, NY 14853, USA \label{38} \and
Institute of Astronomy and Astrophysics, Academia Sinica, 645 N. A'ohoku Place, Hilo, HI 96720, USA \label{39} \and
Shanghai Astronomical Observatory, Chinese Academy of Sciences, 80 Nandan Road, Shanghai 200030, PR China \label{40} \and
Key Laboratory of Radio Astronomy and Technology, Chinese Academy of Sciences, A20 Datun Road, Chaoyang District, Beijing, 100101, PR China \label{41} \and
Korea Astronomy and Space Science Institute, Daedeok-daero 776, Yuseong-gu, Daejeon 34055, Republic of Korea \label{42} \and
Department of Astronomy, Yonsei University, Yonsei-ro 50, Seodaemun-gu, 03722 Seoul, Republic of Korea \label{43} \and
WattTime, 490 43rd Street, Unit 221, Oakland, CA 94609, USA \label{44} \and
Department of Astronomy, University of Illinois at Urbana-Champaign, 1002 West Green Street, Urbana, IL 61801, USA \label{45} \and
Instituto de Astronomía, Universidad Nacional Autónoma de México (UNAM), Apdo Postal 70-264, Ciudad de México, México \label{46} \and
Institute of Astrophysics, Central China Normal University, Wuhan 430079, PR China \label{48} \and
Department of Astrophysical Sciences, Peyton Hall, Princeton University, Princeton, NJ 08544, USA \label{49} \and
NASA Hubble Fellowship Program, Einstein Fellow \label{50} \and
Dipartimento di Fisica ``E. Pancini'', Università di Napoli ``Federico II'', Compl. Univ. di Monte S. Angelo, Edificio G, Via Cinthia, I-80126, Napoli, Italy \label{51} \and
INFN Sez. di Napoli, Compl. Univ. di Monte S. Angelo, Edificio G, Via Cinthia, I-80126, Napoli, Italy \label{52} \and
Wits Centre for Astrophysics, University of the Witwatersrand, Braamfontein, Johannesburg 2050, South Africa \label{53} \and
Department of Physics, University of Pretoria, Hatfield, Pretoria 0028, South Africa \label{54} \and
Centre for Radio Astronomy Techniques and Technologies, Department of Physics and Electronics, Rhodes University, Makhanda 6140, South Africa \label{55} \and
ASTRON, Oude Hoogeveensedijk 4, 7991 PD Dwingeloo, The Netherlands \label{56} \and
LESIA, Observatoire de Paris, Université PSL, CNRS, Sorbonne Université, Université de Paris, 5 place Jules Janssen, F-92195 Meudon, France \label{57} \and
JILA and Department of Astrophysical and Planetary Sciences, University of Colorado, Boulder, CO 80309, USA \label{58} \and
Tsung-Dao Lee Institute, Shanghai Jiao Tong University, Shengrong Road 520, Shanghai, 201210, PR China \label{59} \and
National Astronomical Observatories, Chinese Academy of Sciences, 20A Datun Road, Chaoyang District, Beijing 100101, PR China \label{60} \and
Las Cumbres Observatory, 6740 Cortona Drive, Suite 102, Goleta, CA 93117-5575, USA \label{61} \and
Department of Physics, University of California, Santa Barbara, CA 93106-9530, USA \label{62} \and
National Radio Astronomy Observatory, 520 Edgemont Road, Charlottesville, USA \label{63} \and
Department of Electrical Engineering and Computer Science, Massachusetts Institute of Technology, 32-D476, 77 Massachusetts Ave., Cambridge, MA 02142, USA \label{64} \and
Google Research, 355 Main St., Cambridge, MA 02142, USA \label{65} \and
Institut für Theoretische Physik und Astrophysik, Universität Würzburg, Emil-Fischer-Str. 31, Würzburg 97074, Germany \label{66} \and
Department of History of Science, Harvard University, Cambridge, MA 02138, USA \label{67} \and
Department of Physics, Harvard University, Cambridge, MA 02138, USA \label{68} \and
NCSA, University of Illinois, 1205 W. Clark St., Urbana, IL 61801, USA \label{69} \and
Royal Netherlands Meteorological Institute, Utrechtseweg 297, 3731 GA, De Bilt, The Netherlands \label{70} \and
Institute for Mathematics and Interdisciplinary Center for Scientific Computing, Heidelberg University, Im Neuenheimer Feld 205, Heidelberg 69120, Germany \label{74} \and
Institut f\"ur Theoretische Physik, Universit\"at Heidelberg, Philosophenweg 16, 69120 Heidelberg, Germany \label{75} \and
CP3-Origins, University of Southern Denmark, Campusvej 55, DK-5230 Odense, Denmark \label{76} \and
Instituto Nacional de Astrofísica, Óptica y Electrónica. Apartado Postal 51 y 216, 72000. Puebla Pue., México \label{77} \and
Consejo Nacional de Humanidades, Ciencia y Tecnología, Av. Insurgentes Sur 1582, 03940, Ciudad de México, México \label{78} \and
Key Laboratory for Research in Galaxies and Cosmology, Chinese Academy of Sciences, Shanghai 200030, PR China \label{79} \and
Graduate School of Science, Nagoya City University, Yamanohata 1, Mizuho-cho, Mizuho-ku, Nagoya, 467-8501, Aichi, Japan \label{80} \and
Department of Physics, McGill University, 3600 rue University, Montréal, QC H3A 2T8, Canada \label{82} \and
Trottier Space Institute at McGill, 3550 rue University, Montréal,  QC H3A 2A7, Canada \label{83} \and
NOVA Sub-mm Instrumentation Group, Kapteyn Astronomical Institute, University of Groningen, Landleven 12, 9747 AD Groningen, The Netherlands \label{84} \and
Department of Astronomy, School of Physics, Peking University, Beijing 100871, PR China \label{85} \and
Kavli Institute for Astronomy and Astrophysics, Peking University, Beijing 100871, PR China \label{86} \and
Department of Astronomical Science, The Graduate University for Advanced Studies (SOKENDAI), 2-21-1 Osawa, Mitaka, Tokyo 181-8588, Japan \label{87} \and
Department of Astronomy, Graduate School of Science, The University of Tokyo, 7-3-1 Hongo, Bunkyo-ku, Tokyo 113-0033, Japan \label{88} \and
The Institute of Statistical Mathematics, 10-3 Midori-cho, Tachikawa, Tokyo, 190-8562, Japan \label{89} \and
Department of Statistical Science, The Graduate University for Advanced Studies (SOKENDAI), 10-3 Midori-cho, Tachikawa, Tokyo 190-8562, Japan \label{90} \and
Kavli Institute for the Physics and Mathematics of the Universe, The University of Tokyo, 5-1-5 Kashiwanoha, Kashiwa, 277-8583, Japan \label{91} \and
Leiden Observatory, Leiden University, Postbus 2300, 9513 RA Leiden, The Netherlands \label{92} \and
ASTRAVEO LLC, PO Box 1668, Gloucester, MA 01931, USA \label{93} \and
Applied Materials Inc., 35 Dory Road, Gloucester, MA 01930, USA \label{94} \and
Institute for Astrophysical Research, Boston University, 725 Commonwealth Ave., Boston, MA 02215, USA \label{95} \and
University of Science and Technology, Gajeong-ro 217, Yuseong-gu, Daejeon 34113, Republic of Korea \label{96} \and
Institute for Cosmic Ray Research, The University of Tokyo, 5-1-5 Kashiwanoha, Kashiwa, Chiba 277-8582, Japan \label{97} \and
Joint Institute for VLBI ERIC (JIVE), Oude Hoogeveensedijk 4, 7991 PD Dwingeloo, The Netherlands \label{98} \and
CSIRO, Space and Astronomy, PO Box 76, Epping, NSW 1710, Australia \label{99} \and
Department of Physics, Ulsan National Institute of Science and Technology (UNIST), Ulsan 44919, Republic of Korea \label{100} \and
Department of Physics, Korea Advanced Institute of Science and Technology (KAIST), 291 Daehak-ro, Yuseong-gu, Daejeon 34141, Republic of Korea \label{101} \and
Kogakuin University of Technology \& Engineering, Academic Support Center, 2665-1 Nakano, Hachioji, Tokyo 192-0015, Japan \label{102} \and
Graduate School of Science and Technology, Niigata University, 8050 Ikarashi 2-no-cho, Nishi-ku, Niigata 950-2181, Japan \label{103} \and
Physics Department, National Sun Yat-Sen University, No. 70, Lien-Hai Road, Kaosiung City 80424, Taiwan, R.O.C. \label{104} \and
School of Astronomy and Space Science, Nanjing University, Nanjing 210023, PR China \label{105} \and
Key Laboratory of Modern Astronomy and Astrophysics, Nanjing University, Nanjing 210023, PR China \label{106} \and
Common Crawl Foundation, 9663 Santa Monica Blvd. 425, Beverly Hills, CA 90210 USA \label{108} \and
Instituto de Física, Pontificia Universidad Católica de Valparaíso, Casilla 4059, Valparaíso, Chile \label{109} \and
Department of Physics, National Taiwan University, No. 1, Sec. 4, Roosevelt Rd., Taipei 106216, Taiwan, R.O.C \label{111} \and
Instituto de Radioastronomía y Astrofísica, Universidad Nacional Autónoma de México, Morelia 58089, México \label{112} \and
David Rockefeller Center for Latin American Studies, Harvard University, 1730 Cambridge Street, Cambridge, MA 02138, USA \label{113} \and
Yunnan Observatories, Chinese Academy of Sciences, 650011 Kunming, Yunnan Province, PR China \label{114} \and
Center for Astronomical Mega-Science, Chinese Academy of Sciences, 20A Datun Road, Chaoyang District, Beijing, 100012, PR China \label{115} \and
Key Laboratory for the Structure and Evolution of Celestial Objects, Chinese Academy of Sciences, 650011 Kunming, PR China \label{116} \and
Gravitation and Astroparticle Physics Amsterdam (GRAPPA) Institute, University of Amsterdam, Science Park 904, 1098 XH Amsterdam, The Netherlands \label{118} \and
School of Physics and Astronomy, Shanghai Jiao Tong University, Shanghai, PR China \label{120} \and
Institut de Radioastronomie Millimétrique (IRAM), Avenida Divina Pastora 7, Local 20, E-18012, Granada, Spain \label{121} \and
National Institute of Technology, Hachinohe College, 16-1 Uwanotai, Tamonoki, Hachinohe City, Aomori 039-1192, Japan \label{122} \and
Research Center for Astronomy, Academy of Athens, Soranou Efessiou 4, 115 27 Athens, Greece \label{123} \and
Department of Physics, Villanova University, 800 Lancaster Avenue, Villanova, PA 19085, USA \label{124} \and
Physics Department, Washington University, CB 1105, St. Louis, MO 63130, USA \label{125} \and
Departamento de Matemática da Universidade de Aveiro and Centre for Research and Development in Mathematics and Applications (CIDMA), Campus de Santiago, 3810-193 Aveiro, Portugal \label{126} \and
School of Physics, Georgia Institute of Technology, 837 State St NW, Atlanta, GA 30332, USA \label{127} \and
School of Space Research, Kyung Hee University, 1732, Deogyeong-daero, Giheung-gu, Yongin-si, Gyeonggi-do 17104, Republic of Korea \label{128} \and
Canadian Institute for Theoretical Astrophysics, University of Toronto, 60 St. George Street, Toronto, ON M5S 3H8, Canada \label{129} \and
Dunlap Institute for Astronomy and Astrophysics, University of Toronto, 50 St. George Street, Toronto, ON M5S 3H4, Canada \label{130} \and
Canadian Institute for Advanced Research, 180 Dundas St West, Toronto, ON M5G 1Z8, Canada \label{131} \and
Dipartimento di Fisica, Università di Trieste, I-34127 Trieste, Italy \label{132} \and
INFN Sez. di Trieste, I-34127 Trieste, Italy \label{133} \and
Department of Physics, National Taiwan Normal University, No. 88, Sec. 4, Tingzhou Rd., Taipei 116, Taiwan, R.O.C. \label{134} \and
Center of Astronomy and Gravitation, National Taiwan Normal University, No. 88, Sec. 4, Tingzhou Road, Taipei 116, Taiwan, R.O.C. \label{135} \and
Finnish Centre for Astronomy with ESO, University of Turku, FI-20014 Turun Yliopisto, Finland \label{136} \and
Aalto University Metsähovi Radio Observatory, Metsähovintie 114, FI-02540 Kylmälä, Finland \label{137} \and
Gemini Observatory/NSF NOIRLab, 670 N. A’ohōkū Place, Hilo, HI 96720, USA \label{138} \and
Frankfurt Institute for Advanced Studies, Ruth-Moufang-Strasse 1, D-60438 Frankfurt, Germany \label{139} \and
School of Mathematics, Trinity College, Dublin 2, Ireland \label{140} \and
Department of Physics, University of Toronto, 60 St. George Street, Toronto, ON M5S 1A7, Canada \label{141} \and
Department of Physics, Tokyo Institute of Technology, 2-12-1 Ookayama, Meguro-ku, Tokyo 152-8551, Japan \label{142} \and
Hiroshima Astrophysical Science Center, Hiroshima University, 1-3-1 Kagamiyama, Higashi-Hiroshima, Hiroshima 739-8526, Japan \label{143} \and
Aalto University Department of Electronics and Nanoengineering, PL 15500, FI-00076 Aalto, Finland \label{144} \and
Jeremiah Horrocks Institute, University of Central Lancashire, Preston PR1 2HE, UK \label{146} \and
National Biomedical Imaging Center, Peking University, Beijing 100871, PR China \label{147} \and
College of Future Technology, Peking University, Beijing 100871, PR China \label{148} \and
Tokyo Electron Technology Solutions Limited, 52 Matsunagane, Iwayado, Esashi, Oshu, Iwate 023-1101, Japan \label{149} \and
Netherlands Organisation for Scientific Research (NWO), Postbus 93138, 2509 AC Den Haag, The Netherlands \label{151} \and
Frontier Research Institute for Interdisciplinary Sciences, Tohoku University, Sendai 980-8578, Japan \label{152} \and
Astronomical Institute, Tohoku University, Sendai 980-8578, Japan \label{153} \and
Department of Physics and Astronomy, Seoul National University, Gwanak-gu, Seoul 08826, Republic of Korea \label{154} \and
SNU Astronomy Research Center, Seoul National University, Gwanak-gu, Seoul 08826, Republic of Korea \label{155} \and
University of New Mexico, Department of Physics and Astronomy, Albuquerque, NM 87131, USA \label{156} \and
Physics Department, Brandeis University, 415 South Street, Waltham, MA 02453, USA \label{157} \and
Tuorla Observatory, Department of Physics and Astronomy, University of Turku, FI-20014 Turun Yliopisto, Finland \label{158} \and
Radboud Excellence Fellow of Radboud University, Nijmegen, The Netherlands \label{159} \and
School of Natural Sciences, Institute for Advanced Study, 1 Einstein Drive, Princeton, NJ 08540, USA \label{160} \and
School of Physics, Huazhong University of Science and Technology, Wuhan, Hubei, 430074, PR China \label{161} \and
Mullard Space Science Laboratory, University College London, Holmbury St. Mary, Dorking, Surrey, RH5 6NT, UK \label{162} \and
Center for Astronomy and Astrophysics and Department of Physics, Fudan University, Shanghai 200438, PR China \label{163} \and
Astronomy Department, University of Science and Technology of China, Hefei 230026, PR China \label{164} \and
Department of Physics and Astronomy, Michigan State University, 567 Wilson Rd, East Lansing, MI 48824, USA \label{165}
}

\date{Received Month XX, 2025; accepted Month XX, 2025}

\abstract
 {Sagittarius A* (\sgra), the supermassive black hole at the center of the Milky Way, provides a unique laboratory  to study accretion dynamics and plasma processes near the event horizon.}
  {We investigated the variability and polarization properties of \sgra using ALMA observations during the 2018 Event Horizon Telescope campaign.} 
  {We analyzed high-cadence full-polarization light curves from ALMA at millimeter wavelengths, performed time-series analysis, and investigated the temporal behavior during an X-ray flare observed by \textit{Chandra} on 2018 April 24. 
  The variability characteristics are compared with expectations from standard accretion flow models.}
  {We find low variability in total intensity ($\sigma/\mu < 10\%$), but significantly higher variability in linear and circular polarization ($\sim 30\%$ and $\sim 50\%$, respectively). A time-series analysis reveals red-noise variability, with power spectral densities between $-2$ and $-3$ across all Stokes parameters. Polarized intensity shows stable intra-day timescales, while total intensity exhibits more variable timescales, suggesting distinct emission regions, with polarization likely arising from a coherent structure. On April 24, a statistically significant inter-band delay in polarized intensity coincides with a near-simultaneous X-ray and millimeter peak that deviates from the typical delayed flare scenario. This event also features enhanced millimeter variability and coherent polarization loop evolution. The observed simultaneity challenges standard models of transient synchrotron emission with cooling delays, favoring instead a scenario of continuous energy injection in an optically thin region.
  }
  {Our results offer new constraints on the physical mechanisms driving variability in \sgra, and provide key observational input for refining theoretical models of accretion and plasma behavior in the vicinity of supermassive black holes.}


\authorrunning{Ezequiel Albentosa-Ruiz et al.}
\titlerunning{Full-polarization millimeter variability of \sgra}

\maketitle

\section{Introduction}
\label{sec:introduction}

The Galactic center (GC) is one of the most extensively studied astrophysical environments, hosting a rich and diverse population of radio sources~\citep{Heywood2022}.
At the heart of the GC lies Sagittarius A$^*$ \citep[\sgra;][]{Balick1974}, a supermassive black hole with a mass of approximately $4 \times 10^6 M_\odot$ \citep{Do2019, Gravity2022, EHTC2022a}. 
This source exhibits significant variability at radio frequencies \citep{Brown1982, Iwata2020, Wielgus2022, Mus2022}, with spatial variations in its emission structure observable on timescales shorter than 30 minutes~\citep{Gravity2018b, Gravity2023}.

Particularly striking are the intense flare events detected in the near-infrared (NIR) and X-ray regimes in the vicinity of \sgra~\citep{Genzel2003, Aschenbach2004, Eckart2006, Boyce2019}.
These flares are thought to result from magnetic reconnection events, which dissipate magnetic energy and may produce transient features such as orbiting hot spots of plasma~\citep{Yuan2003,Dexter2020,Porth2021, Ripperda2022, Wielgus2022_orbital}.

Multiwavelength studies of \sgra's variability have provided valuable insights into its radiation mechanisms and spatial emission regions.
Simultaneous X-ray and infrared (IR) flares suggest that these emissions predominantly originate from the same regions, with delays of approximately 10–20 minutes between the two wavebands~\citep{Eckart2004, Marrone2008, Boyce2019}.
In contrast, millimeter and submillimeter flares show more complex behavior, with reported delays, relative to NIR and X-ray emissions, ranging from 20–30 minutes~\citep{Marrone2008, Witzel2021} to several hours~\citep{Yusef2008, Eckart2012,SgraP2}.
Some studies have even reported minimal or negligible delays between millimeter and IR/X-ray flares~\citep{Fazio2018}, or have suggested that previously perceived delays may have been coincidental~\citep{Capellupo2017}.
This inconsistency underscores the need for high-fidelity millimeter light curves, which have recently become accessible through advanced facilities such as the Atacama Large Millimeter/Submillimeter Array (ALMA) and the Submillimeter Array  (SMA; \citealt{Bower2015, Witzel2021}).

During the first Event Horizon Telescope (EHT) observing campaign in April 2017, millimeter light curves of \sgra~were obtained from both ALMA and SMA, as reported by \citet{Wielgus2022}.
On April 11, 2017, an X-ray flare was detected by \textit{Chandra}~\citep{SgraP2}, providing a unique opportunity to investigate the effect of an X-ray flare on millimeter-wavelength light curves.
While the light curves from April 6 and 7, 2017, exhibited a low-variability state, the April 11 light curve showed pronounced variability following the X-ray flare.  

Beyond total intensity, the exceptional sensitivity of ALMA has also enabled detailed studies of the polarimetric properties of \sgra~and other active galactic nuclei (AGN) during the 2017 Very Long Baseline Interferometry (VLBI) campaign~\citep{Goddi2021}.
Polarization measurements offer a powerful, time-resolved probe of the emission and propagation processes in \sgra, complementing the spatially resolved Event Horizon Telescope (EHT) analysis~\citep{SgrAP8, Joshi2024}. Although \sgra\ is powered by a radiatively inefficient accretion flow~\citep{Yuan2014}, fundamental properties such as magnetic field geometry, plasma composition, and turbulence remain poorly constrained. Unlike total intensity, polarization encodes information about magnetic field structure, Faraday rotation, optical depth, and the nature of the emitting electrons~\citep[e.g.,][]{Macquart2006, Johnson2015, Wielgus2024}.
In particular, variations in polarization degree and angle constrain several physical properties: the viewing geometry~\citep[axisymmetric fields tend to depolarize when seen face-on; e.g.,][]{Shcherbakov2012}, the magnetic field strength, number density and temperature of electrons through the Faraday depth~\citep{Quataert2000,Wielgus2024}, and turbulence in the accretion flow~\citep{Bower2005}. ALMA observations have revealed hour- to month-scale variability in linear polarization and persistent circular polarization, consistent with Faraday conversion in an ordered field~\citep{Bower2018}. Temporal features, such as polarization angle swings, depolarization dips, or $Q$–$U$ loops, provide further diagnostics to distinguish between accretion scenarios, with magnetically arrested disk (MAD) models predicting strong organized fields, and standard and normal evolution (SANE) models favoring weaker turbulent configurations. Moreover, frequency-dependent polarization delays constrain the optical depth and cooling timescales of the emitting plasma~\citep{Wielgus2022_orbital, Michail2024}.

The polarimetric variability observed in \sgra after the X-ray flare detected on April 11, 2017, suggests the presence of orbital motion of a hot spot near the black hole following a high-energy flare~\citep{Wielgus2022_orbital}, consistent with earlier findings by GRAVITY~\citep{Gravity2018b, Gravity2020b}.
More recently, \citet{Wielgus2024} conducted a joint analysis of  polarized light curves of \sgra~at 85–101 GHz and 212–230 GHz with ALMA, revealing insights into inefficient accretion flows and an internal Faraday screen through the analysis of rotation measure (RM) and its variability. These results imply that a significant fraction of Faraday rotation occurs in the compact source near the event horizon, and that the magnetic field on this scale is organized and not violently variable.

Building upon these findings, for this work we analyzed \sgra's millimeter light curves obtained from ALMA and SMA observations during the second EHT observing campaign in April 2018, covering four observing days: April 21, 22, 24, and 25.
We investigated the full-polarization variability by characterizing and comparing the light curves from the 2017 and 2018 campaigns using time-series analysis techniques. 
Similarly to April 11, 2017, an X-ray flare was detected by \textit{Chandra} on April 24, 2018~\citep{Mossoux2020}, coinciding with ALMA coverage.
Therefore, in addition to the full-polarization variability of \sgra's light curves, we closely examined this flare event, assessed its imprint on the millimeter light curves, and compared our findings with those from 2017.

This paper is structured as follows. Section~\ref{sec:observations} describes the observations and details the data reduction procedures implemented to recover the compact source emission variability from ALMA data.
In Sect.~\ref{sec:analysis} we present the full-polarization ALMA light curves and compare them with historical data.
Section~\ref{sec:variability} details the complete time-variability analysis of the 2017 and 2018 light curves.
In Sect.~\ref{sec:discussion} we discuss specific properties of the light curves, highlighting persistent clockwise $Q$–$U$ rotation, accretion physics, and X-ray flaring, and compare the variability with GRMHD predictions.
Finally, we summarize our conclusions in Sect.~\ref{sec:conclusions}.

\section{Observations and data calibration}
\label{sec:observations}

\subsection{ALMA observations and data processing}
\sgra was observed with phased-ALMA \citep{Matthews_2018,Crew_2023}  on April 21, 22, 24, and 25 as part of the 2018 EHT+ALMA campaign. A detailed analysis of the EHT VLBI observations of \sgra in 2018 will be presented elsewhere (EHT Collaboration in prep). 
The VLBI observations were carried out while the array was in its most compact configurations and only antennas within a  radius of 180 m (from the array center) were used for phasing.
The observations were  performed in full-polarization mode in order to supply the inputs to the polarization conversion process at the VLBI correlators \citep{Marti_2016}. 

The spectral setup included four spectral windows (SPWs) of 1875~MHz,
two in the lower and two in the upper sideband, 
centered at 213.1, 215.1, 227.1, and 229.1 GHz. 
Table \ref{tab:lcurves_data} summarizes the observational and spectral setup of the April 2018 observations, including four tracks (which ranged from roughly 6 to 9 hours).
An absorption feature in the SPW centered at 227.1~GHz was flagged to ensure consistency in the \sgra\ light curves (see Appendix~\ref{sec:absorption}).

ALMA data acquired during VLBI observations are calibrated  using the Common Astronomy Software Applications (CASA) package \citep{CASA2022}  and the special procedures known as “Quality Assurance Level 2” (QA2) described in \citet{Goddi2019}. 
Bandpass solutions were obtained using 3C 279, which also served as the polarization calibrator on all four observing days. Flux calibration was performed using Titan on April 21 and 25, and 3C 279 on April 22 and 24.
To evaluate the accuracy of our flux-density calibration, we compared the QA2-derived fluxes of VLBI targets with independent Atacama Compact Array (ACA) flux monitoring of Grid Sources, some of which overlap with our VLBI sample. The analysis shows that ALMA fluxes during VLBI observations agree within 10$\%$ in Band 6, consistent with ALMA’s nominal absolute calibration uncertainty \citep{Remijan2019}. This result aligns with similar findings reported in \cite{Goddi2021,Goddi2025} for VLBI observations in Bands 6 and 7.

The QA2 process relies on self-calibration under the assumption of a point source with constant flux density. This approach is effective for most VLBI targets, which are generally stable over the course of an observation. As shown in Appendix~\ref{ap:visib_cal_targets}, the visibility amplitude light curves of both the calibrators and EHT targets confirm this flux stability. However, this method is not well suited for \sgra, which exhibits intrinsic variability on minute timescales. In the following, we describe the specialized procedure required to calibrate a time-variable source like \sgra accurately.

\subsection{Intra-field calibration of \sgra}
\label{subsec:QA2}
\label{subsec:pipeline1}

\begin{figure}[h!]
    \centering
    \includegraphics[width=8.5cm]{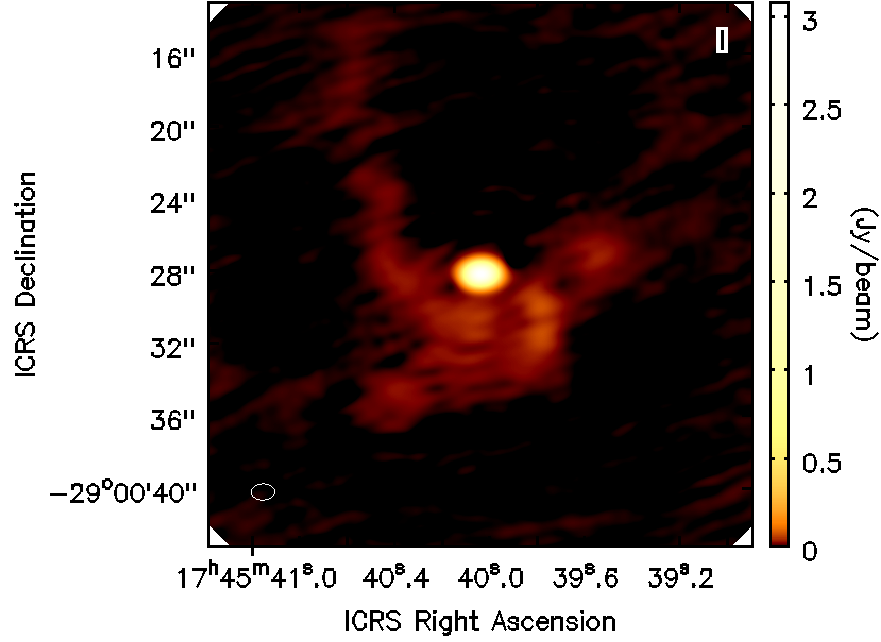}
    \caption{Stokes I CLEAN image of \sgra and the minispiral from visibilities at 213.1 GHz, produced after the QA2 calibration for April 21.}
     \label{fig:CLEAN_SGRA_image}
\end{figure}

\sgra  can be understood as the sum of two components (shown in Fig. \ref{fig:CLEAN_SGRA_image}):
\begin{itemize}
\item An extended component (up to parsec scales), hereby called the minispiral. Given its large extension, it is safe to assume that this component is not variable on the timescales comparable to the duration of the observations (i.e., a few hours).
\item A compact, bright component, which presents high variability: \sgra itself.
\end{itemize}

The assumption of constant flux density in the QA2 calibration results in a core with constant brightness, shifting all variability to the minispiral. To derive the light curves of \sgra (i.e., the compact core), we implemented an algorithm to enhance the QA2 gain calibration for variable sources, building upon previous work presented in \cite{Wielgus2022,Mus2022}.\footnote{This algorithm is incorporated into a CASA script and is described in detail in the EHT Memo \href{https://eventhorizontelescope.org/sites/g/files/omnuum3116/files/2025-03/eht_memo_albentosaruiz_2025-tdwg-01_0.pdf}{2025-TDWG-01}.
The full pipeline is available for download on GitHub: \url{https://github.com/ealruiz/calminispiral}.}
Here we summarize the main procedure, consisting of  four  steps:

\begin{enumerate}
    \item Generate a CLEAN image of the source (i.e., the core and the minispiral), labeled as \texttt{IM0}. Although including artifacts (e.g., negative features simulating partial sidelobes of the PSF around \sgra), this image serves as an initial model for the minispiral.
    \item Subtract the core of \sgra from the \texttt{IM0} image by setting the flux of the pixels corresponding to the compact component to zero, producing an image of only the minispiral (\texttt{IM1}).
    \item Visibility (two-component) model fitting: Construct a Visibility model as the Fourier transform of the minispiral ($\texttt{MOD}=FT(\texttt{IM1})$) scaled by a time-varying factor $S_1$ (which accounts for the minispiral's artificial variability introduced by QA2) and a constant factor $S_2$ (core flux density),\footnote{The model for Stokes $Q,U,V$ assumes a centered point source (i.e., $V^{mod}=S$, where $S$ is the flux density of the core Stokes parameter), as the minispiral is unpolarized at millimeter wavelengths (see \citealt{Goddi2021}).} thus describing the entire structure of \sgra:
    \begin{equation}
        V^{mod} = S_1 \cdot \texttt{MOD} + S_2.
    \end{equation}
    For each integration time $t$, we fit the observed visibilities to our  model by minimizing the $\chi^2$ function
    \begin{equation}
        \chi^2(t)=\sum_{i,t} \omega_i \cdot \abs{ S_1(t) \cdot \texttt{MOD} + S_2 - V_i^{obs}(t)}^2,
    \label{eq:chi2}
    \end{equation}
    where $\omega_i$ are the weights of each visibility.
    \item Transfer the variability from $S_1$ to $S_2$ by scaling the \sgra light curves as
    \begin{equation}
        V_{cal}(t) = \frac{\bar{S_1}}{S_1(t)} \cdot V_{QA2}(t),
    \end{equation}
    where $V_{QA2}$ represents the QA2 visibilities, $V_{cal}(t)$ are the calibrated visibilities, and $\bar{S_1}$ is the median of all minispiral flux densities across all days of the campaign.
\end{enumerate}

\begin{figure}[h!]
    \centering
    \includegraphics[width=8.25cm]{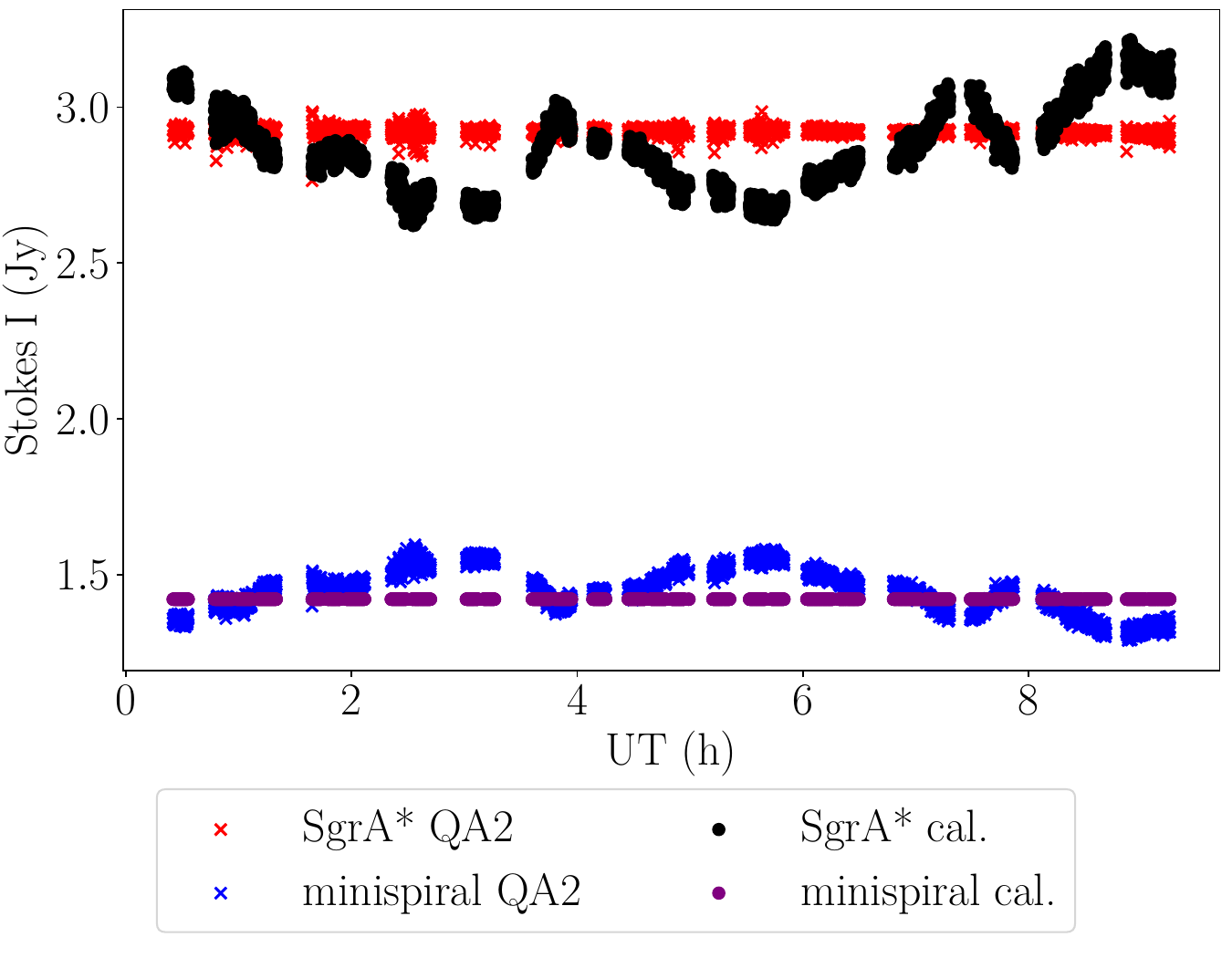}
    \caption{April 22 total flux density of \sgra and the minispiral at 213.1~GHz, before (red and blue crosses) and after (purple and black dots) correcting for variability transfer. The flux density uncertainty, estimated from the covariance matrix, is approximately 0.1\% and is therefore not visible.}
    \label{fig:minispiral_cal_flux}
\end{figure}

This algorithm corrects the amplitude of all visibilities for each integration time, ensuring that the minispiral brightness remains nearly constant (within $\lesssim$1\%), while allowing the core to exhibit the expected flux density variability, as illustrated in Fig. \ref{fig:minispiral_cal_flux}. 
A final round of flagging is typically performed to remove data points that vary significantly between consecutive time intervals, as detailed in Sect.~\ref{sec:ALMA_lcurves}.

A complementary calibration method for the time-variable source \sgra\ is presented in Appendix~\ref{appendix:manual}, where the manual reduction of the ALMA data is described in detail. Additionally, in Appendix~\ref{appendix:sma} we present the SMA observations and data calibration, and compare the resulting light curves of \sgra\ with those from ALMA as a consistency test.

\section{\sgra full-polarization variability}
\label{sec:analysis}

\subsection{ALMA light curves}
\label{sec:ALMA_lcurves}
Following the intra-field calibration of the ALMA data, presented in Sect. \ref{subsec:pipeline1}, we retrieved the full-polarization \sgra light curves with a 4 second cadence, which will be analyzed in the following sections.\footnote{We have also reprocessed the 2017 light curves using the same intra-field calibration pipeline applied to the 2018 data; the results are shown in Appendix \ref{sec:2017lcurves}.} To remove outliers, we fit the Stokes I light curves with a fifth-order spline function and discard data points deviating beyond a 3$\sigma$ threshold. Due to the higher noise in the April 25 light curve, we averaged the data over 16-second intervals, matching the timing of the ALMA phasing loop \citep{Goddi2019}, to mitigate its impact.

\begin{figure*}[!ht]
    \centering
    \includegraphics[width=0.925\textwidth]{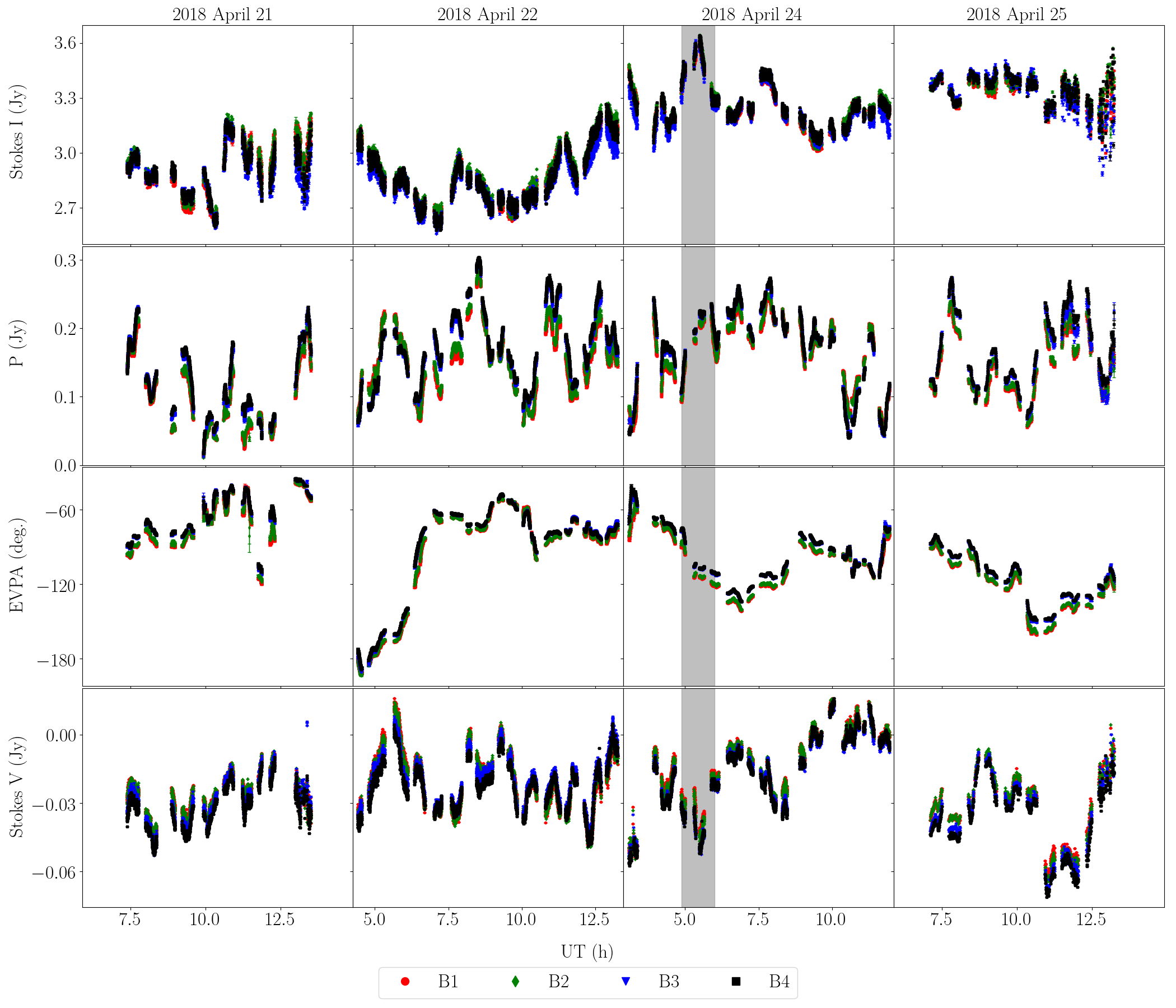}
    \caption{\sgra ALMA light curves of Stokes I, the polarized intensity, the EVPA, and Stokes V (from top to bottom) for the four spectral bands, for all four days (from left to right, April 21, 22, 24, and 25). Stokes V light curves are tentative, as the detected levels fall below ALMA’s guaranteed CP accuracy. The gray-shaded band on April 24 marks the time range of the \textit{Chandra} X-ray flare.} 
    \label{fig:ALMAlcurves}
\end{figure*}

\begin{figure*}[!ht]
    \centering
    \includegraphics[width=0.925\textwidth]{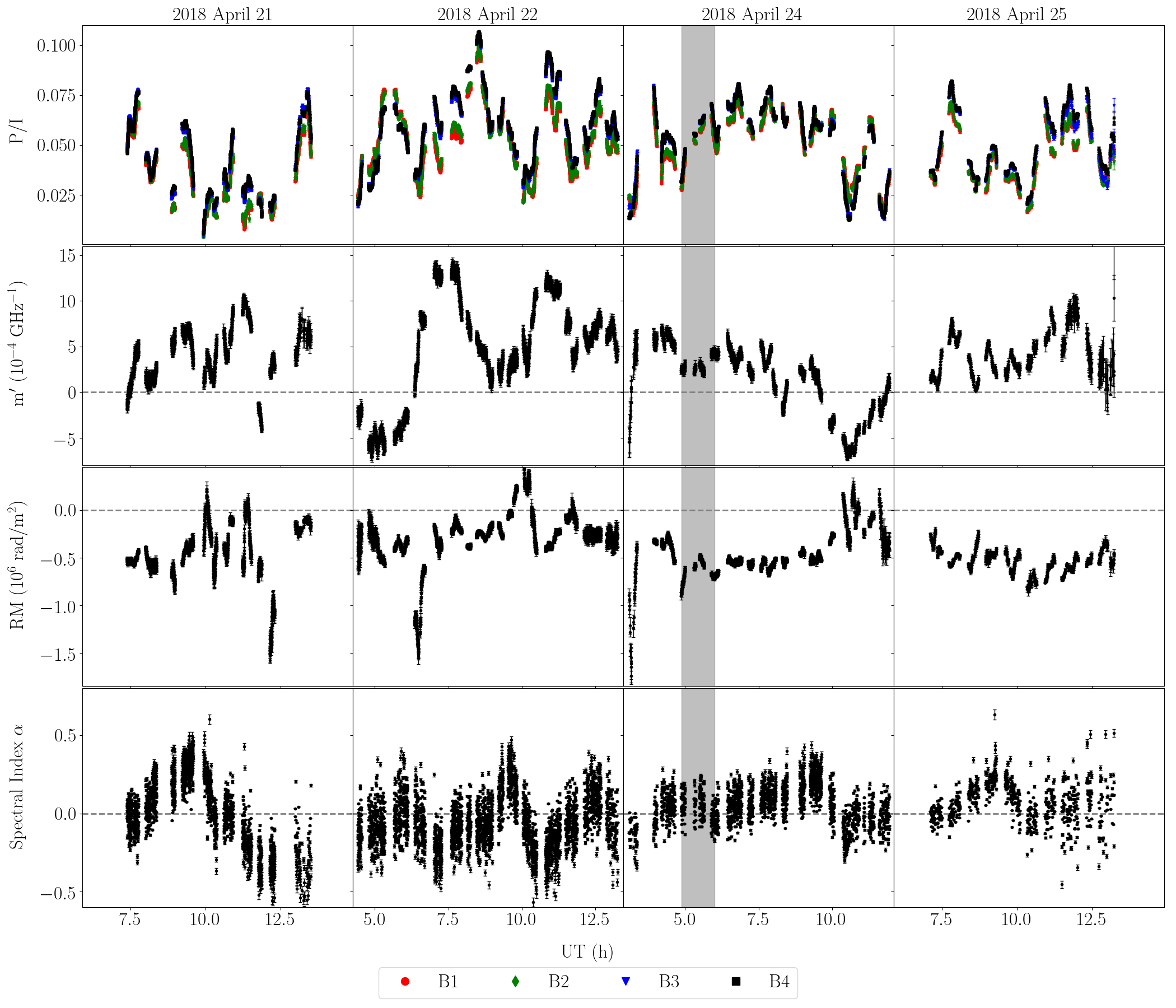}
    \caption{\sgra ALMA light curves of the fractional polarization for the four spectral bands, the depolarization measure, the rotation measure, and the spectral index (from top to bottom) for all four days (from left to right, April 21, 22, 24, and 25). The gray-shaded band on April 24 marks the time range of the \textit{Chandra} X-ray flare.
    }
    \label{fig:ALMAlcurves_products}
\end{figure*}

The ALMA light curves for the four observing days are presented in Appendix \ref{ap:2018_lcurves_table} (Table \ref{tab:lcurves_data}) and plotted in Figs. \ref{fig:ALMAlcurves} and \ref{fig:ALMAlcurves_products}. Table \ref{tab:lcurves_data} summarizes the main variability characteristics of \sgra, reporting for each of the four ALMA spectral windows the average and dispersion of three parameters: total intensity (Stokes I), polarized intensity $P=\sqrt{Q^2+U^2}$, and Stokes V. Variability is quantified using the modulation index, defined as the ratio of the standard deviation to the mean. The numerical values are listed in Table \ref{tab:lcurves_data}.
Comparing the modulation indices of the Stokes I light curves with those of the polarized intensity reveals a difference of an order of magnitude, suggesting significantly greater variability in polarization than in total flux density. This behavior is also evident in Fig. \ref{fig:ALMAlcurves}, which shows the time evolution of total flux density, polarized intensity, polarization angle (EVPA; defined as $\phi=0.5\arg(Q+iU)$), and Stokes V across the four observing nights.

While the Stokes V light curve is predominantly negative, consistent with previous studies \citep{Marrone2006, Bower2018, Wielgus2022_orbital}, the average amplitude of the circular polarization is around 1$\%$ of the total flux, well below the ALMA’s nominal 3$\sigma$ detection threshold of 1.8$\%$. In this dataset, calibration was performed using 3C279, whose intrinsic circular polarization is unknown. As a result, any true Stokes V signal from the calibrator may be absorbed into instrumental terms and subsequently imprinted onto the target source as an artificial signal. In light of these limitations, we do not attempt to analyze the Stokes V data. A dedicated investigation of the circular polarization, following the approach of \citet[][Appendix~G]{Goddi2021}, is planned for a future publication.

Figure \ref{fig:ALMAlcurves_products} presents four additional parameters: 
the fractional linear polarization  $m=P/I$ (LP), 
the Depolarization Measure ($m^\prime$),
the Rotation Measure (RM), 
and the Spectral Index $\alpha$.
The RM is defined by the relation $\phi(\lambda) = \phi_0 + \mathrm{RM} (\lambda^2 - \lambda_0^2)$ \citep[e.g.,][]{Brentjens2005}, where $\phi_0$ is the EVPA at the reference wavelength $\lambda_0=c/\nu_0$. Similarly, the depolarization measure $m^\prime$ quantifies the change in LP per unit frequency (in GHz$^{-1}$), following the relation $m(\nu) = m_0 + m^\prime (\nu - \nu_0)$, where $m_0$ is the LP at the reference frequency $\nu_0$ \citep[e.g.,][]{Goddi2021}.
A more detailed discussion of these polarization properties, their variability, and their implications for the accretion rate is provided in Appendix~\ref{appendix:sgra_polarization}.

Finally, the spectral index $\alpha$ is obtained by fitting the flux density variation across the ALMA frequency range using the model $I = I_0 \cdot (\nu/\nu_0)^\alpha$, where $I_0$ is the flux density at the reference frequency $\nu_0$.
The resulting spectral index exhibits slight oscillations around zero, consistent with the findings from the 2017 light curves presented in \citet{Wielgus2022}. These variations are accompanied by high uncertainty, primarily driven by the short timescale variability of the spectral index and calibration effects. Factors such as the ALMA intra-site antenna configuration, or the calibrator targets and the minispiral model used in QA2 and intra-field calibration, contribute to uncertainties in the absolute flux density across SPWs.

The variability of the full-polarization \sgra light curves, characterized by the modulation index and other advanced time-series analysis tools, are explored in detail in Sect. \ref{sec:variability}.

\subsection{Comparison to historical data}
\begin{figure*}
    \centering
    \includegraphics[width=0.8\linewidth]{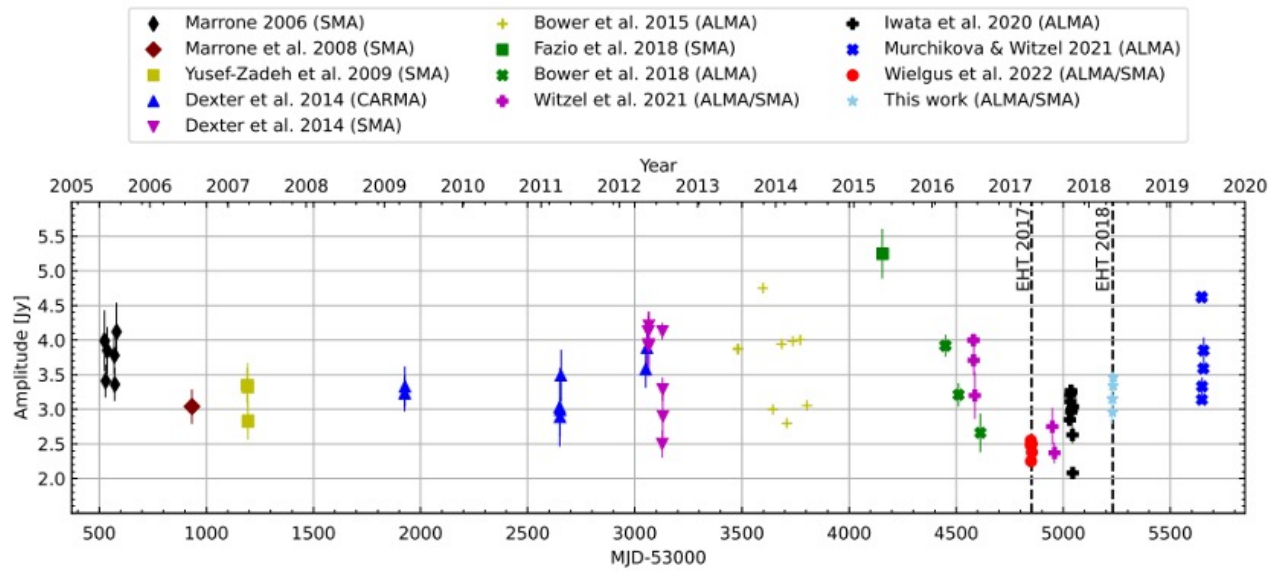}
    \caption{Historical 230 GHz amplitude measurements of \sgra from 2005 to 2019 in Table 4 of \citet{Wielgus2022} and average flux density and standard deviation from Table \ref{tab:lcurves_data}. The 2017 and 2018 EHT observing campaigns are marked by black vertical lines. Standard deviations for both the 2017 and 2018 EHT observations are plotted, but are too small to be visible.} 
    \label{fig:lcbig}
\end{figure*}

To investigate long-term trends in \sgra’s behavior, we analyze historical flux density and polarization data from 2005 to 2019, as compiled in Table 6 of \citet{Wielgus2022}. Figure \ref{fig:lcbig} presents the flux density as a function of time, showing that the daily average values from 2018 fall within the range of previously reported measurements. The data also align with a broader trend of increasing average flux density from 2017 \citep{Wielgus2022} to 2019 \citep{MW2021}.

\begin{figure}
    \centering
    \includegraphics[width=9cm]{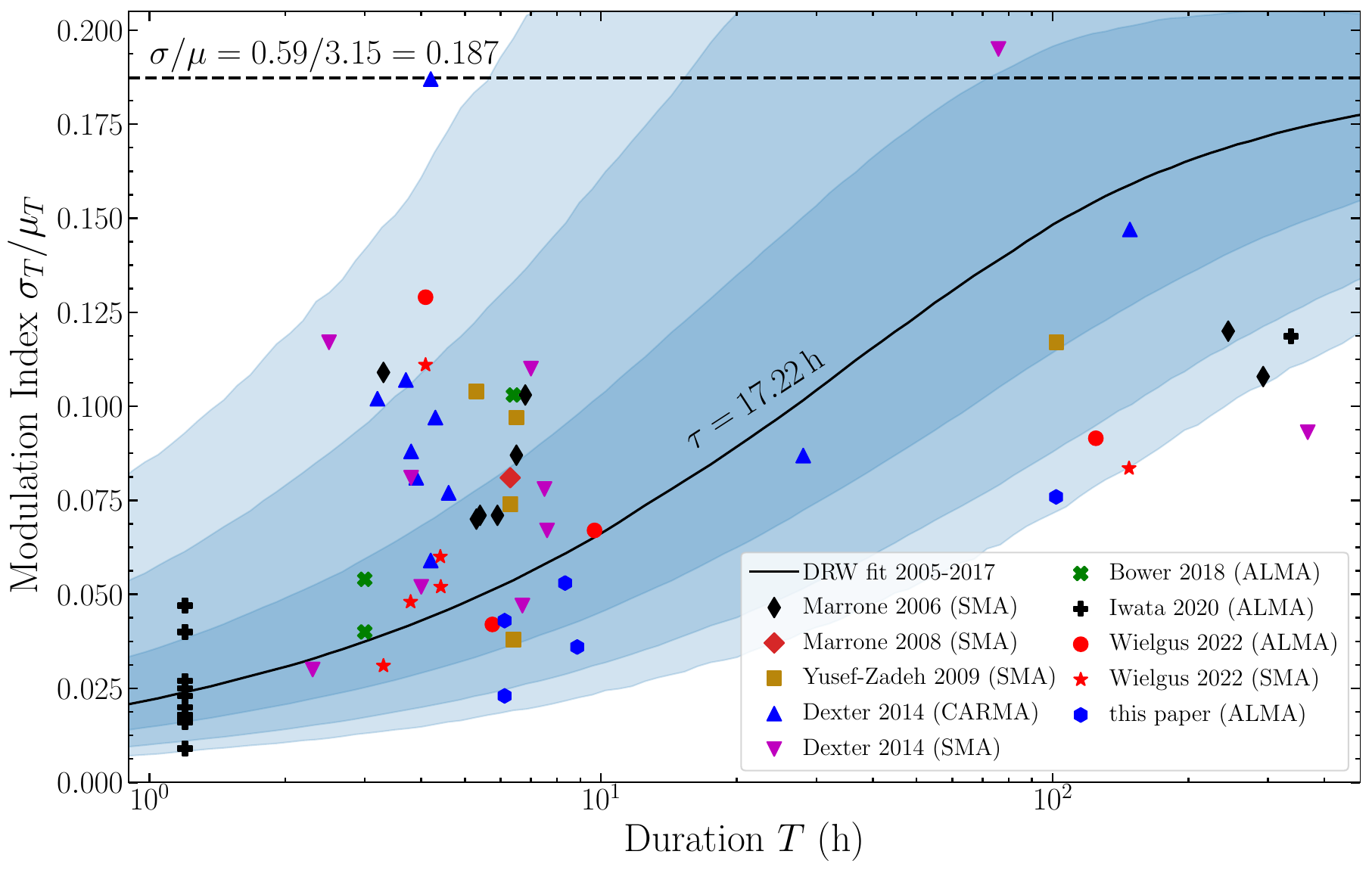}
    
    \caption{Modulation index measured across various observations as a function of observational duration. 
    Each data point from the 2018 ALMA dataset at B3 (227\,GHz) is represented by hexagons, while other data points, as well as the fitted damped random walk (DRW) curve with shaded confidence intervals are derived from \citet{Wielgus2022}.}
    \label{fig:mod_index_history}
\end{figure}

\begin{figure}
    \centering
    \includegraphics[width=8.5cm]{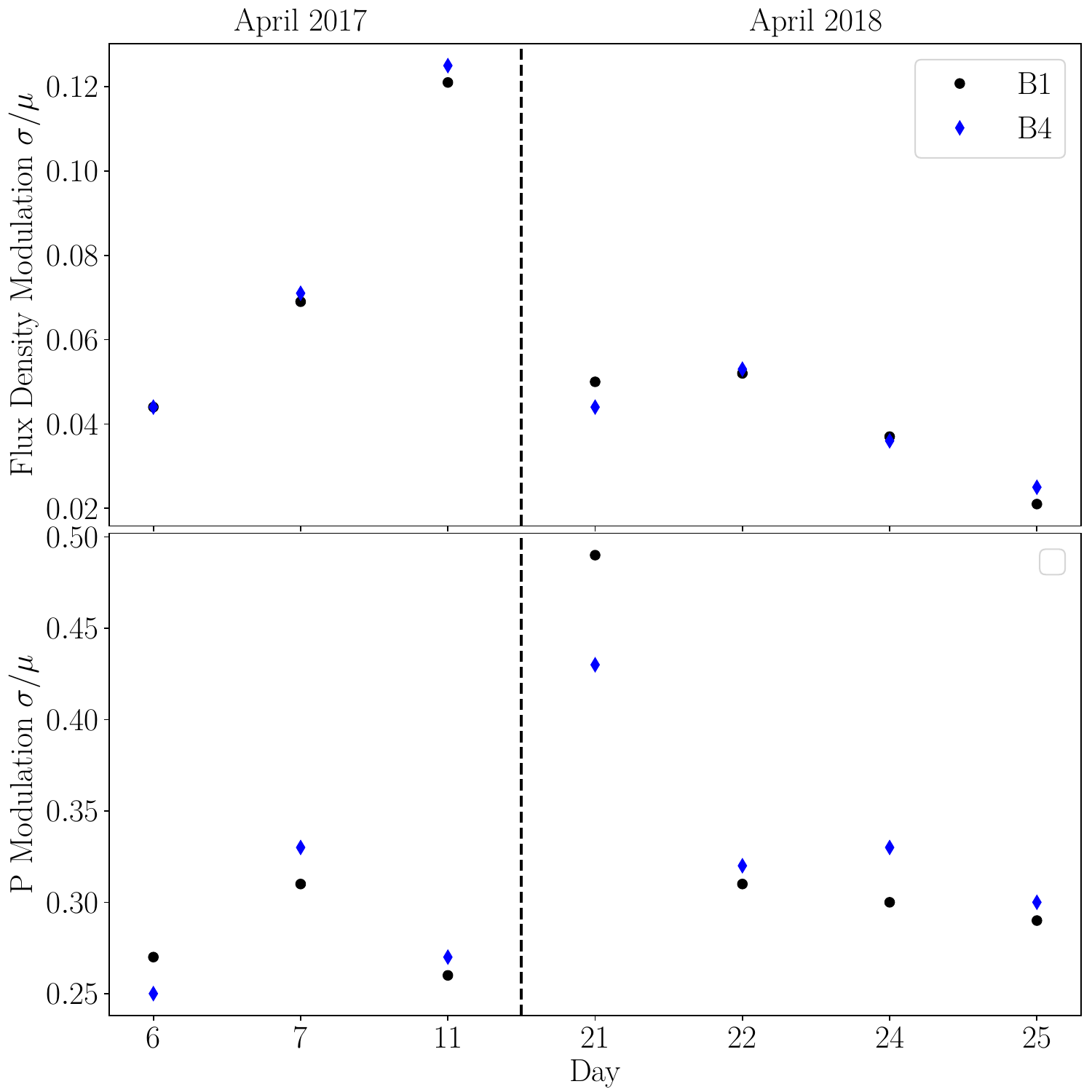}
    \caption{Modulation indices $\sigma/\mu$ of the Stokes I (top) and polarized intensity (bottom) ALMA light curves, for the 2017 and 2018 observations at B1 (black dots) and B4 (blue diamonds), obtained from our analysis.}
    \label{fig:modindx_2017-18}
\end{figure}

An examination of the modulation indices further supports consistency with past observations (Fig. \ref{fig:mod_index_history}). Comparing our derived modulation indices with the damped random walk (DRW) model fitted to historical data by \citet{Wielgus2022}, we find that variability in 2018 was slightly lower than the expected value from the model. However, a direct comparison of the modulation indices from full-polarization ALMA light curves in the 2017 and 2018 campaigns reveals consistent values across both years, as well as notable stability throughout the entire experiment (Fig. \ref{fig:modindx_2017-18}).

The average linear polarization fraction across the four SPWs ranges from 3.8\% to 5.8\%, slightly lower than the 2017 ALMA values of 7.7\%–8.5\% reported in \citet{Wielgus2024}. Nonetheless, these values remain broadly consistent with historical measurements spanning 3.6\%–9.9\% \citep{Bower2003,Bower2005,Marrone2007,Bower2018}. 
The average EVPA across spectral windows ranged from -70.30° to -117.97°. The spread in EVPA values seen in Fig. \ref{fig:ALMAlcurves} reflects short-term variability in \sgra, while the broader range observed over nearly 20 years \citep{Bower2003,Bower2005,Marrone2007,Bower2018,Wielgus2022_orbital} suggests significant long-term evolution in the linear polarization.

For circular polarization, we observe daily averages ranging from -0.41\% to -1.0\% in 2018. The most negative values align with the 2017 ALMA results, which reported an average CP of -1.0\% to -1.6\% \citep{Goddi2021,Wielgus2024}, as well as with \citet{Bower2018}, who found a mean CP of -1.1$\pm$0.2\% at 225 GHz in 2016. We note, however, that all of these circular polarization measurements should be regarded as tentative detections, as the measured levels fall below the official CP accuracy threshold guaranteed by the ALMA observatory.

The daily average RM in our 2018 data ranges from $-5.32\times10^5$ to $-2.63\times10^5$ rad m$^{-2}$, comparable to 2017 ALMA results, which reported daily averages between $-5.04\times10^5$ and $-3.19\times10^5$ rad m$^{-2}$ \citep{Goddi2021,Wielgus2024}. Our values are also consistent with past measurements over the last two decades from ALMA \citep{Bower2018}, SMA \citep{Marrone2007}, and BIMA \citep{Bower2003,Bower2005}. This agreement suggests a relatively stable long-term Faraday rotation, while also reflecting short-term variability in \sgra’s polarization properties.

\section{Time-series analysis of the \sgra light curves}
\label{sec:variability}

\subsection{Cross-correlations between Spectral Windows}
\label{subsec:spws_cross_corr}

To compute the correlation of the \sgra light curves at different SPWs, we used the Locally Normalized Discrete Correlation Function \citep[LNDCF; see][]{Lehar1992}. The LNDCF consists of binning pairs of flux density measurements $(a_i,b_j)$ by a time difference (lag $\Delta t$), and then computing estimates of the correlation between the two signals using the formula
\begin{equation}
    \text{LNDCF} (\Delta t) = \frac{1}{M} \sum_{i,j} \frac{(a_i - \overline{a}_{\Delta t})(b_j - \overline{b}_{\Delta t})}{\sqrt{(\sigma^2_{a,\Delta t} - e^2_a)(\sigma^2_{b,\Delta t} - e^2_b)}},
    \label{eq:LNDCF}
\end{equation}
where $M$ is the number of data pairs in the lag bin $\Delta t$, $e_a$ and $e_b$ represent the measurement errors of the data points of their respective signals, and the signal amplitudes and standard deviations (i.e., $\overline{a}_{\Delta t}$, $\overline{b}_{\Delta t}$, $\sigma^2_{a,\Delta t}$, $\sigma^2_{b,\Delta t}$) are computed for each lag using the flux densities contributing to the LNDCF. Here, we define LNDCF$(0)=$LNDCF$_0$.

\begin{figure}
    \centering
    \includegraphics[width=8cm]{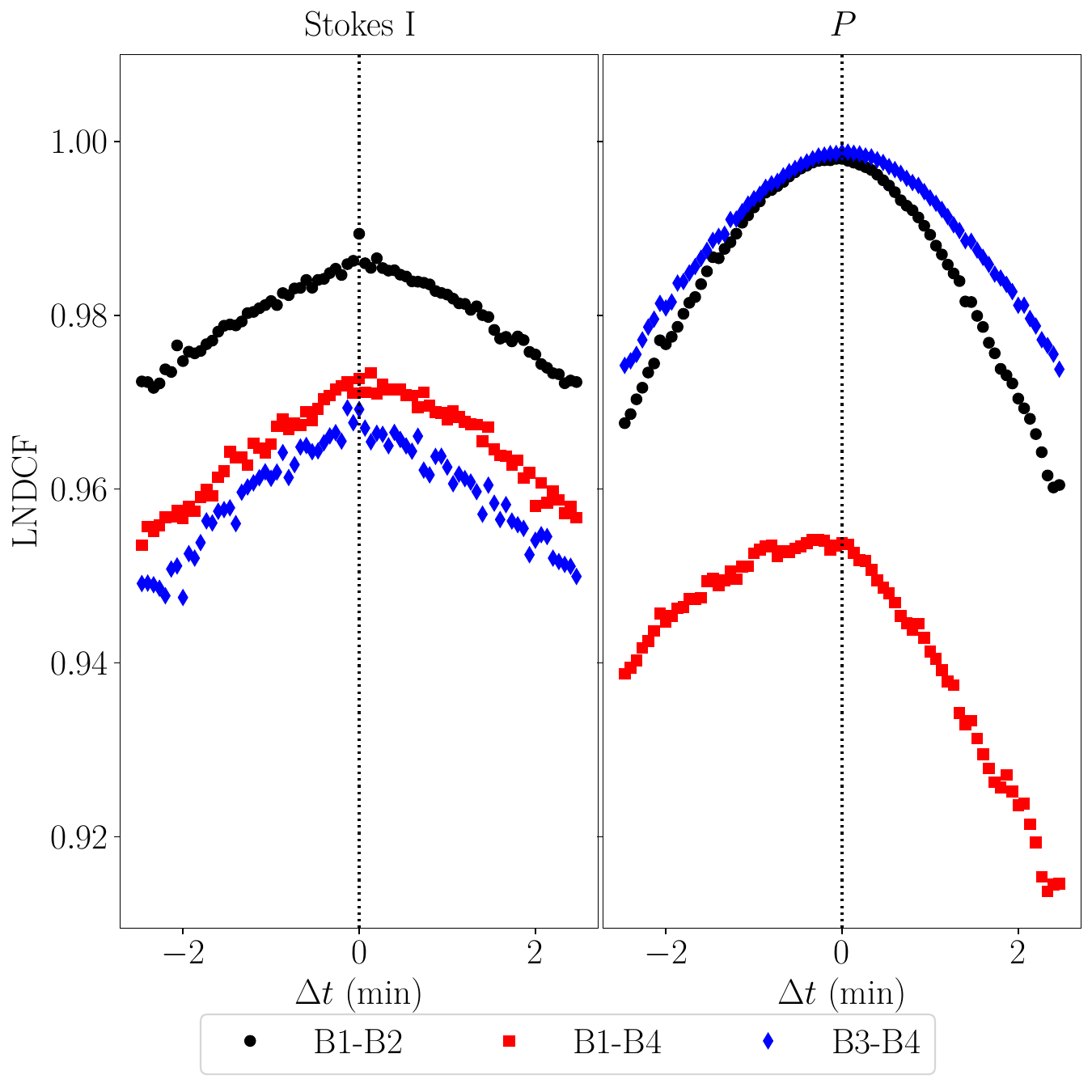}
    \caption{LNDCF between SPWs B1-B2 (black dots), B1-B4 (red squares), and B3-B4 (blue diamonds), for total flux density (left) and polarized intensity (right), for April 24.
    }
    \label{fig:LNDCF_spws}
\end{figure}

Given the abundance of light curves available (one for each parameter, day of observation of \sgra and spectral window), we first computed the LNDCF between SPWs. Strongly correlated signals between SPWs (i.e., LNDCF$_0 \gtrsim 0.95$) suggest similarity in the information provided by the different spectral windows from a physical perspective. As illustrated in Fig. \ref{fig:LNDCF_spws}, a strong correlation exists between the different SPWs, particularly within the same spectral band, both for the total intensity and the polarized intensity. Consequently, focusing our study on one spectral window per band for the subsequent analysis is appropriate, a conclusion further supported by individual inspections of each spectral window.

\begin{figure}
    \centering
    \includegraphics[width=8cm]{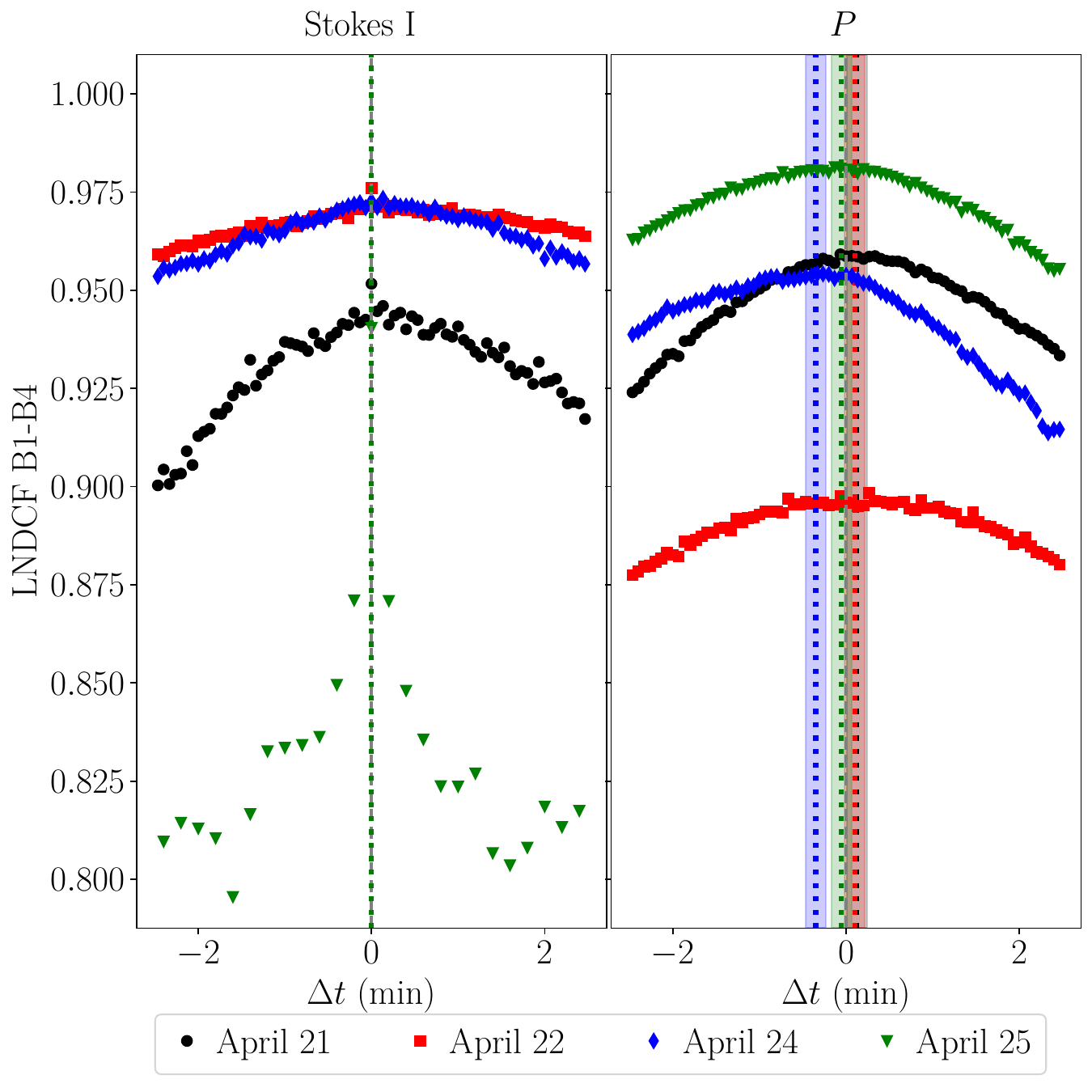}
    \caption{LNDCF between ALMA SPWs B1-B4 for all four days, showing total flux density (left) and polarized intensity (right). The dotted lines indicate the delays between SPWs B1-B4 retrieved from the LNDCF.}
    \label{fig:LNDCF_lag}
\end{figure}

Examining the LNDCF between SPWs across various spectral bands (B1-B4) at minute-scale time lags, as depicted in Fig. \ref{fig:LNDCF_lag}, reveals a clearer shift in the maximum of the LNDCF for polarized intensity on April 24, with a delay of $-21 \pm 13$ seconds, i.e., with the light curve at B1 lagging behind that at B4. This delay, occurring on the day \textit{Chandra} reported a flare, is consistent, although less pronounced than the delay of $-45 \pm 15$ seconds reported in~\citealt[][Appendix G]{Wielgus2022_orbital}, observed on April 11, 2017, when a flare was also detected by \textit{Chandra}. In contrast, the delays observed on other days remain consistent with zero: $7 \pm 9$ seconds on April 21, $6 \pm 10$ seconds on April 22, and $-3 \pm 11$ seconds on April 25. No significant delay is observed for total intensity, remaining consistent with zero across all days.

\subsection{Structure function}
\label{subsec:SF}

The  structure function  (SF) of the polarized flux density serves as a powerful diagnostic of variability in the emission, revealing characteristic timescales and amplitudes that are sensitive to the turbulence within the accretion flow. In particular, short timescale variability is expected to originate from turbulent processes occurring close to the black hole.
Moreover, GRMHD simulations predict that the variability power spectrum depends on the underlying magnetic field configuration, typically modeled as either a MAD or SANE accretion flow~\citep[e.g.,][]{SgraP5,Moscibrodzka2024arXiv241206492M}. As these models differ significantly in their magnetic flux distribution and turbulence levels, the results from the SF analysis provide insight into the plasma conditions near the event horizon.

\begin{figure}[!h]
    \centering
    \includegraphics[width=4.28cm]{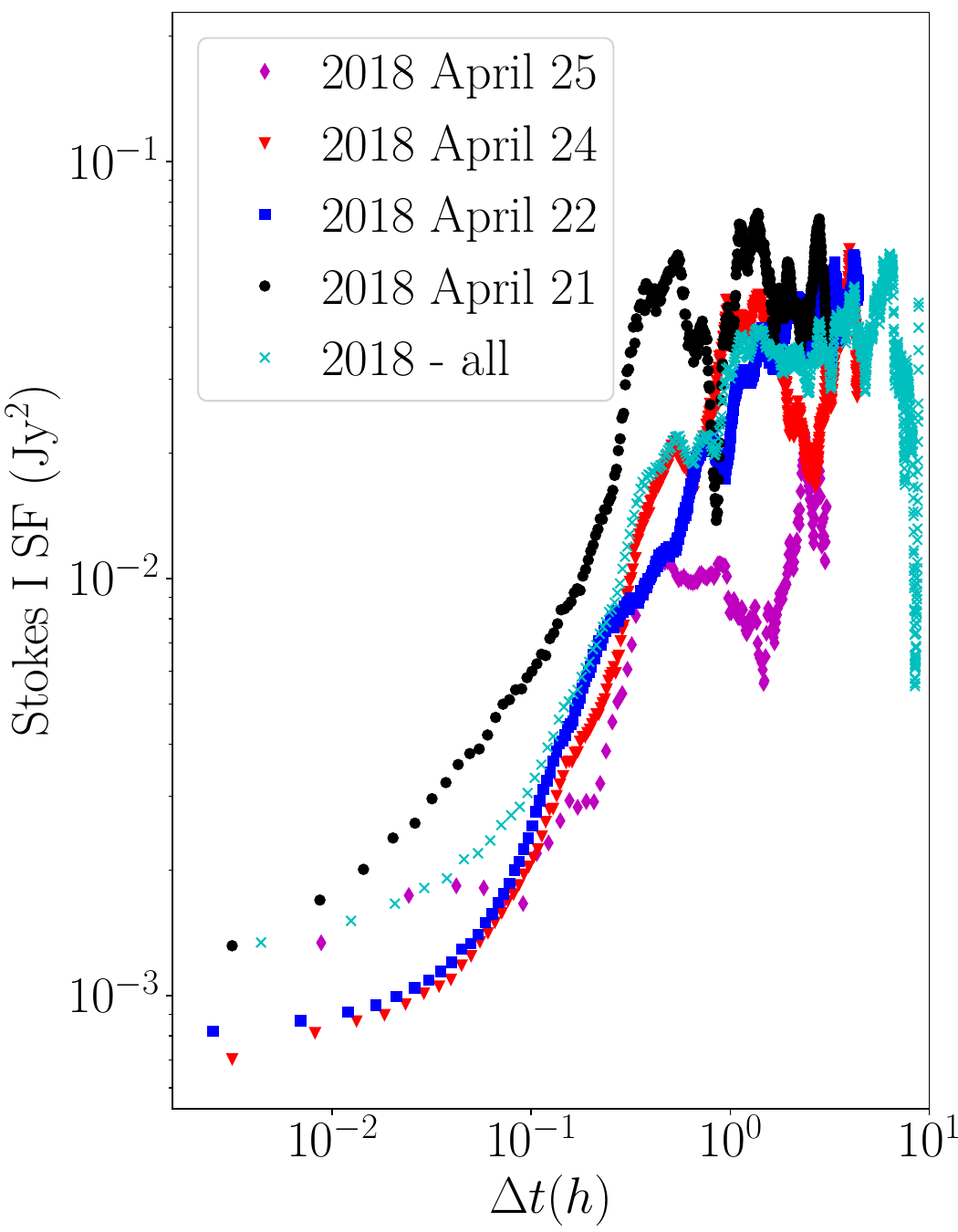}
    \includegraphics[width=4.28cm]{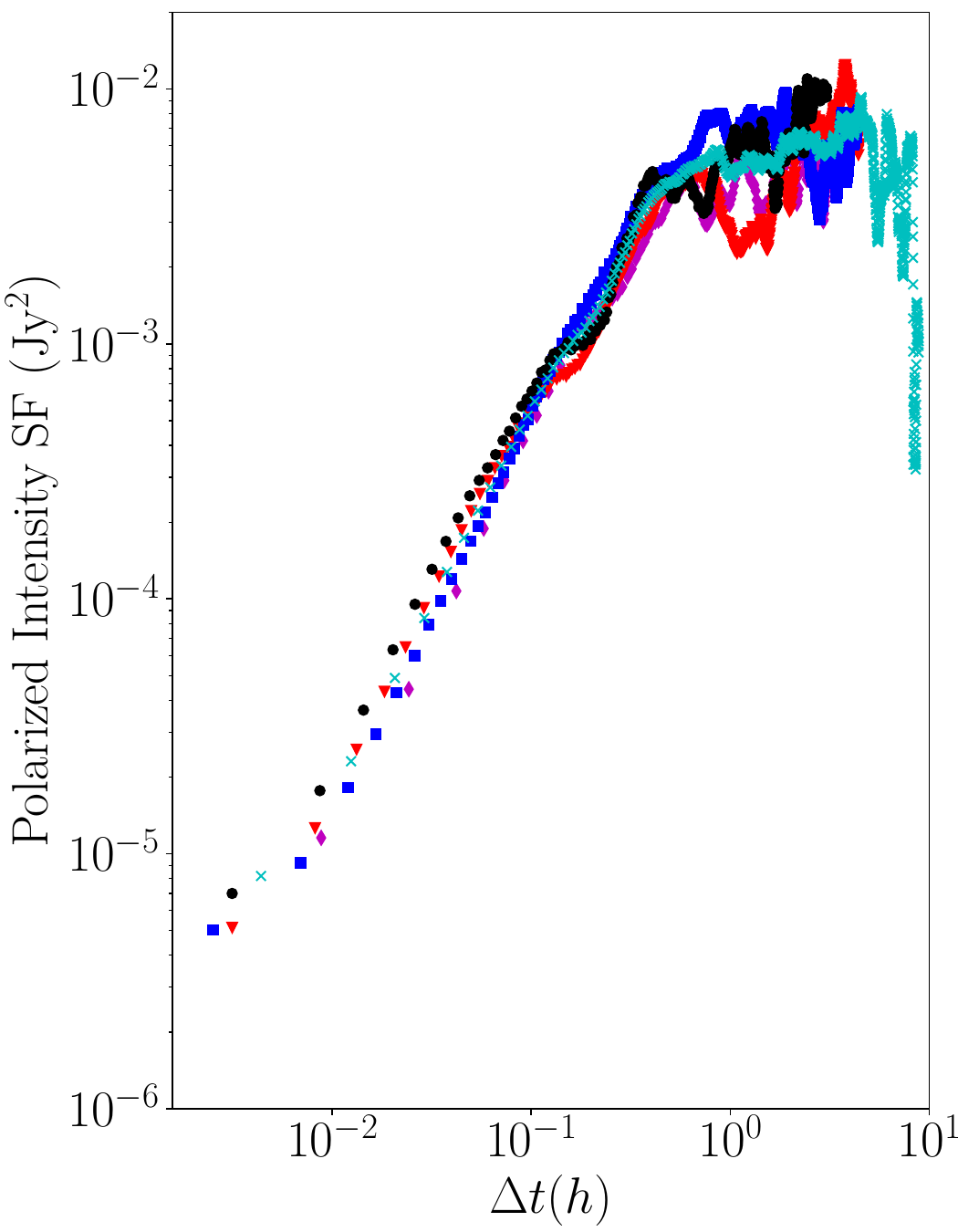}
    \caption{SF plots of the \sgra 213.1 GHz light curves for total intensity (left) and polarized intensity (right), for April 21, 22, 24 and 25, and all combined (black dots, blue squares, red diamonds, magenta triangles, and cyan crosses, respectively).
    }
    \label{fig:SF}
\end{figure}

To characterize the power spectrum and retrieve the characteristic variability timescales of our \sgra light curves, both in total flux density and polarization, we analyzed the behavior of the first-order SF \citep[see][]{Simonetti1985,Wielgus2022}. Given a time series $\{x_i\}=x_1,x_2,\ldots,x_n$ observed at times $\{t_i\}=t_1,t_2,\ldots,t_n$, the SF at a time lag $\Delta t$ is defined as the sum of all the pairs in the time series, $N_{\Delta t}$, for which $(t_j - t_i) \leq \Delta t$,
\begin{equation}
    SF(\Delta t) = \frac{1}{N_{\Delta t}} \sum_{i,j=i+1} \left(x_j - x_i \right)^2.
\end{equation}
The analysis of the SF provides information on the temporal structure of our light curves. The SF of a signal affected by measurement and calibration errors, and random noise, such as the ones analyzed in this work, presents three types of slopes in the data (see Fig. \ref{fig:SF}):

\begin{itemize}
    \item At very short timescales, the slope of the SF is almost flat, and corresponds to the level of the noise, which dominates the amplitude of our signal. This SF plateau level is twice the variance of the noise in the data. The higher the noise, the higher the plateau, which could hide the signal properties. 
    \item A steep increase in the SF, as a result of the random noise becoming less dominant compared with the real signal in our data, indicating real variability in our data. The power spectral density (PSD) slope index $\alpha_{PSD}$ can be calculated using the slope index $\alpha_{SF}$ of this increase in the SF, which follows a power law, as $\alpha_{PSD} = -(1+ \alpha_{SF})$.
    \item A plateau at long time lags, providing an estimate on the characteristic timescale of the light curves' variability.
\end{itemize}

In Fig. \ref{fig:SF} we observe that while the shapes of the total intensity SFs differ greatly across different days, both the shapes and timescales of the polarized intensity SFs are very consistent. 
Moreover, the SF for polarization is highly coherent, in contrast to the total flux density SF. This suggests that polarization arises from a coherent emission region, while total flux density comprises contributions from different, less coherent regions.

\begin{table}[ht]
\centering
\small
\caption{Results from the time-series analysis of the ALMA \sgra light curves in total flux density and polarized intensity.}
\label{tab:SF_HPF_timescales}
\renewcommand{\arraystretch}{1.25}
\begin{tabular}{ccccc}
    \toprule
     Day & SPW & $\alpha_{PSD}$ & $\alpha_{PSD}$ & Timescale \\
     (2018) &  & (SF) & ($P(\omega)$) & (h) \\
    \midrule
    \multicolumn{5}{c}{Stokes I} \\
    \midrule
    April 21 & B1 & $-2.64 \pm 0.04$ & $-2.2 \pm 0.2$ & $0.45^{+0.12}_{-0.08}$ \\
     & B4 & $-2.59 \pm 0.03$ & $-2.2 \pm 0.2$ & $0.36^{+0.07}_{-0.05}$ \\
    April 22 & B1 & $-2.235 \pm 0.012$ & $-2.19 \pm 0.12$ & $0.26^{+0.14}_{-0.12}$ \\
     & & $-1.947 \pm 0.015$ & -- & $1.78^{+0.24}_{-0.18}$ \\
     & B4 & $-2.155 \pm 0.013$ & $-2.19 \pm 0.11$ & $0.25^{+0.16}_{-0.14}$ \\
     & & $-1.842 \pm 0.017$ & -- & $1.9^{+0.4}_{-0.3}$ \\
    April 24 & B1 & $-2.43 \pm 0.03$ & $-2.23 \pm 0.13$ & $0.87^{+0.18}_{-0.13}$ \\
     & B4 & $-2.33 \pm 0.03$ & $-2.22 \pm 0.14$ & $0.87^{+0.19}_{-0.13}$ \\
    April 25 & B1 & $-2.85 \pm 0.02$ & $-2.3 \pm 0.3$ & $0.42^{+0.06}_{-0.05}$ \\
     & B4 & $-2.670 \pm 0.019$ & $-2.2 \pm 0.3$ & $0.43^{+0.05}_{-0.04}$ \\
    All days & B1 & $-2.659 \pm 0.011$ & $-2.42 \pm 0.05$ & $0.64^{+0.06}_{-0.05}$ \\
     & B4 & $-2.609 \pm 0.012$ & $-2.38 \pm 0.06$ & $0.60^{+0.06}_{-0.05}$ \\

    \midrule
    \multicolumn{5}{c}{Polarized Intensity} \\
    \midrule
    April 21 & B1 & $-2.34 \pm 0.03$ & $-2.66 \pm 0.06$ & $0.52^{+0.11}_{-0.08}$ \\
     & B4 & $-2.33 \pm 0.02$ & $-2.59 \pm 0.08$ & $0.51^{+0.10}_{-0.08}$ \\
    April 22 & B1 & $-2.529 \pm 0.008$ & $-2.65 \pm 0.05$ & $0.50^{+0.03}_{-0.03}$ \\
     & B4 & $-2.530 \pm 0.008$ & $-2.61 \pm 0.05$ & $0.50^{+0.03}_{-0.02}$ \\
    April 24 & B1 & $-2.336 \pm 0.015$ & $-2.83 \pm 0.06$ & $0.51^{+0.08}_{-0.06}$ \\
     & B4 & $-2.341 \pm 0.013$ & $-2.76 \pm 0.06$ & $0.53^{+0.06}_{-0.05}$ \\
    April 25 & B1 & $-2.359 \pm 0.018$ & $-2.62 \pm 0.08$ & $0.55^{+0.08}_{-0.06}$ \\
     & B4 & $-2.328 \pm 0.016$& $-2.59 \pm 0.09$ & $0.60^{+0.08}_{-0.06}$ \\
     All days & B1 & $-2.405 \pm 0.010$ & $-2.42 \pm 0.05$ & $0.61^{+0.05}_{-0.04}$ \\
     & B4 &$-2.388 \pm 0.010$ & $-2.61 \pm 0.03$ &$0.60^{+0.05}_{-0.04}$ \\
    \bottomrule
\end{tabular}
\tablefoot{The PSD slopes and timescales are estimated using the SF. For comparison, PSD slopes $\alpha_{PSD}$ from the HPF periodogram $P(\omega)$, described in Appendix~\ref{sec:Periodogram_results}, are also included.}
\end{table}

The results of the SF analysis, described in detail in Appendix~\ref{appendix:SF_inspection}, are summarized in Table \ref{tab:SF_HPF_timescales} (which also includes the results from the high-pass filter periodogram; see Appendix~\ref{sec:Periodogram_results}). We present the results only for the SPWs B1 and B4, as they are representative of the two ALMA lower and upper spectral sidebands. Moreover, there are no significant differences in the SF between SPWs B1-B2 and B3-B4, as evidenced by the strong correlation between different SPWs shown in Fig. \ref{fig:LNDCF_spws}, so analyzing the timescales for all SPWs would yield similar results.

\begin{figure}[ht]
    \centering
    \includegraphics[width=9cm]{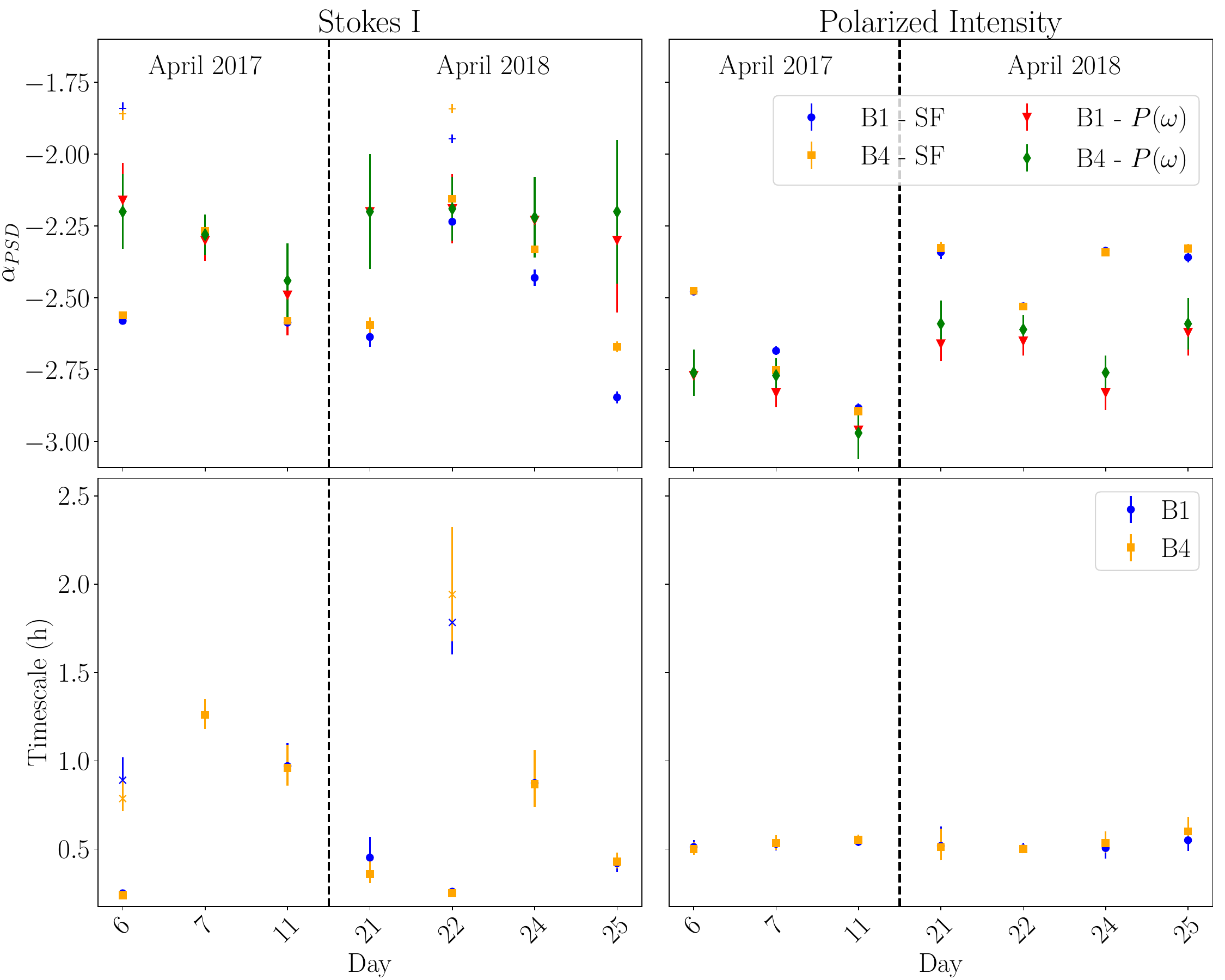}
    
    \caption{PSD slope index (top) and timescales (bottom) estimated from the SF of the total intensity (left) and polarized intensity (right) 2017 and 2018 light curves, for the spectral bands B1 (blue dots) and B4 (orange squares). A second characteristic PSD and timescale, derived from the Stokes I SF, are marked with a cross (x). PSD values are computed from the SF slope as $\alpha_{PSD} = -(1+ \alpha_{SF})$ (blue dots and orange squares for the spectral bands B1 and B4, respectively), and from the HPF periodogram (red triangles and green diamonds for the spectral bands B1 and B4, respectively; see Appendix~\ref{sec:Periodogram_results}).}
    \label{fig:SF_HPF_PSD_Times}
\end{figure}

To conclude the SF analysis, we applied the same procedure used to estimate the PSD and timescales from the SF of the 2018 light curves to the full-polarization light curves obtained from the ALMA April 2017 observations, presented in Fig. \ref{fig:ALMA2017lcurves}. Figure \ref{fig:SF_HPF_PSD_Times} presents a comparison between the results of the SF analysis of the 2017 light curves (shown in Appendix \ref{sec:2017lcurves}) and the 2018 light curves. We observe fluctuations in the timescales for total intensity, while the results for polarized intensity display a stable timescale across both years, around $0.5-0.6$ hours. This consistency further supports the argument that \sgra polarization arises from a coherent emission region.

\section{Discussion}
\label{sec:discussion}

\subsection{Polarimetric loops}
\label{subsec:QUloops}

\begin{figure*}[!ht]
    \centering
    \includegraphics[width=\textwidth]{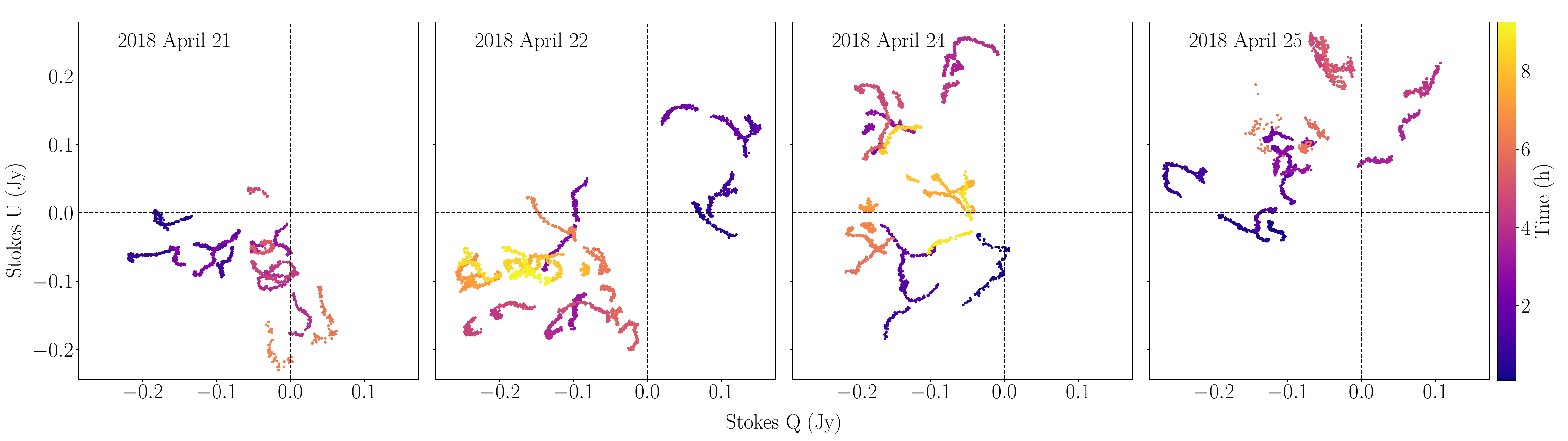}
    \caption{\sgra ALMA polarimetric loops observed in April 2018, for the spectral band B4, for all four days. The colors of the data points represent the time evolution of the $Q-U$ behavior.
    }
    \label{fig:QUloops}
\end{figure*}

Coherent variation of the measured linear polarization, forming loops in the $Q-U$ plane, can serve as a useful tool to constrain the models of the \sgra geometry, as discussed in \citet{Wielgus2022_orbital}. Such patterns may be associated with bubbles of strongly energized electrons forming as a consequence of a rapid release of magnetic energy into plasma, observed as a high-energy flare. Such bubbles (hotspots) could then transiently orbit the central black hole before being destroyed by instabilities, shearing in a differentially rotating flow, and/or radiatively cooling. \citet{Vos2022} and \citet{Vincent2024} provided detailed theoretical discussions on the formation of $Q-U$ loops, similar to the observational signatures identified in the infrared observations of flaring \sgra \citep{Gravity2018b,Gravity2023,Yfantis2024IR}. The millimeter wavelength polarimetric signatures associated with the 2017 April 11 X-ray flare were systematically studied and modeled by \citet{Yfantis2024} and \citet{Levis2024}, both concluding consistency with a clockwise hotspot motion in a compact orbit at a low inclination, as well as the dominance of a vertical magnetic field component. In Fig.~\ref{fig:QUloops}, we present the polarimetric $Q-U$ plane variation observed by ALMA in April 2018, including April 24, when \textit{Chandra} reported an X-ray flare \citep{Mossoux2020}. Some more discussion about the flare is given in Sect.~\ref{subsec:flare}. 

By eye, the 2018 $Q-U$ data do not readily exhibit coherent looping behavior that could be interpreted as orbital motion. This is in contrast to the 2017 April 11 flare, for which the loop is apparent; see Fig. 1 of \citet{Wielgus2022_orbital}. However, \citet{Ricarte2025} recently developed a method to determine $Q-U$ rotation speeds and preferential handedness of the pattern using the differential geometry of planar curves. The method is capable of statistically characterizing the local curvature on the $Q-U$ plane and infer the corresponding mean angular velocity, even in cases for which the coherent looping behavior is not visually apparent. In brief, smoothing splines are fit to each scan, and the signed Gaussian curvature is integrated over these light curves with respect to the arc length to obtain an average $Q-U$ rotation rate and clockwise fraction. In this previous work, for the 2017 data of \citet{Wielgus2022_orbital}, \citet{Ricarte2025} calculated a pattern speed of $\Omega_{QU} = -2.6 \pm 0.6 \ \mathrm{deg}\,\mathrm{t_g^{-1}}$, where $t_g = GMc^{-3}=20\,\mathrm{s}$, and that $65\% \pm 9\%$ of the scans were curved in a clockwise orientation.

We repeated this analysis for the ALMA 2018 light curves presented in this paper, and the results are shown in \autoref{tab:curvatures}.  We considered only the B4 light curves, as the EVPA evolution is almost identical across the four SPWs. This technique is not affected by an overall offset due to the RM. It may, however, be sensitive to the variable internal Faraday effects in \sgra \citep{Wielgus2024}. Systematic error bars were computed by surveying over spline fitting parameters as in \citet{Ricarte2025}.  We consistently recover clockwise motion on all days, similar to the 2017 data. Intriguingly, the most clockwise-biased day is the flaring day, 2018 April 24. Similarly, \citet{Ricarte2025} reported that the flaring period of 2017 April 11 is atypically biased towards clockwise as well. This suggests that $Q-U$ loops may become more coherent during flares, possibly due to the emergence of a dominant polarized hotspot. 

\begin{table}[htbp]
    \centering
    \small
    \caption{$Q-U$ loop speeds and fraction of scans that are clockwise.}
    \begin{tabular}{ccc}
    \hline
    Time Interval        & $\Omega_{QU} \ \mathrm{deg} \, t_g^{-1}$ & Clockwise Fraction \\
    \hline
    April 21          & $-2.7 \pm 0.7$ & $0.60 \pm 0.08$                                           \\
    April 22          & $-0.8 \pm 0.4$  & $0.55 \pm 0.03$                                        \\
    April 24        & $-1.9 \pm  0.4$   & $0.72 \pm 0.05$                                         \\
    April 25 & $-0.9 \pm 0.5$  & $0.61 \pm 0.09$                                          \\
    All Days           & $-1.6 \pm 0.9$  & $0.62 \pm 0.09$ \\ 
    \hline
    \end{tabular}
    \label{tab:curvatures}
    \tablefoot{Negative pattern speeds correspond to clockwise motion, for which there is a clear bias.}
\end{table}

For the 2018 data, our all-day average $Q-U$ rotation rate  of $-1.6 \pm 0.9 \ \mathrm{deg} \, t_g^{-1}$ is consistent within 1$\sigma$ with the 2017 measurement reported by \citet{Ricarte2025}. This agreement suggests a relatively coherent  clockwise accretion flow persisting over a timescale of $1.0 \ \mathrm{yr} \approx 1.6\times10^6 \ t_g$.
Continued monitoring will be important to assess the long-term stability of this behavior.

One possible explanation for the observed differences between the coherent loopy pattern seen during the 2017 April 11 flare and the  more disordered pattern observed on 2018 April 24 is that the 2018 X-ray flare did not actually originate in the immediate vicinity of the event horizon, where the millimeter synchrotron radiation is emitted. However, in Sect.~\ref{subsec:flare} we discuss hints of causal relation between the high-energy flare and the millimeter ALMA light curves.
Another possibility is the formation of a dominant single hotspot on 2017 April 11 and multiple simultaneous hotspots on 2018 April 24. While presence of several orbiting hotspots may scramble the detailed $Q-U$ signatures, the overall clockwise rotation pattern could still be maintained, driven by components moving with a characteristic orbital velocity. Continuous monitoring of Sgr~A* in X-ray and in millimeter is necessary to determine whether formation of coherent $Q-U$ loops associated with high-energy flaring is common.

\subsection{High-energy flare}
\label{subsec:flare}

\begin{figure}[h!]
    \centering
    \includegraphics[width=0.725\linewidth]{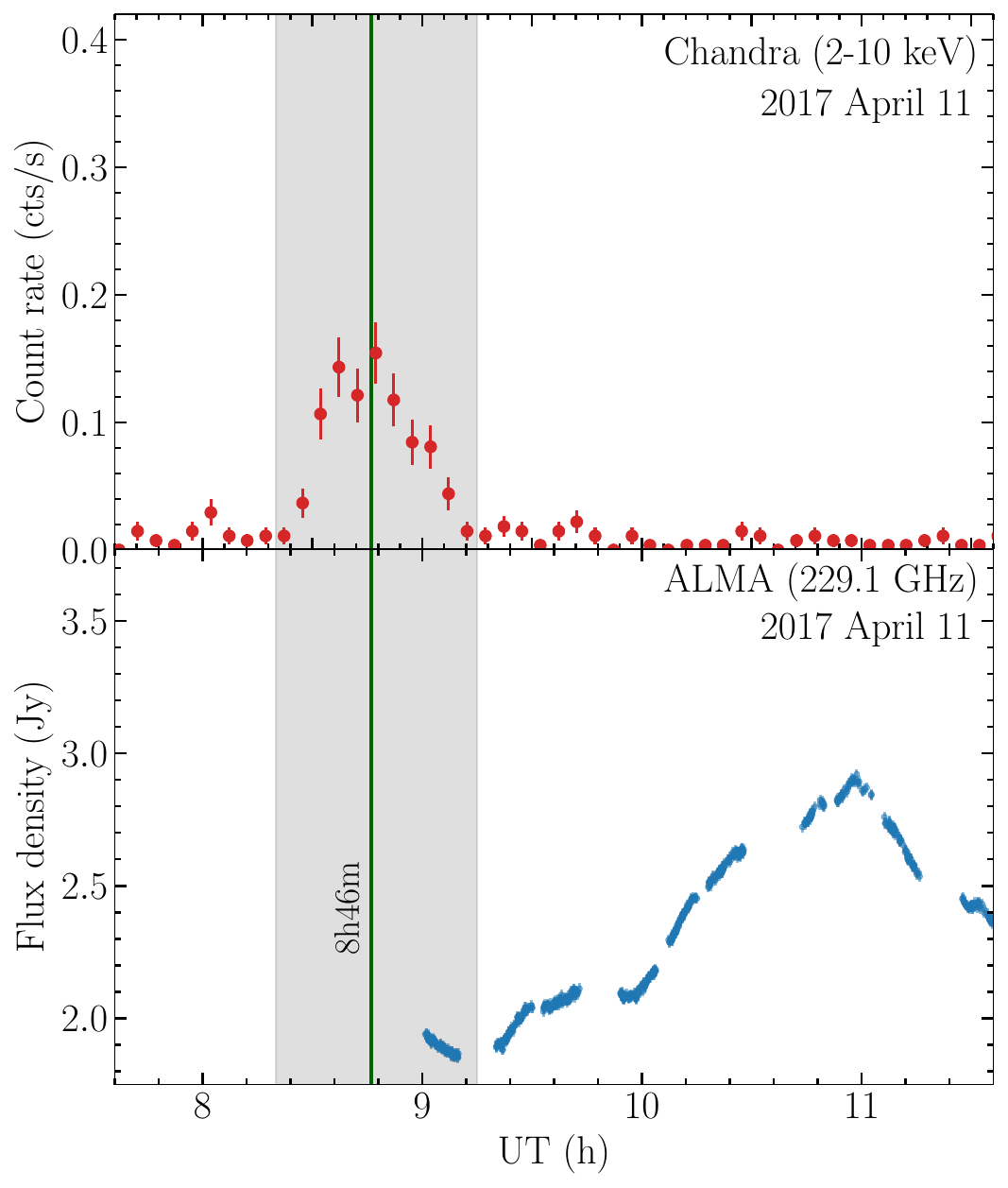}
    \includegraphics[width=0.725\linewidth]{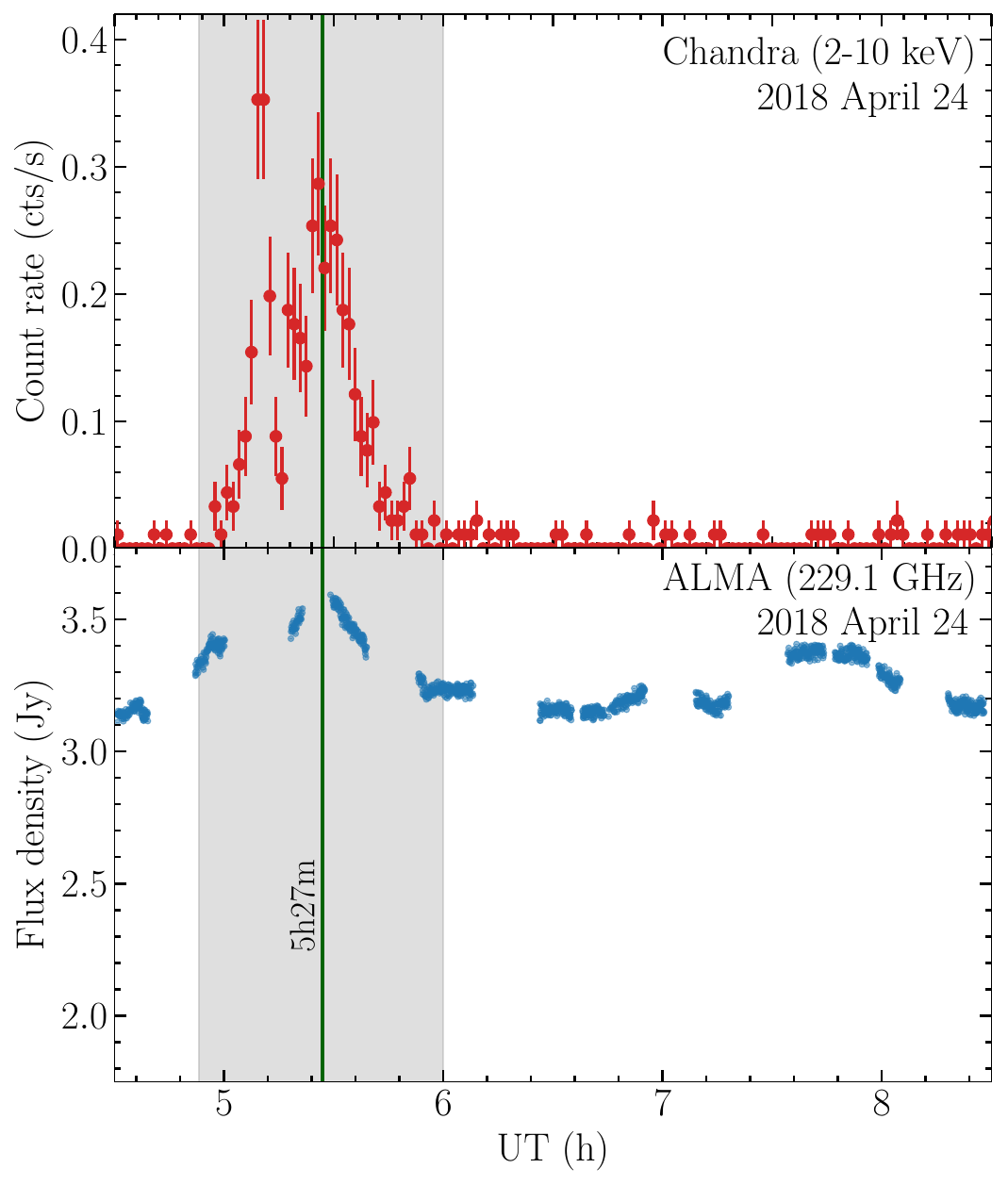}
    \caption{\textit{Chandra} counts (top) and \sgra total intensity ALMA light curves (bottom) for the flares observed in 2017 and 2018 during the EHT campaigns. The gray-shaded band marks the time range of the high-energy flare, with indicated maximum.}
    \label{fig:CHANDRA_vs_ALMA}
\end{figure}

On 2018 April 24 a high-energy flare from \sgra was detected by the \textit{Chandra} X-ray Observatory. Unlike  the X-ray flare observed on  2017 April 11 \citep{SgraP2,Wielgus2022_orbital}, which occurred just before the start of the ALMA observations, the more powerful, double-peaked 2018 flare was captured during the ALMA coverage. A comparison between the \textit{Chandra} and ALMA observations for the two events is shown in Fig.~\ref{fig:CHANDRA_vs_ALMA}. While delays between high-energy and millimeter peaks are commonly observed \citep[e.g.,][]{Yusef2008,Michail2024}, the second peak of the 2018 flare appears to coincide with the maximum of the millimeter radio emission.
This behavior is reminiscent of  the IR/submillimeter flare reported by \citet{Fazio2018}.

Although the apparent alignment between the X-ray and millimeter light curves could occur by chance, given that the millimeter emission exhibits continuous red noise variability with local maxima typically occurring every few hours, there are reasons to suspect a physical connection.
In particular, both magnetic reconnection events and millimeter-wavelength synchrotron emission in \sgra\ are expected to originate in the innermost regions of the accretion flow, near the event horizon. Assuming a causal link between the X-ray and millimeter peaks, the standard interpretation involving a transient, energized component that subsequently cools (producing delayed emission at lower frequencies) could not explain the observed simultaneity. 
An alternative explanation is that the X-ray and millimeter emissions are co-located and co-moving, and that the observed peak results from a Doppler boost associated with the motion of the emitting region. Such simultaneous emission across a broad energy range could arise if the emission region is optically thin and continuously energized, allowing for a mix of electron populations—some cooling while others are still being accelerated.

Further evidence supporting a causal link comes from the polarized light curves. On the flare day, an inter-band delay in the polarization amplitude $|P|$ is detected, with B1 lagging behind B4 by $21 \pm 13$ s (see Fig.~\ref{fig:LNDCF_lag}). This is comparable to the $45 \pm 15$ s lag reported during the 2017 flare \citep{Wielgus2022_orbital}. In both the 2017 and 2018 data sets, delays in the other Stokes parameters, as well as $|P|$ delays on non-flaring days, are consistent with zero. 
Finally, the millimeter light curves exhibit enhanced variability around the time of the flare (see Sect.~\ref{subsec:GRMHD}), and the clockwise coherence of the $Q$–$U$ polarization loop pattern is strongest on the flare day (Sect.~\ref{subsec:QUloops}). Taken together, these findings offer compelling evidence for a causal connection between the April 24 X-ray flare and its millimeter counterpart.

\subsection{Comparison with GRMHD variability}
\label{subsec:GRMHD}

\begin{figure}[h!]
    \centering
    \includegraphics[width=0.8\linewidth]{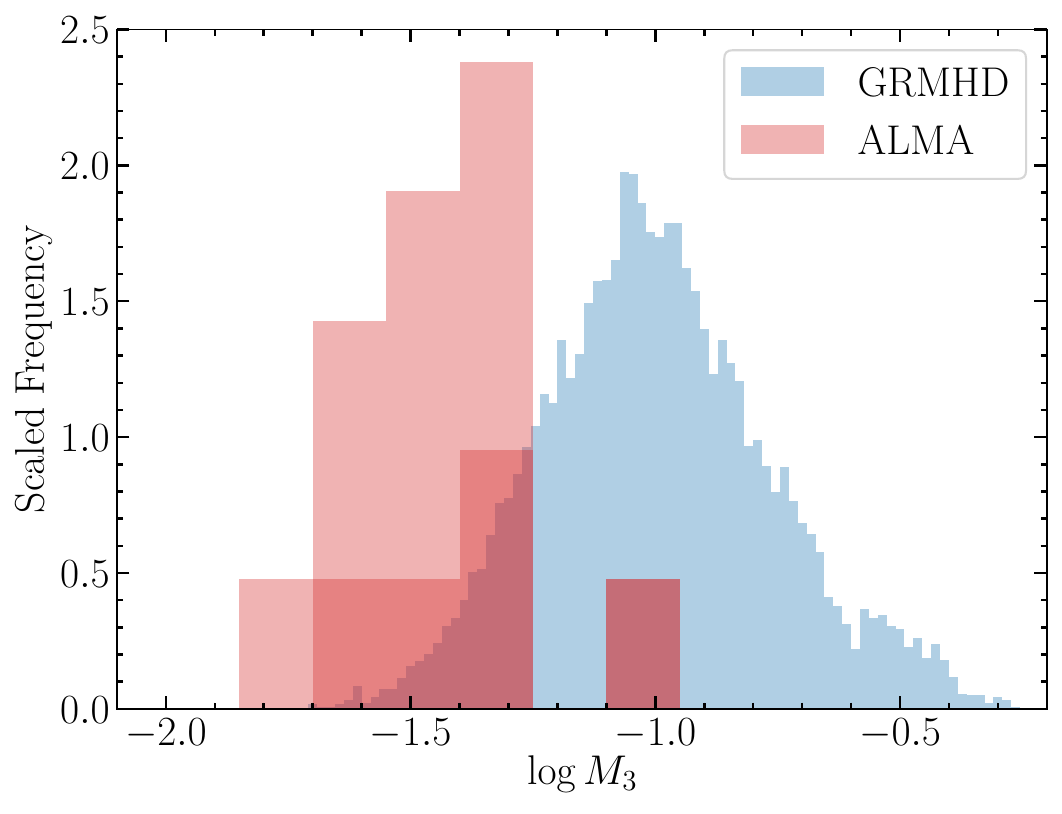}
    \\
    \vspace{0.3cm}
    \includegraphics[width=0.8\linewidth]{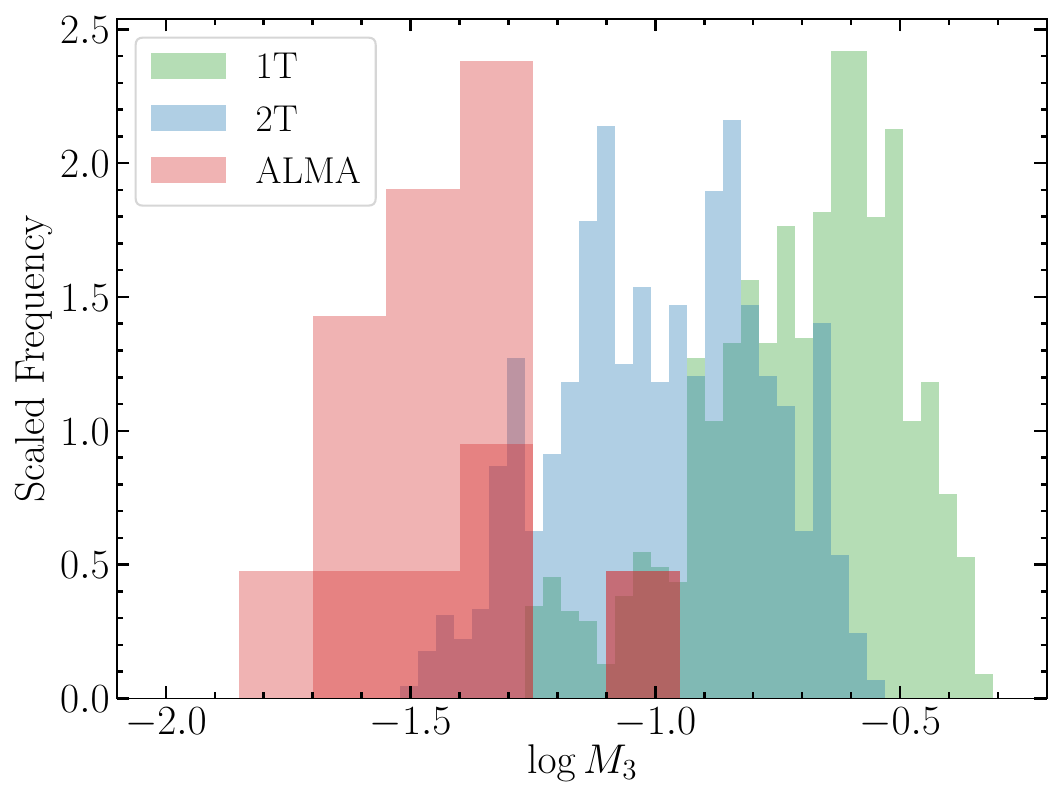}
    \caption{Distribution of $M_3$ in ALMA observations taken during the 2017 and 2018 EHT campaigns, compared to the distributions from different GRMHD models. 
    The top panel shows the comparison with the EHT library of GRMHD models, while the bottom panel shows the comparison with the 1T and 2T GRMHD models from \citet{Salas2025MNRAS.538..698S}. 
    The dark red part of the observational histogram represents the 2017 ALMA data from \citet{Wielgus2022}, and the light red part corresponds to the 2018 results introduced in this paper.}
    \label{fig:M3_GRMHD_combined}
\end{figure}

In \citet{SgraP5} the variability of \sgra, constrained by ALMA observations \citep{Wielgus2022}, was compared to predictions of general relativistic magnetohydrodynamic (GRMHD) simulations in the EHT simulation library \citep{SgraP5,Dhruv2025}. The variability metric used was the $M_3$ parameter, defined as the ratio of standard deviation to mean value (i.e., the modulation index) calculated over 3 hours long independent segments of total intensity light curves. The analysis revealed a systematic mismatch between most GRMHD models, particularly the strongly magnetized ones preferred by the static consistency metrics \citep{SgraP5,SgrAP8}, and the observations. The numerical simulations appear to overproduce variability. Here, we extended the previous work by incorporating 2018 ALMA light curves presented in this paper. In total, we have five independent measurements of $M_3$ from 2017 and nine from 2018, plotted in red in Fig.~\ref{fig:M3_GRMHD_combined} (with the 2017 data in a darker share). A single outlier with $M_3 \approx 0.1$ corresponds to radio observations following an X-ray \textit{Chandra} flare on 2017 April 11 \citep{SgraP2, Wielgus2022_orbital}. Apart from this case, both 2017 and 2018 measurements indicate consistently low amount of variability; see also Fig.~\ref{fig:modindx_2017-18}. The variability does not increase significantly on 2018 April 24, when another X-ray flare was detected by \textit{Chandra} \citep{Mossoux2020}, although it is slightly elevated earlier on that day relative to later time, with $M_3 =(0.047,0.029, 0.019)$ over three subsequent 3\,h long observing periods (the X-ray flare occurred near the end of the first 3\,h period).

We compared the $M_3$ values measured in the Stokes I observations with those derived from GRMHD simulations in the EHT library. We included 9720 synthetic light curves, each of a duration of 540\,$GMc^{-3} \approx 3$\,h. The synthetic light curves were generated from 10 independent GRMHD simulations (strongly or weakly magnetized accretion state and 5 black hole spin values), each with 36 different radiative transfer choices for thermal relativistic distribution of energy of electrons (nine inclination angles times four values of the ion-to-electron temperature ratio parameter $R_{\rm high}$). Additional details are given in \citet{SgraP5}.

The mismatch between the observed variability and that predicted by the simulations persists, suggesting that standard fluid models may be inadequate for describing the properties of turbulent, collisionless astrophysical plasmas. The accretion flow surrounding \sgra is most likely collisionless  \citep{mahadevan_quataert_19997_adafs,SgraP5}, where the electron-ion collision timescale is much longer than the accretion timescale. Under these conditions, the ions and electrons likely to maintain different temperatures \citep[][]{Shapiro1976ApJ...204..187S,Rees1982Natur.295...17R}.

Most GRMHD models consider a single-temperature ion (1T) plasma, where the electron density and temperature are not considered in the evolution equations \citep[][]{Gammie2003ApJ...589..444G,Tchekhovskoy2011MNRAS.418L..79T}. In these 1T simulations, the ion-to-electron temperature ratio $T_{\rm i}/T_{\rm e}$ is set by the $R(\beta)$ prescription, governed by the parameter $R_{\rm high}$ \citep{Moscibrodzka2016A&A...586A..38M}, which constitutes one of the main uncertainties in EHT modeling. The discrepancy in 230 GHz variability may partially stem from not self-consistently modeling the evolution of $T_{\rm e}$ when using the $R(\beta)$ prescription. In reality, $T_{\rm e}$ is determined by microphysical plasma processes and radiation interactions, rather than simply by $T_{\rm i}$. A first-principles kinetic approach is required to completely model these collisionless effects \citep[][]{Parfrey2019PhRvL.122c5101P,Crinquand2022PhRvL.129t5101C,Galishnikova2023PhRvL.130k5201G}.

Nonetheless, it is possible to effectively model the electron thermodynamics with two-temperature (2T) treatments in GRMHD simulations by describing a gas consisting of ions and electrons that share the same dynamical equations but have independent thermodynamical evolution \citep[e.g.,][]{Ressler2015MNRAS.454.1848R,Sadowski2017MNRAS.466..705S,Chael2018MNRAS.478.5209C}. 2T treatments in strongly magnetized simulations more accurately predict 230 GHz variability in \sgra compared to 1T simulations; see Fig.~\ref{fig:M3_GRMHD_combined}. Moreover, including radiative synchrotron cooling of electrons in 2T treatments further decreases $M_3$ relative to uncooled simulations \citep{Salas2025MNRAS.538..698S}. These results are consistent with theoretical expectations that the difference in adiabatic indices\footnote{The adiabatic index $\gamma$ characterizes the fluid response to compression, relating gas pressure $p_g$ and density $\rho$ via $p_g \propto \rho^\gamma$.} between relativistic electrons and non-relativistic ions effectively suppress fluctuations in the electron temperature \citep{Gammie2025ApJ...980..193G,Salas2025MNRAS.538..698S}. 
\cite{Moscibrodzka2024arXiv241206492M} demonstrated that 2T strongly magnetized models exhibit less variability at 230 GHz compared to 1T models. They find that $M_3$ increases with black hole spin but decreases slightly when non-thermal electron physics are included in the ray-tracing.

The collisionless nature of low-luminosity accretion flow such as in \sgra highlights the importance of non-ideal physics stemming from long mean free paths of particles. \cite{Dhruv2025_2} consider one such weakly collisional model \citep{chandra_emhd_2015}, which includes viscosity and heat conduction. They find that incorporating these low-collisionality effects systematically lowers the 230 GHz variability in all MAD models. Additionally, \cite{Nathanail2025} explore how the inclusion of explicit resistivity reduces the variability of GRMHD simulations, potentially helping to reconcile the different accretion flow models.

Apart from the amount of variability, we are interested in its power distribution across frequencies, as discussed using SF in Sect.~\ref{subsec:SF}. As demonstrated by \citet{Wielgus2022}, GRMHD models generally indicate a steep power law of the short timescale variability $\alpha_{PSD}$ between -2.5 and -2.9. This is steeper than the DRW process, characterized by  $\alpha_{PSD} = -2$. The power law in the 2017 ALMA total intensity light curves was estimated by \citet{Wielgus2022} to be $\alpha_{PSD} \approx -2.6$, which broadly agrees with the predictions of numerical models. We see consistent estimates of $\alpha_{PSD}$ in the 2018 data set, as presented in Fig.~\ref{fig:SF_HPF_PSD_Times} and in Table~\ref{tab:SF_HPF_timescales}. It is also worth noticing that we consistently see a sub-hour decorrelation timescale through the PSD and SF analysis, Table~\ref{tab:SF_HPF_timescales}. While these findings seem to suggest that such an analysis is sensitive to dynamical timescales in the \sgra system, these conclusions must be carefully tested and verified, particularly since the ALMA light curves data can be fitted well with stochastic Gaussian process models of significantly longer correlation timescales \citep{Wielgus2022}.

\section{Conclusions}
\label{sec:conclusions}

This work presents a comprehensive analysis of high-cadence, high signal-to-noise full-polarization light curves of \sgra, obtained with ALMA during the April 2018 EHT campaign. Notably, during the same week, the \textit{Chandra} X-ray Observatory reported a flare on April 24 (between 4:53 and 6:00 UT), enabling a joint analysis of the millimeter and X-ray light curves on the day of the flare.

We first characterized the overall variability in total intensity, which remains low, with $\sigma / \mu < 10\%$, consistent with previous EHT campaigns and earlier observations. The estimated variability remains below the levels predicted by standard accretion flow models, though recent GRMHD simulations yield similarly reduced variability \citep[][Dhruv et al., in prep.]{Moscibrodzka2024arXiv241206492M,Salas2025MNRAS.538..698S,Nathanail2025}. In contrast, the polarized intensity shows stronger variability, with $\sigma / \mu \sim 30\%$.

To quantify the polarization variability, we employed advanced time-series analysis tools. Cross-correlations between the four spectral windows (B1–B4) reveal strong inter-band coherence, with LNDCF$_0 \gtrsim 0.95$. On minute timescales, we detect no measurable delays between B1 and B4 for total intensity, consistent with optically thin synchrotron emission at 1.3 mm, an interpretation supported by the 2017 campaign as well.
For the polarized intensity, delays are consistent with zero on most days, though with marginal positive shifts on April 21–22 and marginal negative ones on April 25. On April 24—the day of the X-ray flare—we detect a statistically significant delay of $21 \pm 13$ s, with B1 lagging B4. A similar delay was reported for the 2017 flare event, further strengthening the association between polarization structure and high-energy activity.

The high quality of the ALMA light curves enabled variability analysis on short timescales. Both the SF and high-pass filter (HPF) periodogram analyses reveal red-noise behavior spanning timescales from minutes to hours. The derived power spectral densities are consistent across methods: $-2.4 \pm 0.3$ for total flux density (matching the 2017 value) and $-2.6 \pm 0.1$ for polarized intensity. Structure function analysis further reveals intra-day variability timescales of $\sim$20 minutes to 1.5 hours in total intensity, while the polarized intensity remains stable around $\sim$30 minutes—suggesting a more coherent emission region for the polarized component.

The April 24 X-ray flare offers a rare opportunity to probe the connection between X-ray and millimeter-wavelength emission. While previous flares (e.g., April 11, 2017) exhibited millimeter counterparts delayed by several hours, the 2018 event reveals near-simultaneous peaks in X-ray and millimeter emission, within a five-minute window. This is accompanied by a $\sim$20\% increase in millimeter flux density.
This simultaneity is further supported by an inter-band delay in polarized intensity ($21 \pm 13$,s), an enhanced coherence in the $Q$–$U$ polarization loops (clockwise direction), and an increased intra-day millimeter variability during the flare (with a subsequent decline thereafter).
These findings challenge the standard scenario of delayed synchrotron emission from a cooling, expanding component. Instead, they support a scenario in which the emission region is optically thin and continuously energized, allowing both electron cooling and re-acceleration to occur concurrently.

A detailed analysis of the $Q$–$U$ loop rotation rate reveals a persistent clockwise pattern, consistent with 2017 observations \citep{Ricarte2025}. This suggests a coherent structure in the underlying accretion flow on year-long timescales—corresponding to $\sim 1.6 \times 10^6 \ t_g$. This persistence provides a non-trivial constraint for GRMHD simulations and warrants comparison with wind-fed accretion flow models \citep[e.g.,][]{Ressler2020}. Continued monitoring of \sgra\ will be essential to assess the long-term stability of this signal.

Finally, similar to the essential role played by the 2017 ALMA light curve in constraining the temporal variability of \sgra and supporting the data calibration for horizon-scale imaging \citep{Blackburn2019, Wielgus2022,SgraP2}, the characteristics of the 2018 ALMA light curve exert a similarly critical influence on the imaging based on the 2018 EHT observations (EHT Collaboration in prep). In addition, it provides complementary information on the source’s variability, contributing to a more comprehensive understanding of its temporal behavior.

\begin{acknowledgements}
The Event Horizon Telescope Collaboration thanks the following organizations and programs: 
the Academia Sinica; the Academy of Finland (projects 274477, 284495, 312496, 315721); 
the Agencia Nacional de Investigación y Desarrollo (ANID), Chile via NCN19\_058 (TITANs), Fondecyt 1221421, BASAL FB210003; 
the Alexander von Humboldt Stiftung; an Alfred P. Sloan Research Fellowship; 
Allegro, the European ALMA Regional Centre node in the Netherlands, NOVA, and the astronomy institutes of the Universities of Amsterdam, Leiden, and Radboud; 
the ALMA North America Development Fund; the Black Hole Initiative (John Templeton Foundation grants 60477, 61497, 62286; Gordon and Betty Moore Foundation grant GBMF-8273); 
the Brinson Foundation; ``la Caixa'' Foundation (ID 100010434, LCF/BQ/DI22/11940027, LCF/BQ/DI22/11940030); 
the Canada Research Chairs program; \textit{Chandra} grants DD7-18089X, TM6-17006X; 
the China Scholarship Council; the China Postdoctoral Science Foundation (2020M671266, 2022M712084); 
Conicyt Fondecyt Postdoctorado (3220195); 
Consejo Nacional de Humanidades, Ciencia y Tecnología (CONAHCYT, Mexico, projects U0004-246083, U0004-259839, F0003-272050, M0037-279006, F0003-281692, 104497, 275201, 263356, CBF2023-2024-1102, 257435); 
Colfuturo Scholarship; 
the Junta de Andalucía (P18-FR-1769), the Consejo Superior de Investigaciones Científicas (2019AEP112); 
the Delaney Family John A. Wheeler Chair at Perimeter Institute; 
DGAPA-UNAM (IN112820, IN108324); 
the Dutch Research Council (NWO) VICI award (639.043.513), grant OCENW.KLEIN.113, Dutch Black Hole Consortium (NWA 1292.19.202); 
Dutch Supercomputers Cartesius and Snellius (NWO 2021.013); 
the EACOA Fellowship (ASIAA, NAOJ, CAS, KASI); 
the ERC Synergy Grants ``BlackHoleCam'' (610058), ``BlackHolistic'' (10107164); 
EU Horizon 2020 grants RadioNet (730562), M2FINDERS (101018682), FunFiCO (777740); 
ERC Advanced Grant ``JETSET'' (884631); 
EU Horizon Europe SE programme NewFunFiCO (10108625); 
Horizon ERC 2021 programme (101040021); 
FAPESP (2021/01183-8); Fonds de Recherche du Québec – Nature et Technologies (FRQNT); 
Fondo CAS-ANID CAS220010; Gordon and Betty Moore Foundation (GBMF-3561, GBMF-5278, GBMF-10423); 
the Institute for Advanced Study; the ICSC – Centro Nazionale di Ricerca in HPC, Big Data and Quantum Computing (NextGenerationEU); 
INFN Napoli TEONGRAV; the International Max Planck Research School for Astronomy and Astrophysics at Bonn and Cologne; 
EU NextGenerationEU RRF M4C2 1.1 project 2022YAPMJH; 
DFG research grant ``Jet physics on horizon scales and beyond'' (443220636); 
Joint Columbia/Flatiron Postdoctoral Fellowship (Simons Foundation); 
the Japan MEXT (JPMXP1020200109); JSPS Fellowship (JP17J08829); 
Joint Institute for Computational Fundamental Science, Japan; 
CAS Frontier Sciences Program (QYZDJ-SSW-SLH057, QYZDJSSW-SYS008, ZDBS-LY-SLH011); 
Leverhulme Trust Early Career Fellowship; 
the Max-Planck-Gesellschaft (MPG); the Max Planck Partner Group of MPG and CAS; 
MEXT/JSPS KAKENHI (18KK0090, JP21H01137, JP18H03721, JP18K13594, 18K03709, JP19K14761, 18H01245, 25120007, 19H01943, 21H01137, 21H04488, 22H00157, 23K03453);
the Generalitat Valenciana GenT Project CIDEGENT/2018/021 and grant ASFAE/2022/018;
MICINN Projects PID2019-108995GB-C22, PID2022-140888NB-C22; 
the Astrophysics and High Energy Physics programme by MCIN, with funding from NextGenerationEU (PRTR-C17I1).
MIT International Science and Technology Initiatives (MISTI); 
the Ministry of Science and Technology (MOST) of Taiwan (various grants 103–110 listed); 
NSTC of Taiwan (111-2124-M-001-005, 112-2124-M-001-014, 112-2112-M-003-010-MY3); 
Taiwan MoE Yushan Young Scholar Program; National Center for Theoretical Sciences of Taiwan; 
NASA grants 80NSSC23K1508, 80NSSC20K0527, 80NSSC20K0645; 
NASA Hubble Fellowship Program (Einstein Fellowship; grants HST-HF2-51431.001-A, HST-HF2-51482.001-A, HST-HF2-51539.001-A, HST-HF2-51552.001A); 
NINS of Japan; National Key R\&D Program of China (2016YFA0400704, 2017YFA0402703, 2016YFA0400702); 
NSTC (111-2112-M-001-041, 111-2124-M-001-005, 112-2124-M-001-014); 
the US NSF (AST-0905844, AST-0922984, AST-1126433, OIA-1126433, AST-1140030, DGE-1144085, AST-1207704, AST-1207730, AST-1207752, MRI-1228509, OPP-1248097, AST-1310896, AST-1440254, AST-1555365, AST-1614868, AST-1615796, AST-1715061, AST-1716327, AST-1726637, OISE-1743747, AST-1743747, AST-1816420, AST-1935980, AST-1952099, AST-2034306, AST-2205908, AST-2307887); 
NSF Astronomy and Astrophysics Postdoctoral Fellowship (AST-1903847);
NSERC (Canada); NRF Korea (grants NRF-2015H1D3A1066561, RS-2024-00407499, 2019R1F1A1059721, 2021R1A6A3A01086420, 2022R1C1C1005255, 2022R1F1A1075115); 
NOVA Virtual Institute of Accretion fellowships; NOIRLab (AURA/NSF); 
Onsala Space Observatory (VR grant 2017-00648); 
Perimeter Institute (Government of Canada and Province of Ontario); 
Portuguese FCT (CEEC 5th edition, UIDB/04106/2020, UIDP/04106/2020, PTDC/FIS-AST/3041/2020, CERN/FIS-PAR/0024/2021, 2022.04560.PTDC); 
Princeton Gravity Initiative; 
Ministerio de Ciencia e Innovación, Spain (PGC2018-098915-B-C21, AYA2016-80889-P, PID2019-108995GB-C21, PID2020-117404GB-C21, RYC2023-042988-I); 
University of Pretoria and SuperMicro SEEDING GRANT 2020; 
Shanghai Municipality grant 22JC1410600; Shanghai CAS Pilot Program JCYJ-SHFY-2021-013; Spanish MCIN/AEI grant CEX2021-001131-S; 
the Spinoza Prize SPI 78-409; 
South African Research Chairs Initiative through SARAO (NRF grant 77948); 
Swedish Research Council (VR); the Taplin Fellowship; the Toray Science Foundation; 
UK STFC (ST/X508329/1); 
US DOE via LANL (contract 89233218CNA000001); 
YCAA Prize Postdoctoral Fellowship; 
NRF Korea/MSIT grant RS-2024-00449206; 
CAPES Brazil PROEX 88887.845378/2023-00; 
Millenium Nucleus NCN23\_002 (TITANs); Comité Mixto ESO-Chile.

This research has made use of NASA's Astrophysics Data System. We gratefully acknowledge the support provided by the extended staff of the ALMA, from the inception of the ALMA Phasing Project through the observational campaigns of 2017 and 2018.

This work has been supported by the grant PRE2020-092200 funded by MCIN/AEI/10.13039/501100011033 and by ESF invest in your future.

The Submillimeter Array is a joint project between the Smithsonian Astrophysical Observatory and the Academia Sinica Institute of Astronomy and Astrophysics and is funded by the Smithsonian Institution and the Academia Sinica.
We recognize that Maunakea, location of the SMA, is a culturally important site for the indigenous Hawaiian people; we are privileged to study the cosmos from its summit.

This paper makes use of the following ALMA data: 2017.1.00797.V ALMA is a partnership of ESO (representing its member states), NSF (USA) and NINS (Japan), together with NRC (Canada), MOST and ASIAA (Taiwan), and KASI (Republic of Korea), in cooperation with the Republic of Chile. The Joint ALMA Observatory is operated by ESO, AUI/NRAO and NAOJ. We thank Frederic Vincent for his help.

\end{acknowledgements}

\bibliographystyle{aa}
\bibliography{references}

\begin{appendix}

\section{Absorption line at 227 GHz}
\label{sec:absorption}

\begin{figure}[!h]
    \centering
    \includegraphics[width=8.5cm]{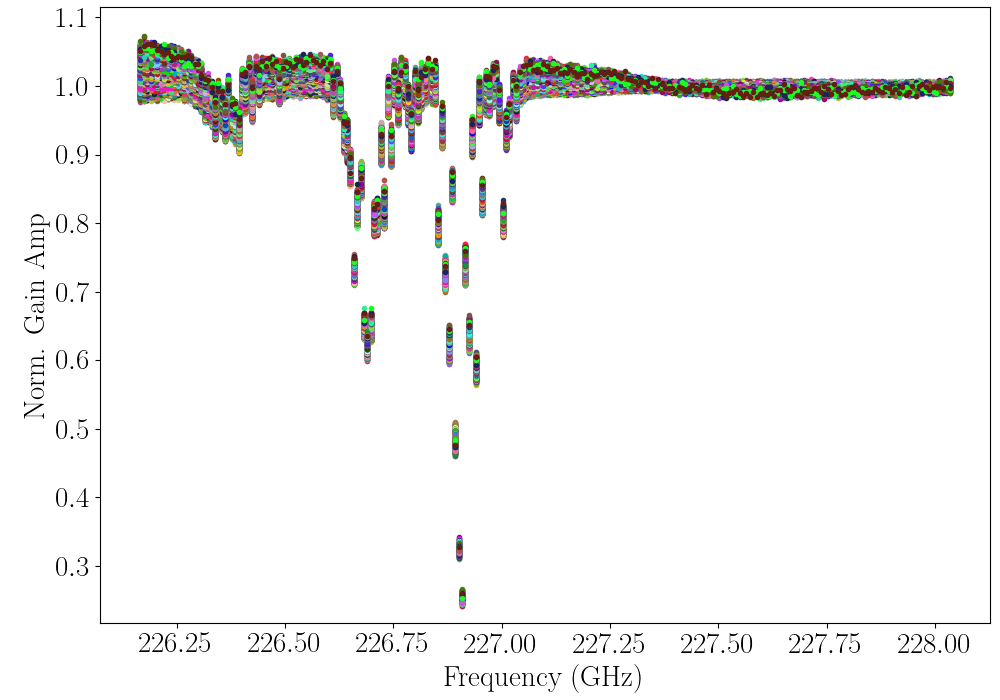}
    \caption{Bandpass of \sgra for the April 22 observations indicating the channels impacted by absorption at B3. The bandpass remains virtually unchanged across all days.
    }
    \label{fig:absorption_chan_spw2}
\end{figure}

Figure~\ref{fig:absorption_chan_spw2} shows the spectrum of \sgra\ in the B3 spectral window on April 22, revealing clear absorption features in channels 17–33 and 56–112. These absorption features are consistently observed across all days of the campaign. Foreground absorption toward \sgra\ at 226.91 GHz was previously reported in the ALMA 2017 data by \citet[][]{Goddi2021}. This absorption causes a flux suppression of approximately 4\% in the B3 spectral window. To ensure consistency across the \sgra\ light curves from all spectral windows, we flagged the frequency channels affected by absorption prior to further analysis.

\section{Light curves of EHT targets}
\label{ap:visib_cal_targets}

\begin{figure*}[!h]
    \centering
    \includegraphics[width=16.1cm]{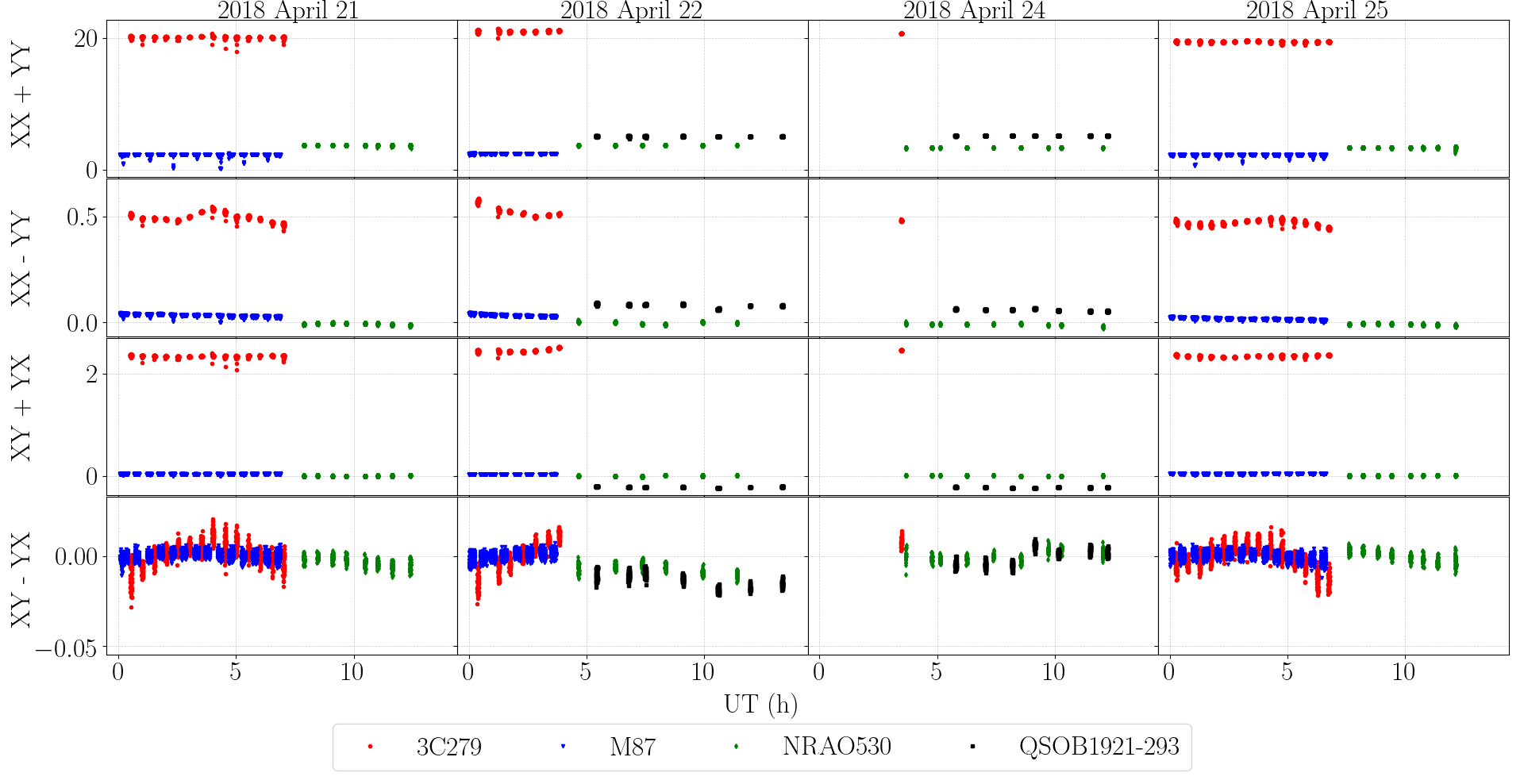}
    \caption{Reconstructed visibility-amplitude light curves for the EHT targets obtained with ALMA during the 2018 EHT VLBI campaign:  3C 279 (red circles), M87 (blue triangles), NRAO 530 (green diamonds), and QSO B1921-293 (black squares).}
    \label{fig:visib_cal_targets}
\end{figure*}

Figure~\ref{fig:visib_cal_targets} shows the visibility amplitudes per integration time for each of the EHT targets observed alongside \sgra, 
on April 21, 22, 24, and 25. We display both the sum and difference of the parallel-hand correlations ($XX$ and $YY$) and the cross-hand correlations ($XY$ and $YX$), which correspond to the Stokes I, Q, U, and V light curves, respectively, according to the Radio Interferometer Measurement Equation formalism~\citep{Smirnov2011}.
The visibility amplitude light curves confirm that, unlike \sgra, the other AGN targets exhibit stable flux densities throughout each night of observation. 
This stability is consistent with the modulation indices, at least an order of magnitude lower than those of \sgra, indicating that \sgra’s variability is intrinsic.

\section{Manual reduction of ALMA data}
\label{appendix:manual}
We employed a second independent method to derive the ALMA light curves. This method utilizes the SEFD-based data reduction process of "ALMA A2" described in \citet{Wielgus2022} to produce calibrated data by correcting for atmospheric water vapor and antenna system temperature effects before performing a bandpass, flux density, and polarization calibration. 
Unlike ALMA A2, where visibilities on short baselines are flagged, this method (similar to that described in Sect. \ref{subsec:QA2}) accounts for the contributions of the minispiral to the observed variability and flux density, correcting for these effects. 

\begin{figure}[!h]
    \centering
    \includegraphics[width=8.75cm]{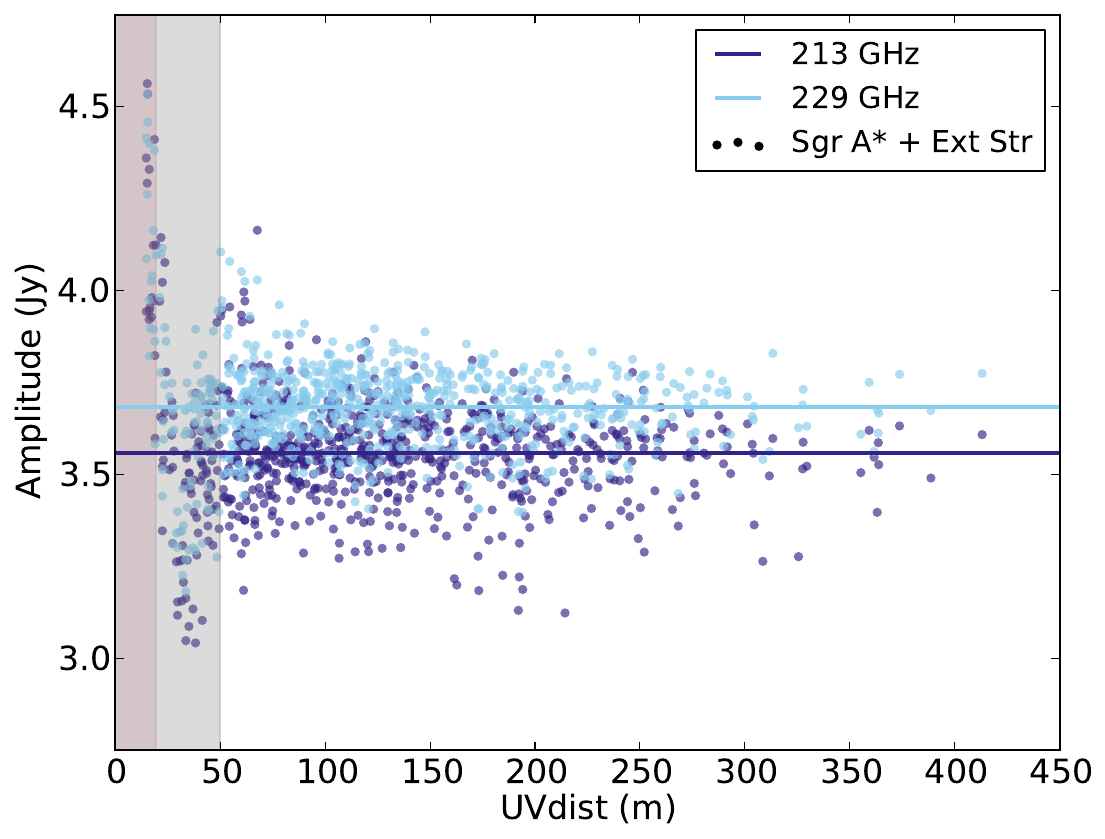}
    \caption{Snapshot of the calibrated data for \sgra at 7:23 UTC on April 25 from the SEFD-based data reduction. The colored circles represent the combined visibilities of the minispiral and \sgra. The solid colored lines are the mean flux density in each spectral window. The gray-shaded area represents the range of visibilities we consider for the minispiral (0-50\,m). The red-shaded area represents the flagging cutoff for the extended structure used by SMA (0-30 k$\lambda$).}    \label{fig:uvdist}
\end{figure}

Visibilities representing the point source were defined as those at a $uv$-distance greater than 50 m, while those at less than 50 m describe the minispiral plus time-variable point source. We used 50 meters rather than 50 k$\lambda$ as a cutoff to ensure independence from  the central wavelength of the spectral windows. Figure \ref{fig:uvdist} shows that most of the extended structure flux is seen on baselines shorter than 50 m, with the data at longer baselines predominantly corresponding to  an unresolved point source as the angular resolution increases. 

As the first step of the analysis, phase self-calibration was performed on all data using the extended baselines ($>$50 m) to remove residual phase errors, correcting each antenna to produce updated visibilities that describe a central point source. 

Next, a central point source was fitted to all data for all baselines using the software package UVMULTIFIT \citep{uvmultifit}. The residual of this fit is the minispiral, which was then used for amplitude self calibration within each observing block of \sgra between calibrators. 

Due to a lack of sufficient data at short baselines ($<$50 m), some antennas were absent from the gain amplitude table generated from the minispiral. The gain amplitude from the minispiral self-calibration was applied to both the full (combined \sgra point source and minispiral) data and the minispiral-only data. The minispiral-only data was then subtracted from the full data, leaving behind a central point source. 

This  point-source-only data was subsequently  used for amplitude self-calibration of the antennas that did not receive amplitude gains from the minispiral self-calibration. Finally, a last  amplitude self-calibration was performed using all baselines. The resulting visibilities were used to calculate the total intensity of \sgra, which was then  averaged for all baselines. Finally, the total intensity was plotted as a function of time to create the light curve.

\section{SMA observations and data processing}
\label{appendix:sma}
SMA observed \sgra for roughly four hours (between 10 and 15 UT) in full-polarization mode, starting as the source was setting in Chile and rising in Hawaii. 
During these observations,  SMA had either six or seven antennas on sky, with an integration time of 9.8 s and a total spectral bandwidth of 16 GHz (8 GHz per sideband, covering 208.1--216.1 and  224.1--232.1 GHz in the lower and upper sidebands, respectively). The array was configured in its extended mode, with typical baseline lengths ranging from 50 to 200 m.

SMA data were initially processed through the \emph{COMPASS} pipeline (Keating et al., \emph{in prep.}), which performs flagging and calibration. After applying bandpass corrections, the data were spectrally averaged by a factor of 128, yielding a final spectral resolution of 17.825 MHz. Bandpass solutions were obtained using 3C 279, while flux calibration was derived using Callisto and the Butler-JPL-Horizons 2012 models.\footnote{ALMA Memo \#594} Amplitude gains were solved using NRAO\,530 and J1924--2914 (the latter observed during two tracks) and applied on an elevation-based basis, with typical amplitude corrections of order 5--10\% below $25^\circ$ in elevation. Due to strong line absorption features, spectral channels between 226.6 and 227.0\,GHz were flagged (see Appendix \ref{sec:absorption}).

After applying the calibration gain solutions, self-calibration gains  were derived under the assumption of a point-source model for \sgra, solving for a 10 s interval. Once self-calibration gains were applied, the SMA light curve   was generated using a naturally weighted average of all visibilities for each time interval and spectral window. A final round of flagging was performed, removing data points with significant window-to-window or interval-to-interval variations. Finally, the data   were averaged over seven integration intervals (equivalent to 70 s) to improve  the signal-to-noise ratio.

\begin{figure*}[!h]
    \centering
    \includegraphics[width=0.9\linewidth]{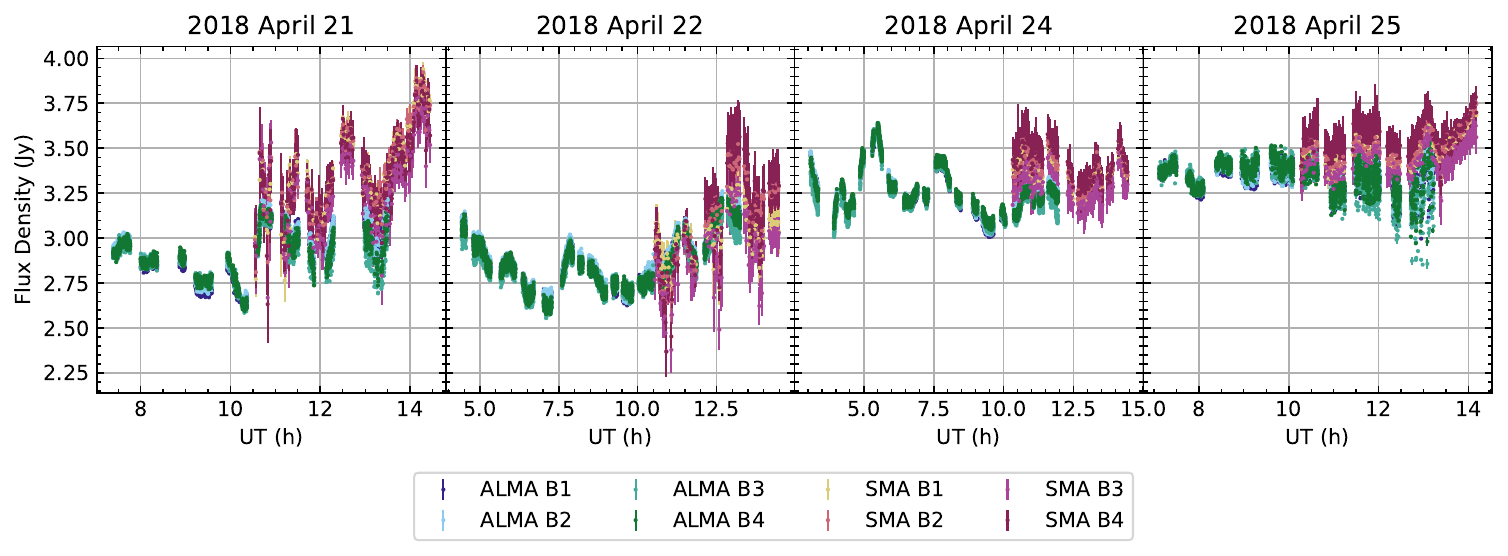}
    \caption{\sgra light curves for ALMA and SMA at all four available bands.} 
    \label{fig:bothsites}
\end{figure*}

Figure \ref{fig:bothsites} presents the QA2 ALMA and SMA light curves for each day across the four available bands. Visually, the two datasets exhibit similar structures during overlapping periods and display flux densities consistent within our estimated flux scale uncertainty of $\lessapprox 10\%$. 
To quantitatively assess the agreement between the SMA and ALMA light curves, we applied the Locally Normalized Discrete Correlation Function \citep[LNDCF;][]{Lehar1992}, as detailed in Sect. \ref{subsec:spws_cross_corr}. We analyzed the correlation during overlapping time intervals, matching spectral windows between the two datasets. The cross-correlation values are strong, reaching approximately 0.8$\pm$0.1 on April 22 and April 24, and 0.7$\pm$0.1 on April 25. On April 21, the LNDCF is slightly lower, around 0.6$\pm$0.1. 
The reduced correlation on April 21 and 25 is likely due to the lower elevation of \sgra during ALMA observations coinciding with SMA coverage, leading to lower observed flux densities and increased noise in the light curves. Given this overall consistency, also in agreement with the findings of~\citet{Wielgus2022}, we based the subsequent variability analysis on ALMA data alone.

\section{Characteristics of the 2018 ALMA light curves}
\label{ap:2018_lcurves_table}

Table \ref{tab:lcurves_data} summarizes the main variability characteristics of the full-polarization \sgra 2018 ALMA light curves presented in this work and plotted in Figs. \ref{fig:ALMAlcurves} and \ref{fig:ALMAlcurves_products}.

\begin{table*}[!ht]
    \centering
    \small
    \caption{ALMA \sgra light curves,  in total flux density, polarized intensity, and Stokes V, presented in this paper.
    }
    \label{tab:lcurves_data}
    \begin{tabular}{ccccccccc}
    \toprule
    Day & UT & Duration & Parameter & \multicolumn{2}{c}{Spectral Window} & Samples & Flux density & Modulation  \\
    (2018) & (h) & (h) & & label & Frequency (GHz) & & $\mu\pm\sigma$ (Jy) & $\sigma/\mu$  \\
    \midrule
    April 21 & 7:23 - 13:30 & 6.12 & Stokes I & B1 & $212.1 - 214.1$ & 1934 & $2.89 \pm 0.15$ & 0.050  \\
    & & & & B2 & $ 214.1 - 216.1 $ & 1934 & $2.90 \pm 0.14$ & 0.049  \\
    & & & & B3 & $ 226.1 - 228.1 $ & 1894 & $2.88 \pm 0.12$ & 0.042  \\
    & & & & B4 & $ 228.1 - 230.1 $ & 1911 & $2.89 \pm 0.13$ & 0.044  \\
    
    & & & P & B1 & $212.1 - 214.1$ & 1934 & $0.10 \pm 0.05$ & 0.49  \\
    & & & & B2 & $ 214.1 - 216.1 $ & 1934 & $0.10 \pm 0.05$ & 0.50  \\
    & & & & B3 & $ 226.1 - 228.1 $ & 1894 & $0.12 \pm 0.05$ & 0.44  \\
    & & & & B4 & $ 228.1 - 230.1 $ & 1911 & $0.12 \pm 0.05$ & 0.43  \\
    
    & & & Stokes V & B1 & $212.1 - 214.1$ & 1934 & $-0.027 \pm 0.009$ & 0.33  \\
    & & & & B2 & $ 214.1 - 216.1 $ & 1934 & $-0.027 \pm 0.009$ & 0.33  \\
    & & & & B3 & $ 226.1 - 228.1 $ & 1894 & $-0.030 \pm 0.010$ & 0.33  \\
    & & & & B4 & $ 228.1 - 230.1 $ & 1911 & $-0.032 \pm 0.009$ & 0.29  \\
    
    \midrule
    April 22 & 4:25 - 13:15 & 8.83 & Stokes I & B1 & $212.1 - 214.1$ & 3436 & $2.86 \pm 0.15$ & 0.052  \\
    & & & & B2 & $ 214.1 - 216.1 $ & 3464 & $2.87 \pm 0.15$ & 0.052  \\
    & & & & B3 & $ 226.1 - 228.1 $ & 3471 & $2.83 \pm 0.13$ & 0.046  \\
    & & & & B4 & $ 228.1 - 230.1 $ & 3483 & $2.85 \pm 0.15$ & 0.053  \\
    
    & & & P & B1 & $212.1 - 214.1$ & 3436 & $0.15 \pm 0.05$ & 0.31  \\
    & & & & B2 & $ 214.1 - 216.1 $ & 3464 & $0.16 \pm 0.05$ & 0.31  \\
    & & & & B3 & $ 226.1 - 228.1 $ & 3471 & $0.17 \pm 0.06$ & 0.32  \\
    & & & & B4 & $ 228.1 - 230.1 $ & 3483 & $0.18 \pm 0.06$ & 0.32  \\
    
    & & & Stokes V & B1 & $212.1 - 214.1$ & 3436 & $-0.020 \pm 0.013$ & 0.65  \\
    & & & & B2 & $ 214.1 - 216.1 $ & 3464 & $-0.020 \pm 0.013$ & 0.65  \\
    & & & & B3 & $ 226.1 - 228.1 $ & 3471 & $-0.020 \pm 0.011$ & 0.55  \\
    & & & & B4 & $ 228.1 - 230.1 $ & 3483 & $-0.022 \pm 0.010$ & 0.46  \\

    \midrule
    April 24 & 3:05 - 11:56 & 8.86 & Stokes I & B1 & $212.1 - 214.1$ & 3030 & $3.24 \pm 0.12$ & 0.037  \\
    & & & & B2 & $ 214.1 - 216.1 $ & 3029 & $3.25 \pm 0.12$ & 0.037  \\
    & & & & B3 & $ 226.1 - 228.1 $ & 2995 & $3.24 \pm 0.12$ & 0.037  \\
    & & & & B4 & $ 228.1 - 230.1 $ & 3014 & $3.25 \pm 0.12$ & 0.036  \\
    
    & & & P & B1 & $212.1 - 214.1$ & 3030 & $0.17 \pm 0.05$ & 0.30  \\
    & & & & B2 & $ 214.1 - 216.1 $ & 3029 & $0.17 \pm 0.05$ & 0.30  \\
    & & & & B3 & $ 226.1 - 228.1 $ & 2995 & $0.18 \pm 0.06$ & 0.33  \\
    & & & & B4 & $ 228.1 - 230.1 $ & 3014 & $0.18 \pm 0.06$ & 0.33  \\
    
    & & & Stokes V & B1 & $212.1 - 214.1$ & 3030 & $-0.011 \pm 0.015$ & 1.33  \\
    & & & & B2 & $ 214.1 - 216.1 $ & 3029 & $-0.012 \pm 0.015$ & 1.28  \\
    & & & & B3 & $ 226.1 - 228.1 $ & 2995 & $-0.015 \pm 0.016$ & 1.07  \\
    & & & & B4 & $ 228.1 - 230.1 $ & 3014 & $-0.015 \pm 0.016$ & 1.07  \\
    
    \midrule
    April 25 & 7:07 - 13:14 & 6.12 & Stokes I & B1 & $212.1 - 214.1$ & 744 & $3.34 \pm 0.07$ & 0.021  \\
    & & & & B2 & $ 214.1 - 216.1 $ & 744 & $3.36 \pm 0.07$ & 0.021  \\
    & & & & B3 & $ 226.1 - 228.1 $ & 744 & $3.36 \pm 0.10$ & 0.029  \\
    & & & & B4 & $ 228.1 - 230.1 $ & 743 & $3.36 \pm 0.08$ & 0.025  \\
    
    & & & P & B1 & $212.1 - 214.1$ & 744 & $0.14 \pm 0.04$ & 0.29  \\
    & & & & B2 & $ 214.1 - 216.1 $ & 744 & $0.15 \pm 0.04$ & 0.28  \\
    & & & & B3 & $ 226.1 - 228.1 $ & 744 & $0.16 \pm 0.05$ & 0.31  \\
    & & & & B4 & $ 228.1 - 230.1 $ & 743 & $0.17 \pm 0.05$ & 0.30  \\
    
    & & & Stokes V & B1 & $212.1 - 214.1$ & 744 & $-0.032 \pm 0.014$ & 0.44  \\
    & & & & B2 & $ 214.1 - 216.1 $ & 744 & $-0.032 \pm 0.015$ & 0.47  \\
    & & & & B3 & $ 226.1 - 228.1 $ & 744 & $-0.036 \pm 0.015$ & 0.42  \\
    & & & & B4 & $ 228.1 - 230.1 $ & 743 & $-0.038 \pm 0.016$ & 0.42  \\
    
    \bottomrule
    \end{tabular}
    \tablefoot{Stokes V values are reported for completeness, but should be considered tentative as their levels fall below ALMA’s guaranteed accuracy for CP measurements.}
\end{table*}

\section{Details of the structure function analysis}
\label{appendix:SF_inspection}

Section~\ref{subsec:SF} presents the main results of the SF analysis of the \sgra 2018 ALMA light curves. In this appendix we examine in greater detail the behavior of the SF and describe the methodology employed to estimate PSD slopes and variability timescales.

In Fig. \ref{fig:SF} the SF of total intensity for April 21 presents a steep increase at a shorter time lag than the other days, and a noticeable dip at long timescales, where a clear plateau is usually expected. Inspection of the light curves presented in Fig. \ref{fig:ALMAlcurves} reveals that this feature arises from data sampling, particularly the sudden increase in flux density around 3.5 hours, where there is a gap in observation that does not capture the complete increase in flux density (i.e., it is a consequence of an extreme event in our data). The sampling effect on the SF, together with the limited duration of our light curves, also explains the oscillations of the plateau at long timescales.

\begin{figure}[!h]
    \centering
    \includegraphics[width=4.28cm]{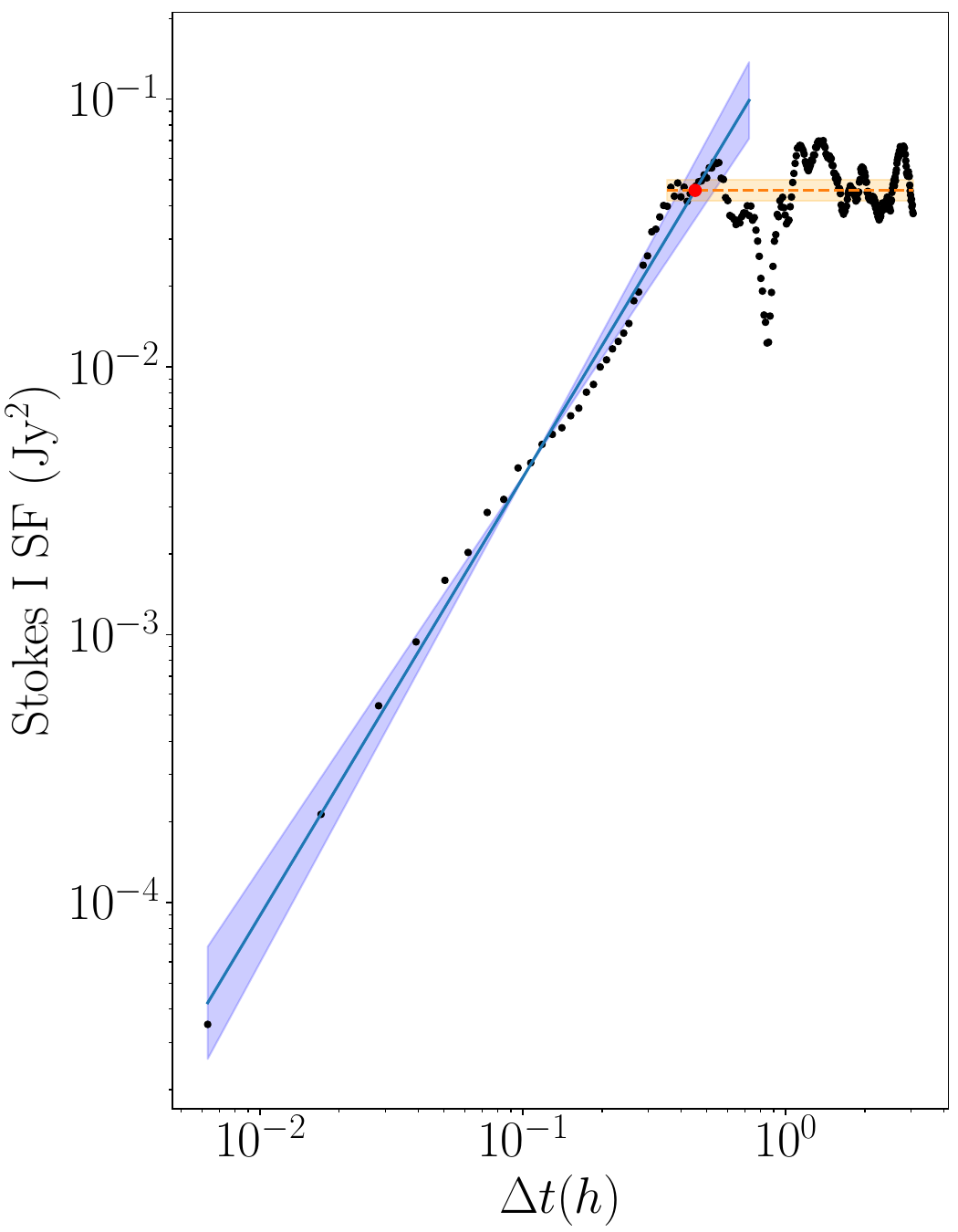}
    \includegraphics[width=4.28cm]{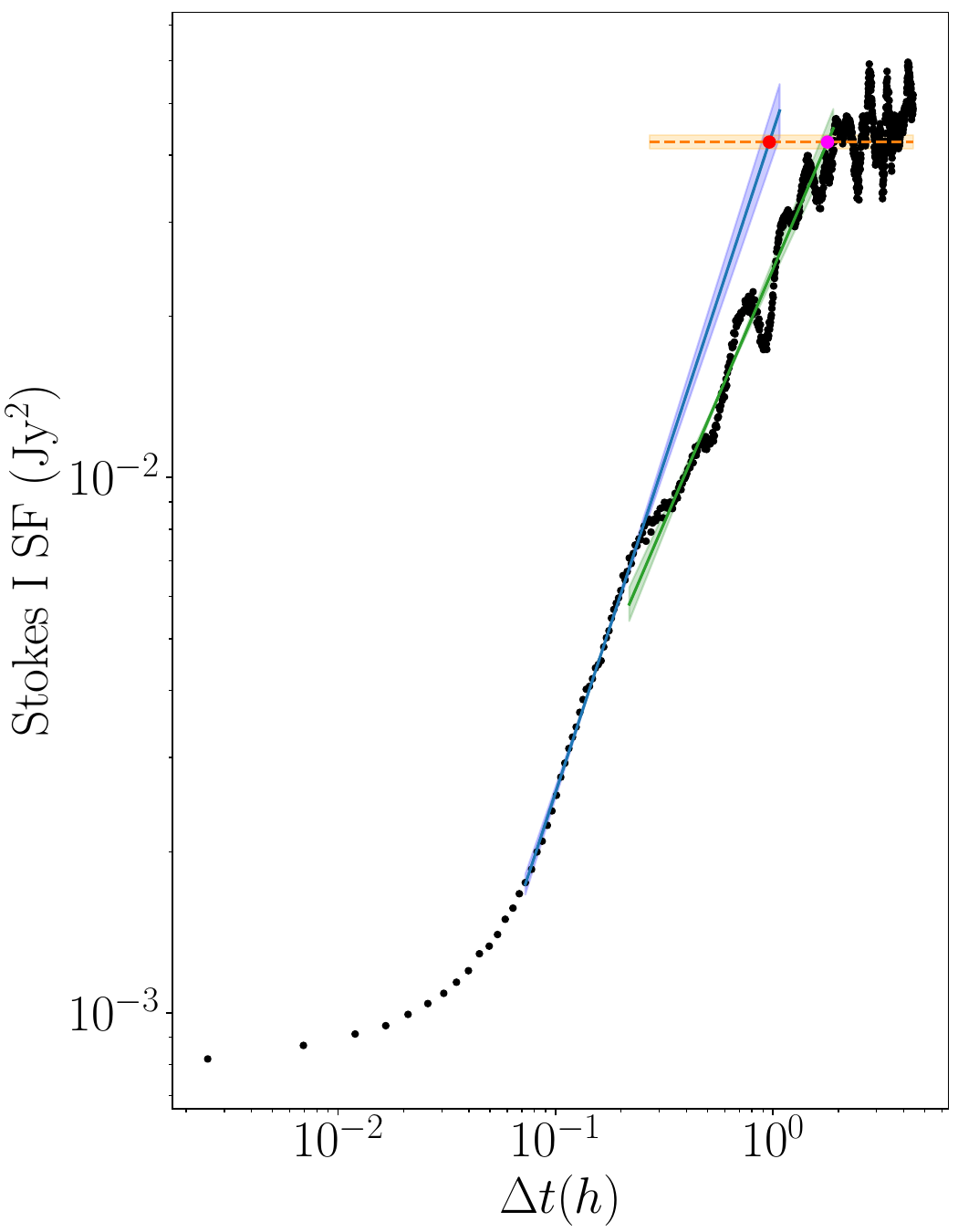}
    \caption{Sample of SF fitted to retrieve the slope (blue line, with an area corresponding to the $3\sigma$ level), the plateau level (orange line) and the timescales (red point, marking the intersection of the slope and the plateau). Left: SF of the April 21 Stokes I denoised light curve. Right: SF of the April 22 Stokes I light curve; we note two distinct slopes, fitted independently (the second slope corresponds to the green line).}
    \label{fig:SF_fit_timescale}
\end{figure}

To compute the timescales of the light curves, we first fitted a slope to the values of the SF at the steep increase time lags, and retrieved the plateau level (as illustrated in some examples in Fig. \ref{fig:SF_fit_timescale}). The intersection of the slope and plateau level gives us the timescales. Since the SF presents a clear plateau at long timescales (despite some oscillations caused by the sampling), these timescales are characteristic of our light curves.

The timescales of the \sgra total flux density (except for April 22) were computed after denoising the light curve. This process involved binning the data using a window approximately two times the observational cadence and fitting a spline to the resulting light curve. This effectively reduced noise while preserving the main features of the signal, mitigating noise that could affect the SF slope estimates for the total flux density. In the SF of the April 22 total flux density, two distinct slopes are observed, revealing two characteristic timescales (see Fig. \ref{fig:SF_fit_timescale}).

Denoising was not applied to the April 22 data because, unlike the other days, the shape of the SF enables a good estimate of the two slopes after removing the time lags affected by white noise in the signal (as shown in Fig. \ref{fig:SF_fit_timescale}). Moreover, denoising would affect the estimate of the first slope, since short time lags are more affected by white noise in our signals, and the presence of two timescales restricts the sampling used to estimate the first slope, making this estimate more sensitive to denoising.

\section{Periodogram results}
\label{sec:Periodogram_results}

To characterize the variability of the full-polarization 2018 \sgra light curves, we employed the  SF  in Sect. \ref{sec:variability}, which yields both the characteristic timescale and the Power Spectral Density (PSD) slope of the light curves.

An alternative method for characterizing the PSD slope of a signal is the Lomb-Scargle periodogram, commonly used in astronomy to account for non-uniform sampling and gaps in time-series data, such as those present in our light curves (see Fig.~\ref{fig:ALMAlcurves}). The Lomb-Scargle periodogram at a given frequency $\omega$ is defined following the formulation stated in \citet{Scargle1982}, as 
\begin{equation*}
    P_X (\omega) = \frac{1}{2} \Bigg\{ \frac{ \left[ \sum_j X_j \cos \omega (t_j - \tau) \right]^2 } { \sum_j X_j \cos^2 \omega (t_j - \tau) } + \frac{ \left[ \sum_j X_j \sin \omega (t_j - \tau) \right]^2 }{ \sum_j X_j \sin^2 \omega (t_j - \tau) } \Bigg\},
\end{equation*}
where $X$ is the physical signal measured at a set of times $t_j$, resulting in  time-series data $\{ X_j=X(t_j), j=1,2, \ldots, N \}$, and $\tau$ is obtained by
\begin{equation*}
    \tan (2\omega \tau) = \frac{\sum_j \sin 2 \omega t_j }{\sum_j \cos 2 \omega t_j }.
\end{equation*}
The periodogram can also be used to identify periodicities within our full-polarization light curves, through statistical analysis of the periodogram distribution evaluated at the set of frequencies $\{ \omega_n = 2 \pi n / \Delta t, n = 1, 2, \ldots, N/2 \}$, where $N$ is the number of data points in the time series, and $\Delta t$ is the time length of the light curve, as proposed in \citet{Scargle1982}.

The estimation of the PSD and the search for periodicities in radio astronomical light curves present challenges, as these datasets are often characterized by non-uniform sampling. Additionally, identifying and assessing the reliability of a periodicity in a light curve is more difficult in the presence of red-noise signals, which are common in many astronomical sources. Unfortunately, these challenges render the estimates of the PSD slopes, obtained after applying the periodogram to our signals, unreliable. This unreliability is evidenced by the inconsistency with the PSD slopes estimated using the SF, as presented in Table \ref{tab:PSD_estimates}. No periodicities were identified in the periodogram distributions of any of our light curves.

\begin{table}[!h]
    \centering
    \small
    \caption{PSD slope estimates of the 2018 ALMA light curves, using the Lomb-Scargle periodogram $P_{\text{LS}}(\omega)$ and the SF.}
    \begin{tabular}{cccccc}
        \toprule
        \multicolumn{2}{c}{Data Set} & PSD ($P_{\text{LS}}(\omega)$) & PSD (SF) &\\
        \midrule
        \multicolumn{4}{c}{Stokes I} \\
        \midrule
        2018 Apr 21 & B1 & $-1.547 \pm 0.057$ & $-2.64 \pm 0.04$ & \\
         & B4 & $-1.573 \pm 0.060$ & $-2.59 \pm 0.03$ & \\
        2018 Apr 22 & B1 & $-1.665 \pm 0.047$ & $-2.235 \pm 0.012$ & \\
         & B4 & $-1.650 \pm 0.048$ & $-2.155 \pm 0.013$ & \\
        2018 Apr 24 & B1 & $-1.729 \pm 0.050$ & $-2.43 \pm 0.03$ & \\
         & B4 & $-1.637 \pm 0.046$ & $-2.33 \pm 0.03$ & \\
        2018 Apr 25 & B1 & $-1.144 \pm 0.055$ & $-2.85 \pm 0.02$ & \\
         & B4 & $-1.231 \pm 0.060$ & $-2.670 \pm 0.019$ & \\
        \midrule
        \multicolumn{4}{c}{Polarized Intensity} \\
        \midrule
        2018 Apr 21 & B1 & $-1.614 \pm 0.054$ & $-2.34 \pm 0.03$ & \\
         & B4 & $-1.688 \pm 0.054$ & $-2.33 \pm 0.02$ & \\
        2018 Apr 22 & B1 & $-1.832 \pm 0.045$ & $-2.529 \pm 0.008$ & \\
         & B4 & $-1.910 \pm 0.049$ & $-2.530 \pm 0.008$ & \\
        2018 Apr 24 & B1 & $-1.595 \pm 0.048$ & $-2.336 \pm 0.015$ & \\
         & B4 & $-1.744 \pm 0.050$ & $-2.341 \pm 0.013$ & \\
        2018 Apr 25 & B1 & $-1.723 \pm 0.064$ & $-2.359 \pm 0.018$ & \\
         & B4 & $-1.718 \pm 0.062$ & $-2.328 \pm 0.016$ & \\
        \bottomrule
    \end{tabular}
    \label{tab:PSD_estimates}
\end{table}

\subsection{The high-pass filter periodogram}

To reduce the sampling-induced noise that negatively impacts PSD estimates in the classic periodogram formulation, we employed the recently proposed HPF periodogram (see \cite{HPFPeriodogram}). This novel periodogram implementation applies a frequency-dependent HPF to the signal, suppressing the variability components of frequencies lower than a given frequency, $\omega$, in the time domain using a data de-trending algorithm before calculating the periodogram value at that frequency.\footnote{The HPF periodogram tool (DOI: \href{https://doi.org/10.5281/zenodo.13917829}{10.5281/zenodo.13917829}) is available for download on GitHub: \url{https://github.com/ealruiz/HPF_Periodogram}.}

The PSD estimates obtained from the HPF periodogram are presented in Table~\ref{tab:SF_HPF_timescales} of the main text, for comparison with the SF results. With this new periodogram implementation, we observe a significant improvement in the results, which are now more consistent with the PSD values obtained from the SF analysis. However, due to the higher noise levels in the total intensity light curves on April 21 and especially April 25, retrieving accurate PSD values remains challenging. This is reflected in the high uncertainty of the estimates and the greater discrepancy with the SF results. Notably, we now observe more stable PSD values across the entire campaign for both total and polarized intensity, compared to the SF estimates, as illustrated in Fig. \ref{fig:SF_HPF_PSD_Times}.

\section{Polarization properties and accretion rate of \sgra}
\label{appendix:sgra_polarization}

The RM and $m^\prime$ observables provide insight into the structure of the Faraday depth across the source \citep[e.g.,][]{Sokoloff1998}. In Table \ref{tab:avg_lcurves_data}, we present the polarized observables averaged for each day. For the total flux density, polarized intensity, EVPA and Stokes V, we report the daily averages for one spectral window (spw) from each ALMA sideband: B1 and B4. Additionally, we provide the average spectral index, RM, and depolarization values, derived from the full-Stokes light curves.

\begin{table*}
    \centering
    \caption{Averaged polarization properties and their measured dispersion across the duration  ($\mu \pm \sigma$) of the ALMA 2018 \sgra light curves.}
    \label{tab:avg_lcurves_data}
    \begin{tabular}{c||cccc||ccc}
    \toprule
    Day & spw & I & P & EVPA & $\alpha$ & RM & Depolarization \\
    (2018) & & (Jy) & (Jy) & (deg.) & & ($10^5$ rad$\cdot$m$^{-2}$) & ($10^{-4}$ GHz$^{-1}$) \\
    \midrule
    April 21 & B1 & $2.89 \pm 0.15$ & $0.10 \pm 0.05$ & $-74 \pm 19$ & $0.0 \pm 0.2$ & $-5.0 \pm 1.4$ & $4.6 \pm 4.6$ \\
    & B4 & $2.89 \pm 0.13$ & $0.12 \pm 0.05$ & $-66 \pm 17$ & & &  \\
    \midrule
    April 22 & B1 & $2.86 \pm 0.15$ & $0.15 \pm 0.05$ & $-94 \pm 40$ & $-0.05 \pm 0.15$ & $-2.7 \pm 1.6$ & $6.0 \pm 6.3$ \\
    & B4 & $2.85 \pm 0.15$ & $0.18 \pm 0.06$ & $-90 \pm 39$ & & &  \\
    \midrule
    April 24 & B1 & $3.24 \pm 0.12$ & $0.17 \pm 0.05$ & $-105 \pm 20$ & $0.05 \pm 0.12$ & $-4.4 \pm 1.7$ & $5.8 \pm 3.6$ \\
    & B4 & $3.25 \pm 0.12$ & $0.18 \pm 0.06$ & $-98 \pm 20$ & & &  \\
    \midrule
    April 25 & B1 & $3.34 \pm 0.07$ & $0.14 \pm 0.04$ & $-122 \pm 22$ & $0.07 \pm 0.15$ & $-6.3 \pm 1.2$ & $3.4 \pm 2.3$ \\
    & B4 & $3.36 \pm 0.08$ & $0.17 \pm 0.05$ & $-114 \pm 21$ & & &  \\
    \bottomrule
    \end{tabular}
\end{table*}

\begin{figure}[!h]
    \centering
    \includegraphics[width=8.5cm]{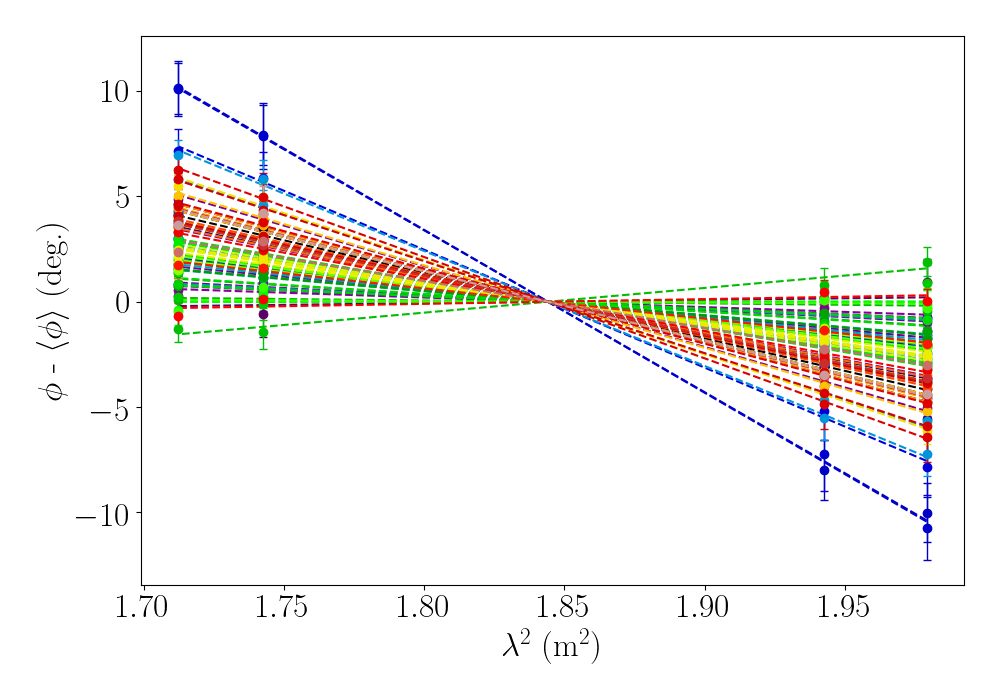}
    \caption{Sample of \sgra\ EVPA measurements across the four ALMA spectral bands (B1–B4) from the four observing days of the 2018 EHT campaign. The EVPA values for each time sample have been mean-subtracted to emphasize the relative variation across frequency. The dashed lines represent the best-fit linear models of the form $\phi(\lambda) = \phi_0 - \mathrm{RM} (\lambda^2 - \lambda_0^2)$, used to estimate the RM evolution curves.}
    \label{fig:EVPA_RMfit}
\end{figure}

Following the methodology described in \cite{Goddi2021}, we estimated that the EVPA values are accompanied by an uncertainty of $\pm$0.2--0.3 degrees. This level of uncertainty, smaller than the EVPA rotation introduced by the measured RM values across the ALMA frequency coverage (see Fig. \ref{fig:ALMAlcurves_products} and Table \ref{tab:avg_lcurves_data}), ensures the robustness of the reported RM evolution curves, as the estimated RM values are not dominated by measurement noise. Figure \ref{fig:EVPA_RMfit} illustrates the frequency dependence of the EVPA following a $\lambda^2$ law, validating the reliability of the derived RM evolution curves. 

The RM values inferred in this study introduce an EVPA rotation of approximately 0.5 to 1 degree within each 2 GHz spectral window, and between 5 to 10 degrees across the full ALMA frequency coverage. This rotation can be visualized from the average EVPA in SPWs B1 and B4 (shown in Table \ref{tab:avg_lcurves_data}). A large EVPA rotation within the observing frequency bandwidth could decrease the measured fractional polarization, resulting in a bandwidth depolarization. Since the EVPA values of \sgra do not exhibit such large rotations, we would expect lower depolarization values. However, the RM can arise from three distinct regions, each with a different spatial scale and, as a result, potentially different variability:
\begin{itemize}
    \setlength{\parskip}{0pt}
    \item internal, hot and turbulent plasma from which synchrotron emission originates. Since Faraday rotation decreases significantly as particles become more energetic \citep{Quataert2000}, this contribution to the RM is sometimes neglected. However, \citet{Wielgus2024} argued that the internal component may be very important in the context of \sgra,
    \item thermal plasma in the vicinity of the black hole, surrounding the emission region, which has a greater impact on the RM and contributes to short-term variability. This region would also introduce stronger depolarization effects,
    \item colder plasma in the most extended and external regions to the black hole, whose effect on RM manifests as long-term variability.
\end{itemize}
Additionally, \citet{Goddi2021} suggests that, beyond bandwidth depolarization caused by strong magnetic fields, which would manifest as a high RM, three other mechanisms may contribute to the depolarization observed in Fig. \ref{fig:ALMAlcurves_products}: Faraday depolarization, beam depolarization, and thermal non-synchrotron emission. The low EVPA rotation due to Faraday rotation observed for \sgra suggests that bandwidth depolarization in \sgra is likely minimal, and that the observed depolarization may instead arise from these alternative mechanisms.

Analyzing both the rotation measure (RM) and depolarization offers valuable insight into the plasma structure in the immediate vicinity of the black hole. As shown in Fig.~\ref{fig:ALMAlcurves}, both quantities exhibit significant variability, further underscored by the large dispersion relative to their daily averages reported in Table~\ref{tab:avg_lcurves_data}. Properly characterizing the plasma properties, however, requires modeling that can disentangle the contributions of Faraday rotation from those of depolarization. While a detailed investigation of the underlying depolarization mechanisms is beyond the scope of this paper, we do explore the RM and depolarization light curves (see Fig.~\ref{fig:ALMAlcurves_products}) using the time-series analysis techniques introduced in this work. The lack of significant correlation between the RM and depolarization curves, as indicated by the correlation function analysis, suggests that bandwidth depolarization is unlikely to be the dominant mechanism.

An analysis of variability timescales using the SF reveals distinct timescales in the April 22 light curves. The RM exhibits a characteristic timescale of $32 \pm 4$ minutes, with a corresponding power spectral density (PSD) slope of $-2.26 \pm 0.016$. This rapid RM variability is consistent with the behavior reported for 2017 data in \citet{Wielgus2024}, and supports the interpretation that Faraday rotation originates from an internal screen co-spatial with the compact synchrotron-emitting region near \sgra.
In contrast, the depolarization light curve displays two prominent timescales: a short one at $15 \pm 2$ minutes (PSD slope $-2.338 \pm 0.013$), and a longer one at $1.16 \pm 0.11$ hours (PSD slope $-2.485 \pm 0.018$). This dual-timescale behavior closely resembles that seen in the SF of the Stokes I light curve (see middle panel of Fig.~\ref{fig:SF_fit_timescale}).
If bandwidth depolarization were the dominant mechanism, one would expect the depolarization and RM timescales to closely match. The presence of two distinct timescales for depolarization instead suggests that additional or alternative mechanisms may be contributing.

\begin{figure*}
    \centering
    \includegraphics[width=0.9\linewidth]{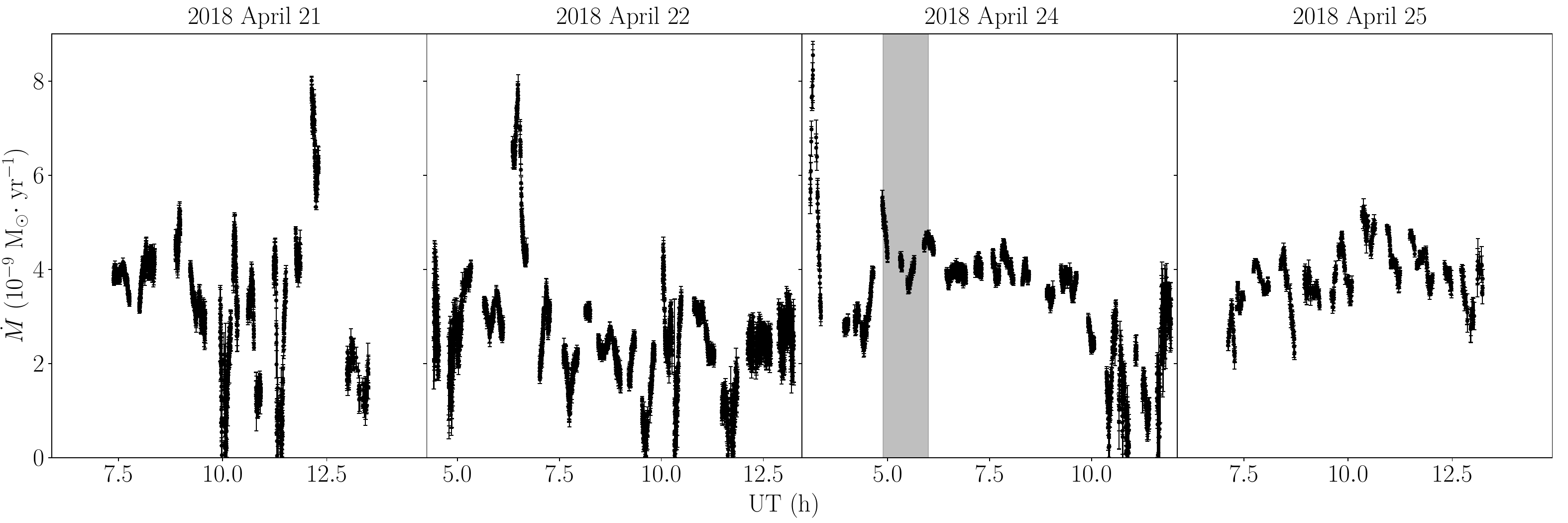}
    \caption{\sgra accretion rate evolution through the ALMA 2018 observations, estimated from the RM light curve presented in Fig. \ref{fig:ALMAlcurves_products}, using the accretion flow model presented in~\cite{Marrone2006}. The gray-shaded band on April 24 marks the time range of the \textit{Chandra} X-ray flare.}
    \label{fig:accRate_ev}
\end{figure*}

Additionally, we estimated the mass accretion rate $\dot{M}$ of \sgra\ (in units of $M_{\odot}\,\mathrm{yr}^{-1}$) using our RM measurements and the expression provided in \citet{Marrone2006}:
\begin{align*}
    \dot{M} = & 2.2\cdot10^{-9} \left[1-\left(\frac{r_\mathrm{out}}{r_\mathrm{in}}\right)^{-(3\beta-1)/2}  \right]^{-2/3} \\ & \times \left(\frac{M_{BH}}{6.6\cdot10^9M_\odot} \right)^{4/3} \left(\frac{2}{3\beta-1} \right)^{-2/3} r_\mathrm{in}^{7/6} RM^{2/3},
\end{align*}
where $\beta$ is a parameter that depends on the accretion flow model, ranging between $1/2$ and $3/2$;
$r_\mathrm{in}$ and $r_\mathrm{out}$ represent the inner and outer edges of the Faraday screen (in units of Schwarzschild radius, $r_s$);
$M_{BH}$ is the mass of \sgra, expressed in $M_{\odot}$;
and RM is given in units of rad$\cdot$m$^{-2}$. 
We adopted $r_\mathrm{in} = 3\, r_s$, following the estimated angular size of the emission region reported in \citet{EHTC2022a}. For simplicity, we took $r_\mathrm{out} \rightarrow \infty$, as its exact value is poorly constrained and has negligible impact on the accretion rate estimate under these assumptions.
To derive an upper limit on $\dot{M}$, we use the average RM values measured across the four observing days, ranging from $-3 \times 10^5$ to $-5 \times 10^5$ rad$\cdot$m$^{-2}$ (see Table~\ref{tab:avg_lcurves_data}), and assume $\beta = 3/2$. This yields an accretion rate of $\dot{M} \approx (2.6$–$3.9) \times 10^{-9}\ M_{\odot} \,\mathrm{yr}^{-1}$, which is consistent with expectations from magnetically arrested disk (MAD) models for \sgra\ \citep[e.g.,][]{SgraP5}. The evolution of $\dot{M}$ estimated from the RM values at different integration times is shown in Fig.~\ref{fig:accRate_ev}.

A more comprehensive analysis, including comparisons with theoretical models and more sophisticated treatments of the Faraday screen geometry, is required to further constrain the plasma conditions around \sgra\ and to better understand the structure of the Faraday-active region near the Galactic Center black hole.

\section{Full polarization 2017 ALMA light curves}
\label{sec:2017lcurves}

\begin{figure*}[!h]
    \centering
    \includegraphics[width=16.1cm]{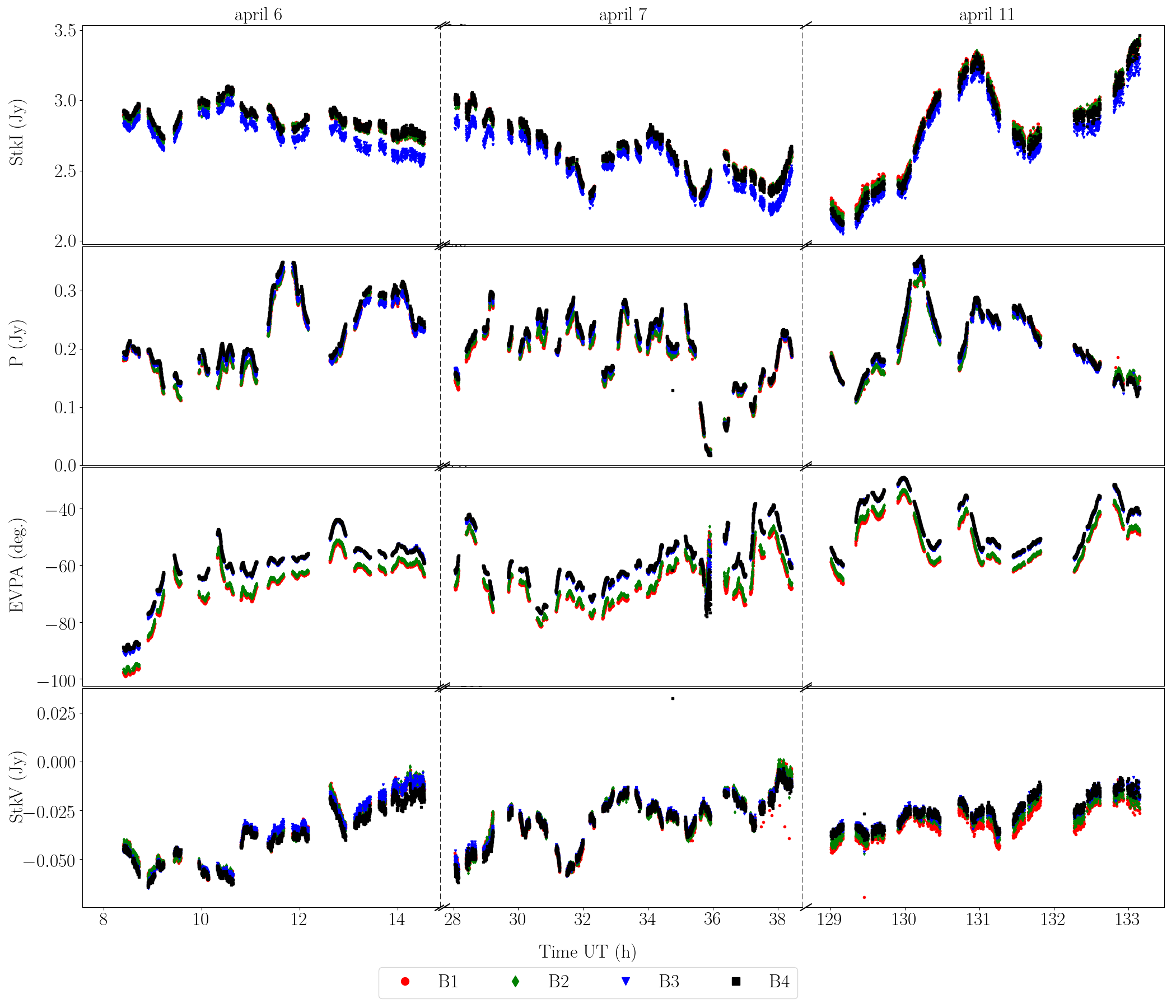}
    \caption{\sgra ALMA light curves of Stokes I, the polarized intensity, the EVPA, and Stokes V (from top to bottom) for the four spectral bands, for all three days (from left to right, 2017 April 6, 7, and 11). Stokes V light curves are tentative, as the detected levels fall below ALMA’s guaranteed CP accuracy.}
    \label{fig:ALMA2017lcurves}
\end{figure*}

The \sgra 2017 light curves from the ALMA observations conducted on April 6, 7, and 11 as part of the 2017 EHT campaign, originally published in \citet{Wielgus2022,Wielgus2022_orbital, Wielgus2024}, were also reprocessed following the updated intra-field calibration presented in Sect. \ref{subsec:pipeline1}. We retrieved the full-polarization \sgra light curves, for comparison with the 2018 light curves. Figure \ref{fig:ALMA2017lcurves} shows the complete ALMA light curves of the total flux, polarized intensity, the EVPA, and Stokes V. A summary of the main characteristics of these light curves is given in Table \ref{tab:2017_lcurves_data}.
The presented results demonstrate a high degree of consistency with the original intra-field reduction A1, described in \citet{Wielgus2022}, although the flux density is higher, as a result of an updated QA2 with different calibrators.
The modulation indices indicate similar variability, further supported by the PSD estimates and the timescales derived from the SF and periodogram analyses (see Fig. \ref{fig:SF_HPF_PSD_Times}).

\begin{table*}[!h]
\centering
\caption{ALMA 2017 \sgra light curves,  in total flux, polarized intensity, and Stokes V.}
\label{tab:2017_lcurves_data}
    \resizebox{\textwidth}{!}{
    \begin{tabular}[c]{cccccccccc}
    \toprule
    Day & UT & Duration & Parameter & \multicolumn{2}{c}{Spectral Window} & Samples & Flux (Jy) & Modulation & max-min \\
    (2017) & (h) & (h) & & label & Frequency (GHz) & & & & (Jy) \\
    \midrule
    April 6 & 8:24 - 14:33 & 6.15 & Stokes I & B1 & $212.1 - 214.1$ & 2324 & $2.88 \pm 0.13$ & 0.044 & 0.41 \\
    & & & & B2 & $ 214.1 - 216.1 $ & 2331 & $2.87 \pm 0.12$ & 0.043 & 0.41 \\
    & & & & B3 & $ 226.1 - 228.1 $ & 2319 & $2.89 \pm 0.13$ & 0.044 & 0.48 \\
    & & & & B4 & $ 228.1 - 230.1 $ & 2329 & $2.89 \pm 0.13$ & 0.044 & 0.42 \\
    
    & & & P & B1 & $212.1 - 214.1$ & 2324 & $0.22 \pm 0.06 $ & 0.27 & 0.22 \\
    & & & & B2 & $ 214.1 - 216.1 $ & 2331 & $0.22 \pm 0.06$ & 0.27 & 0.22 \\
    & & & & B3 & $ 226.1 - 228.1 $ & 2319 & $0.22 \pm 0.05$ & 0.25 & 0.21 \\
    & & & & B4 & $ 228.1 - 230.1 $ & 2329 & $0.23 \pm 0.06$ & 0.25 & 0.22 \\
    
    & & & Stokes V & B1 & $212.1 - 214.1$ & 2324 & $-0.037 \pm 0.015$ & 0.41 & 0.055 \\
    & & & & B2 & $ 214.1 - 216.1 $ & 2331 & $-0.037 \pm 0.015$ & 0.40 & 0.056 \\
    & & & & B3 & $ 226.1 - 228.1 $ & 2319 & $-0.037 \pm 0.016$ & 0.42 & 0.058 \\
    & & & & B4 & $ 228.1 - 230.1 $ & 2329 & $-0.040 \pm 0.015$ & 0.38 & 0.053 \\
    
    \midrule
    April 7 & 4:02 - 14:25 & 10.37 & Stokes I & B1 & $212.1 - 214.1$ & 3771 & $2.61 \pm 0.18$ & 0.069 & 0.73 \\
    & & & & B2 & $ 214.1 - 216.1 $ & 3765 & $2.61 \pm 0.18$ & 0.068 & 0.73 \\
    & & & & B3 & $ 226.1 - 228.1 $ & 3721 & $2.61 \pm 0.18$ & 0.069 & 0.71 \\
    & & & & B4 & $ 228.1 - 230.1 $ & 3727 & $2.62 \pm 0.19$ & 0.071 & 0.75 \\
    
    & & & P & B1 & $212.1 - 214.1$ & 3771 & $0.18 \pm 0.06$ & 0.31 & 0.27 \\
    & & & & B2 & $ 214.1 - 216.1 $ & 3765 & $0.18 \pm 0.06$ & 0.31 & 0.27 \\
    & & & & B3 & $ 226.1 - 228.1 $ & 3721 & $0.19 \pm 0.06$ & 0.31 & 0.28 \\
    & & & & B4 & $ 228.1 - 230.1 $ & 3727 & $0.20 \pm 0.06$ & 0.33 & 0.29 \\
    
    & & & Stokes V & B1 & $212.1 - 214.1$ & 3771 & $-0.030 \pm 0.013$ & 0.43 & 0.058 \\
    & & & & B2 & $ 214.1 - 216.1 $ & 3765 & $-0.030 \pm 0.013$ & 0.43 & 0.060 \\
    & & & & B3 & $ 226.1 - 228.1 $ & 3721 & $-0.029 \pm 0.012$ & 0.41 & 0.056 \\
    & & & & B4 & $ 228.1 - 230.1 $ & 3727 & $-0.030 \pm 0.012$ & 0.41 & 0.059 \\

    \midrule
    April 11 & 9:00 - 14:03 & 2.82 & Stokes I & B1 & $212.1 - 214.1$ & 1378 & $2.73 \pm 0.33$ & 0.119 & 1.15 \\
    & & & & B2 & $ 214.1 - 216.1 $ & 1377 & $2.72 \pm 0.33$ & 0.121 & 1.16 \\
    & & & & B3 & $ 226.1 - 228.1 $ & 1370 & $2.69 \pm 0.34$ & 0.125 & 1.20 \\
    & & & & B4 & $ 228.1 - 230.1 $ & 1362 & $2.69 \pm 0.34$ & 0.125 & 1.19 \\
    
    & & & P & B1 & $212.1 - 214.1$ & 1378 & $0.21 \pm 0.05$ & 0.26 & 0.23 \\
    & & & & B2 & $ 214.1 - 216.1 $ & 1377 & $0.21 \pm 0.05$ & 0.26 & 0.23 \\
    & & & & B3 & $ 226.1 - 228.1 $ & 1370 & $0.21 \pm 0.06$ & 0.26 & 0.24 \\
    & & & & B4 & $ 228.1 - 230.1 $ & 1362 & $0.22 \pm 0.06$ & 0.27 & 0.25 \\
    
    & & & Stokes V & B1 & $212.1 - 214.1$ & 1378 & $-0.033 \pm 0.007$ & 0.20 & 0.032 \\
    & & & & B2 & $ 214.1 - 216.1 $ & 1377 & $-0.030 \pm 0.007$ & 0.23 & 0.031 \\
    & & & & B3 & $ 226.1 - 228.1 $ & 1370 & $-0.028 \pm 0.006$ & 0.23 & 0.029 \\
    & & & & B4 & $ 228.1 - 230.1 $ & 1362 & $-0.027 \pm 0.007$ & 0.25 & 0.030 \\
    
    \bottomrule
    \end{tabular}
    }
    \tablefoot{Stokes V reported values are tentative, as the detected levels fall below ALMA’s guaranteed CP accuracy.}
\end{table*}

\clearpage
\end{appendix}

\end{document}